\documentclass[10pt,twocolumn,aps,prb,balancelastpage,longbibliography]{revtex4-1}
\usepackage[latin9]{inputenc}
\usepackage{color}
\usepackage{verbatim}
\usepackage{amsmath}
\usepackage{amssymb,ulem}
\usepackage{graphicx,cellspace,booktabs}
\usepackage{esint}
\usepackage[unicode=true,pdfusetitle,
 bookmarks=true,bookmarksnumbered=false,bookmarksopen=false,
 breaklinks=false,pdfborder={0 0 1},backref=false,colorlinks=true]{hyperref}
\usepackage{breakurl}
\usepackage{epstopdf}
\usepackage{feynmp-auto}
\usepackage{youngtab}
\usepackage{ytableau}

\makeatletter

\@ifundefined{textcolor}{}
{%
 \definecolor{BLACK}{gray}{0}
 \definecolor{WHITE}{gray}{1}
 \definecolor{RED}{rgb}{1,0,0}
 \definecolor{GREEN}{rgb}{0,1,0}
 \definecolor{BLUE}{rgb}{0,0,1}
 \definecolor{CYAN}{cmyk}{1,0,0,0}
 \definecolor{MAGENTA}{cmyk}{0,1,0,0}
 \definecolor{YELLOW}{cmyk}{0,0,1,0}
}

\usepackage{subfigure}\usepackage{bm}

\newcommand{\ket}[1]{|#1\rangle}

\newcommand{\bv}[1]{\mathbf{#1}}

\renewcommand{\(}{\left(}
\renewcommand{\)}{\right)}

\def\l@subsubsection#1#2{}
\makeatother

\begin{document}

\title{The Higgs Mechanism in Higher-Rank Symmetric $U(1)$ Gauge Theories}

\author{Daniel Bulmash}
\author{Maissam Barkeshli}

\affiliation{Condensed Matter Theory Center and Joint Quantum Institute, Department of Physics, University of Maryland, College Park, Maryland 20472 USA}

\date{\today}
\begin{abstract}

We use the Higgs mechanism to investigate connections between higher-rank symmetric $U(1)$ 
gauge theories and gapped fracton phases. We define two classes of rank-2 symmetric $U(1)$ gauge theories:
the $(m,n)$ scalar and vector charge theories, for integer $m$ and $n$, which respect the symmetry of the square (cubic)
lattice in two (three) spatial dimensions. We further provide local lattice rotor models whose low energy dynamics are described
by these theories. We then describe in detail the Higgs phases obtained when the $U(1)$ gauge symmetry is spontaneously broken to
a discrete subgroup. A subset of the scalar charge theories indeed have X-cube fracton order as their Higgs phase, although we find that
this can only occur if the continuum higher rank gauge theory breaks continuous spatial rotational symmetry. 
However, not all higher rank gauge theories have fractonic Higgs phases; other Higgs phases possess conventional topological
order. Nevertheless, they yield interesting novel exactly solvable models of conventional topological order, somewhat reminiscent of
the color code models in both two and three spatial dimensions. We also investigate phase transitions in these models and 
find a possible direct phase transition between four copies of $\mathbb{Z}_2$ gauge theory in three spatial dimensions and X-cube fracton order.
\end{abstract}
\maketitle

\tableofcontents

\section{Introduction}
\label{sec:intro}

The recent discovery of ``fracton" phases of matter\cite{ChamonGlass,BravyiChamonModel,HaahsCode,YoshidaFractal,VijayFractons,VijayNonAbelianFractons} 
has led to considerable recent research provided a new class of three-dimensonal gapped phases of matter beyond conventional topological order, and which apparently cannot be described using standard gauge
field theories. In addition to subextensive ground state degeneracy on topologically nontrivial manifolds, fracton phases 
are defined by possessing excitations whose motion is restricted to subdimensional manifolds. These theories have prompted a great deal of recent excitement\cite{BravyiHaahSelfCorrection, SivaBravyiMemory, EmergentPhasesFractons, ShiFractonEntanglement, FractonEntanglement, RecoverableInformation, FractonCorrFunctions}.

Recently, certain gapless versions of fracton phases have also been found in terms of higher-rank symmetric $U(1)$ gauge 
theory\cite{XuFractons1,XuFractons2,RasmussenFractons, PretkoSubdimensional, PretkoElectromagnetism,PretkoWitten}. These theories
possess gapless ``photon'' modes, together with ``matter'' whose motion is confined to subdimensional manifolds.\cite{PretkoSubdimensional}
These higher-rank symmetric gauge theories have received little attention in the field theory literature, as they 
inherently break Lorentz symmetry, although some of them are closely related to studies of Lifshitz gravity.\cite{XuFractons1,XuFractons2,XuHorava}

While fracton phases share some features of conventional topological order,
 they appear to require geometric data in addition to topological data\cite{PretkoElasticity,GromovElasticity,SlagleGenericLattices,ShirleyXCubeFoliations}. Constructions involving gauging subsystem symmetries\cite{VijayGaugedSubsystem,Williamson2016}, layering two-dimensional conventional topological order\cite{VijayLayer,MaLayer,RegnaultLayer}, coupled 
chains\cite{HalaszFractons}, or partons\cite{HsiehFractonsPartons} are known to produce fracton phases, but it is unclear whether these 
relationships are essential or incidental. In particular, the general relationships between gapped fracton phases, gapless higher-rank 
gauge theories, and conventional topological order are not well-understood. In order to understand these relationships, one natural approach 
is to consider the Higgs mechanism in the higher rank $U(1)$ theories, since the Higgs mechanism is known to relate gapless phases 
of conventional gauge theories to gapped, topologically ordered ones. 

An intriguing feature of gapped fracton phases is that it is also unclear whether, and to what extent, they can be described using continuum field theories. 
A natural guess is that a Higgs phase of a continuum higher rank gauge theory might provide such a continuum field theoretic description. 
Recently, Ref. \onlinecite{SlagleFieldTheory} found a field theoretic representation of the X-Cube model\cite{VijayGaugedSubsystem}, which is a
an example of a gapped fracton model. This further raises the question of whether such a field theory can be related to a Higgs phase of a gapless higher rank gauge theory. 

In this paper, we present a general analysis of a wide class of higher-rank symmetric $U(1)$ lattice gauge theories and their Higgs phases. This helps us
elucidate the relation between fracton orders, conventional topological order, and gapless higher-rank symmetric gauge theories. Our results are briefly 
summarized below. 

\subsection{Structure of paper and summary of main results}

We begin in Sec. \ref{sec:U1theories} by defining a set of rank-2 symmetric lattice gauge theories which have the symmetry of the square (cubic) lattice in $d=2$ ($d=3$). 
These theories are defined by a set of rotor variables $A_{ij}({\bf r}) \sim A_{ij}({\bf r}) + 2\pi$ on the sites and faces of a square (cubic) lattice. The conjugate
momenta are the ``electric fields'' $E_{ij}({\bf r})$. There are two classes of theories, the $(m,n)$ scalar and $(m,n)$ vector charge theories, where $m$ and $n$
are relatively prime integers. These theories are defined by their Gauss' Law constraints, which induce a set of gauge transformations, summarized in Table \ref{tab:theoryList}.
The scalar and vector charge theories studied in Refs. \onlinecite{XuFractons1,XuFractons2,RasmussenFractons, PretkoSubdimensional, PretkoElectromagnetism,PretkoWitten}, 
whose continuum limit is invariant under continuous spatial rotations, correspond in our notation to the $(1,1)$ scalar charge and $(2,1)$ vector charge theories.  

\begin{table*}
\renewcommand{\arraystretch}{1.3}
\begin{tabular}{@{}lclcl}
\toprule[2pt]
\textbf{Theory} & \phantom{ab} & \textbf{Gauss' Law} & \phantom{ab} & \textbf{Gauge transformation}  \\
\hline
$(m,n)$ scalar && $m \sum_i \Delta_i^2 E_{ii} + n \sum_{i\neq j} \Delta_i \Delta_j E_{ij} = \rho$ && $A_{ij} \rightarrow A_{ij} - \begin{cases}
m \Delta_i^2 \alpha & i=j\\
n \Delta_i \Delta_j \alpha & i \neq j
\end{cases} $ \\
\vspace{-6pt}\\
$(m,n)$ vector && $m\Delta_j E_{jj} + 2n\sum_{i\neq j}\Delta_i E_{ij} = \rho_j$ && $A_{ij} \rightarrow A_{ij}- \begin{cases}
m\Delta_i \alpha_i & i=j\\
n(\Delta_i \alpha_j + \Delta_j \alpha_i) & i \neq j
\end{cases}$ \\
	\bottomrule[2pt]	
\end{tabular}
\caption{Gauss' Laws and gauge transformations for $U(1)$ rank-2 symmetric gauge theories that obey the symmetry of the square (cubic) lattice. $\Delta_i$ denotes a lattice derivative
in the $i$ direction. The normalization has been chosen so that in the compact theories, $(m,n)$ relatively prime integers leads to integer charge lattices.  The $(1,1)$ scalar charge and $(2,1)$ vector charge theories have continuous rotational invariance.}
\label{tab:theoryList}
\end{table*}
Section \ref{sec:U1theories} also provides details of these theories and presents local lattice rotor models whose 
low-energy subspace is the gauge theory coupled to charge-$p$ matter fields. With the exception of the 
$(0,1)$ scalar charge theory in $d=2$, the $(1,0)$ and $(0,1)$ vector charge theories in $d = 2$ and $d = 3$, these models yield
a well-defined class of field theories. 

An important question is the extent to which these gapless theories are stable to arbitrary perturbations. 
The models that we consider correspond to compact $U(1)$ higher-rank gauge theories, and thus 
an understanding of stability requires ruling out non-perturbative instanton processes. We expect that the 
$d = 2$ theories that we consider are unstable to proliferation of instantons, and are thus not stable phases.
As such, they may be thought of as multi-critical points, and it is an interesting question to understand how many relevant
operators exist at their respective fixed points. In $d = 3$, most of the theories do appear to correspond to stable phases. 
In particular, the $(m,n)$ vector charge models are self-dual for $m,n>0$, which, using extensions of arguments for the $(2,1)$ theory\cite{XuFractons1,XuFractons2,RasmussenFractons}, implies stability of the theory. Similarly, the $(1,0)$ scalar charge 
theory also has a self-duality and is stable. On the other hand, the other $(m,n)$ scalar 
charge theories do not have a clear self-duality. Nevertheless, consideration of the magnetic sector of these theories 
suggests that they are stable for $m,n > 0$, although we leave a detailed analysis for future work. 
The stability of the $(0,1)$ scalar charge theory in $d= 3$ also requires further study, as it does possess non-trivial 
instanton process whose relevance must be carefully analyzed. 

In Section \ref{sec:intuition} we explain the physical intuition for how the Higgs mechanism 
affects subdimensional charges in the $U(1)$ theories. Specifically, we demonstrate how condensation of charge $p$ excitations in the $(1,1)$ scalar and $(2,1)$ vector charge theories necessarily renders the charged excitations fully mobile in all directions. Thus the Higgs phases of such theories possess conventional topological order, described by a conventional discrete gauge theory. 

We then explicitly study the Higgs transitions and phase diagrams for general $(m,n)$ scalar and vector charge theories in Secs. \ref{sec:Scalar2DHiggs}-\ref{sec:evenOddVector_d3}. Our main results for the topological order of the Higgs phases are summarized in Table \ref{tab:HiggsPhases}. In particular, we find that in $d=3$, the $(2r, 2s+1)$ scalar charge theories, for $r, s \geq 0$, 
yield the X-cube fracton phase\cite{VijayGaugedSubsystem} upon condensation of charge 2 particles. This is a remarkable result, as we see that the $(2r, 2s+1)$ scalar charge theories 
form a class of gapless higher rank gauge theories that appear to be stable phases of matter (at least for $r > 0$), and thus can be thought of as gapless ``parent'' phases of the 
X-cube fracton phases. As seen in  Table \ref{tab:HiggsPhases}, the Higgs phases of many of the theories that we study simply 
reduce to conventional topological order; nevertheless, in several cases they correspond to exactly solvable models of topological order 
that are quite different from the usual well-known toric code models, and thus they may be of independent interest. In particular,
these models are reminiscent of the color codes.\cite{ColorCodes}  

We further find that the $(2r+1,2s+1)$ scalar and vector Higgsed charge models, for integer $r$ and $s$, in $d=3$ have 
particularly interesting phase diagrams; we study them in Sections \ref{sec:oddOddScalar_d3} and \ref{sec:oddOddVector_d3} 
respectively. The $(2r+1,2s+1)$ scalar charge theory has $\mathbb{Z}_2^4$ topological order in its Higgs phase. 
Upon adding a certain strong Zeeman field, the effective Hamiltonian is the X-cube model; this raises the interesting possibility of a direct 
confinement-like transition between conventional topological order and fracton order in three spatial dimensions. The Higgs phase 
of the $(2r+1,2s+1)$ vector charge theory has $\mathbb{Z}_2^7$ topological order and 
can be driven to $\mathbb{Z}_2$ topological order in a suitable strong Zeeman field limit.

\begin{table*}
\renewcommand{\arraystretch}{1.3}
\begin{tabular}{@{}lclclclcl@{}}
\toprule[2pt]
$U(1)$ \textbf{Charge Type} & \phantom{ab} &$(m,n)$ & \phantom{ab}  & \textbf{Higgs Phase}\\
\hline
$d=2$ scalar \\
&& ($2r+1$, $2s+1$) && $\mathbb{Z}_2^3$ topological order\\
&& $(2r,2s+1)$ && Trivial\\
&& $(2r+1, 2s+2)$ && Trivial\\
&& $(1,0)$ && $\mathbb{Z}_2^4$ topological order\\
$d=2$ vector \\
&& ($2r+1$, $2s+1$) && $\mathbb{Z}_2^3$ topological order\\
&& ($2r+2$, $2s+1$) && $\mathbb{Z}_2^4$ topological order\\
&& ($2r+1$, $2s$) && Trivial\\
&& $(0,1)$ && Trivial\\
$d=3$ scalar \\
&& ($2r+1$, $2s+1$) && $\mathbb{Z}_2^4$ topological order\\
&& $(2r,2s+1)$ && X-Cube fracton order \\
&& $(2r+1, 2s+2)$ && Trivial \\
&& $(1,0)$ && $\mathbb{Z}_2^8$ topological order\\
$d=3$ vector \\
&& ($2r+1$, $2s+1$) && $\mathbb{Z}_2^7$ topological order\\
&& ($4r+2$, $2s+1$) && $\mathbb{Z}_2$ topological order\\
&& ($4r$, $2s+1$) && Trivial\\
&& ($2r+1$, $2s$) && Trivial\\
	\bottomrule[2pt]	
\end{tabular}
\caption{List of $\mathbb{Z}_2$ Higgs phases. Here $r, s \geq 0$ are nonnegative integers; the nomenclature of the $U(1)$ theories is explained in Sec. \ref{sec:U1theories}. The labeling $\mathbb{Z}_2^n$ in $d$ spatial dimensions means that the topological order is $n$ copies of the $d$-dimensional $\mathbb{Z}_2$ toric code. By trivial, we mean that the phase is topologically trivial; interesting patterns of global
symmetry breaking may be present, as we discuss. }
\label{tab:HiggsPhases}
\end{table*}

\section{Higher-Rank Symmetric $U(1)$ Lattice Gauge Theories}
\label{sec:U1theories}

In this section, we carefully define the set of rank-2 symmetric $U(1)$ lattice gauge theories. The set of theories with continuous rotational symmetry is known\cite{RasmussenFractons}, but for later purposes we will also write down the taxonomy of rank-2 theories respecting the symmetries of the square (cubic) lattice in $d=2$ ($d=3$).

\subsection{General Setup}
\label{subsec:ClassesOfTheories}

\begin{figure}
\subfigure[]{
	\includegraphics[width=5cm]{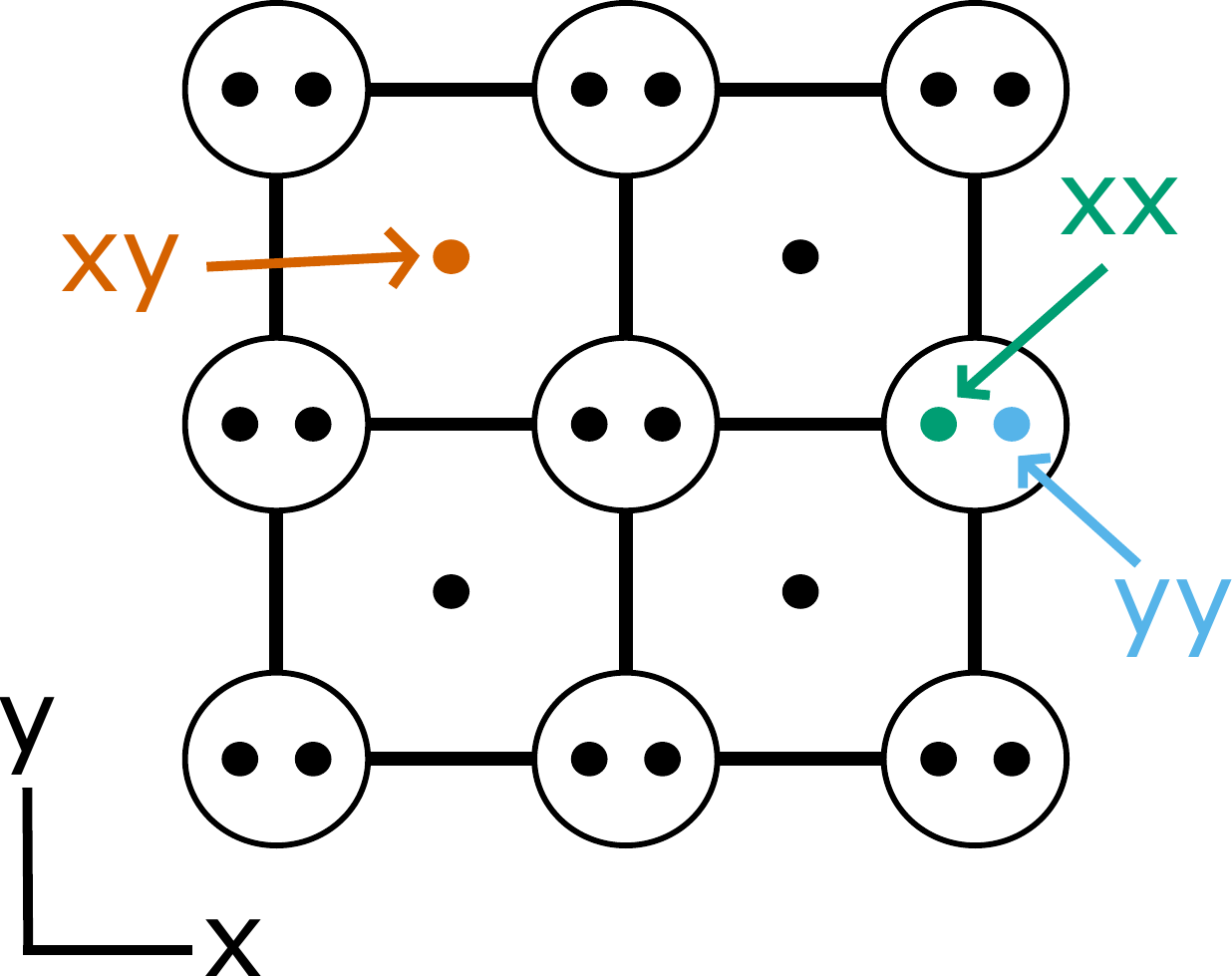}
	\label{fig:2DLatticeSetup}
}
\subfigure[]{
	\includegraphics[width=5cm]{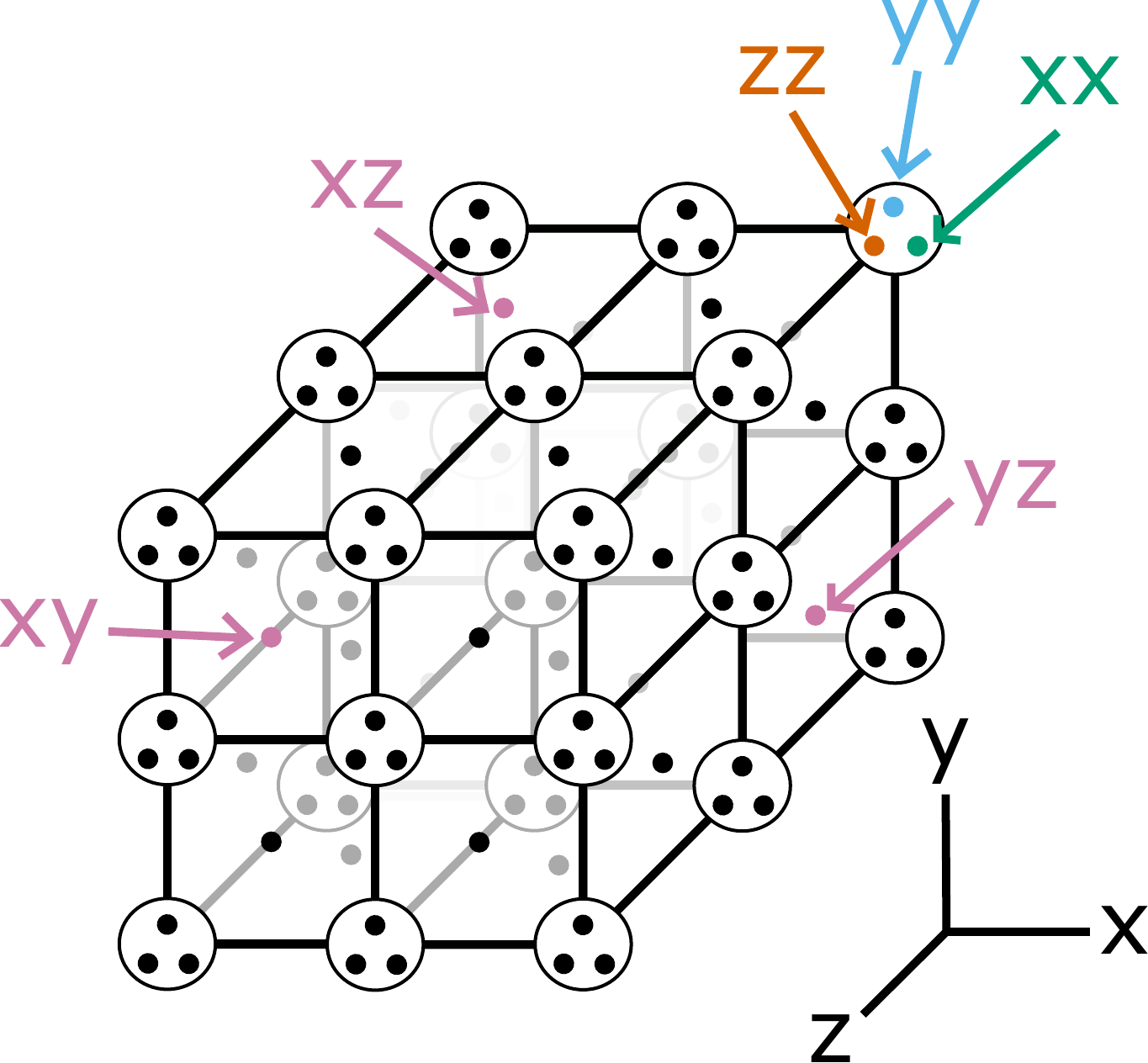}
	\label{fig:3DLatticeSetup}
}
\caption{Arrangement of the gauge field degrees of freedom for a $U(1)$ symmetric lattice gauge theory in (a) $d=2$ and (b) $d=3$. Each dot is a rotor variable, and the circles indicate that $d$ spins live on each site.}
\label{fig:latticeSetups}
\end{figure}

The starting point of these gauge theories is a set of gauge field variables $A_{ij}$, where we define $A$ to be symmetric so that $A_{xy}$ and $A_{yx}$ are just relabelings of the same degree of freedom. The construction may be defined on a lattice, in which case the diagonal components $A_{ii}$ live on the sites of the lattice and the off-diagonal components $A_{ij}$ for $i \neq j$ live on the faces of the square (cubic) lattice in $d=2$ ($d=3$), as shown in Fig. \ref{fig:latticeSetups}. Alternatively, the theory may be constructed in the continuum. We will generally use lattice notation, but we will specify when the lattice is important.

We then define a symmetric tensor $E_{ij}$ of momenta such that
\begin{align}
[A_{ij}(\bv{r}),E_{kl}(\bv{r}')] &= \frac{i}{2}\delta_{\bv{r},\bv{r}'}\(\delta_{ik}\delta_{jl}+\delta_{il}\delta_{jk}\) \label{eqn:comms}
\end{align}

$A_{ij}$ and $E_{ij}$ to both transform as tensors under spatial rotations, that is, (for example) 
$E_{ij} \rightarrow \sum_{k,l} R_{ik}R_{jl}E_{kl}$ where $R$ is a rotation matrix. The form of Eq. \eqref{eqn:comms} 
guarantees rotational invariance of the commutation relations. The factor of $1/2$ ensures that 
$A_{ii}$ is canonically conjugate to $E_{ii}$: $[A_{ii},E_{ii}] = i$. However, for $i \neq j$, 
we have $[A_{ij},E_{ij}] = i/2$. This non-standard normalization for canonical conjugates must be tracked carefully in what follows. 

Next, we demand the existence of local Gauss' Law constraints, which take the form:
\begin{equation}
G_a(E;\bv{r}) = \rho_a(\bv{r}),
\end{equation}
where $G_a$ is some local linear differential operator acting on the $E_{ij}$ tensor and $\rho_a$ is the corresponding matter field. 
The subscript $a$ allows for multiple Gauss' Laws corresponding to multiple species of charges. 

Finally, we compute the gauge transformation rule arising from the Gauss' Law constraint. This is done most clearly in the zero-charge sector, where we demand that a state $\ket{A_{ij}}$ which is an eigenstate of all the $A_{ij}$ operators obey
\begin{equation}
e^{i\alpha_a(\bv{r}) G_a(E;\bv{r})}\ket{A_{ij}} = \ket{A_{ij}}
\end{equation}
for each $a$, each $\bv{r}$, and arbitrary $\alpha_a$. But $e^{i\alpha_a(\bv{r})E_{ij}(\bv{r})}$ shifts the eigenvalue of $A_{ij}(\bv{r})$ according to the commutation relations Eq. \eqref{eqn:comms}, which means that the two configurations of $A_{ij}$ related by $e^{i\alpha G_a}$ must represent the same state, i.e. these states are gauge-equivalent. 

\subsection{Hamiltonians}

The general structure of the Hilbert space and the Hamiltonians we consider is as follows. The Hilbert space of the theory consists of two pieces. First, we have the rotor variables $A_{ij} \sim A_{ij} + 2\pi$ and their conjugates, as discussed in Sec. \ref{subsec:ClassesOfTheories}. Second, we have a set of compact rotor variables $\theta_a \sim \theta_a + 2\pi$ and their canonical conjugates $L_a$ with commutation relations
\begin{equation}
[\theta_a(\bv{r}), L_b(\bv{r}')] = i\delta_{\bv{r},\bv{r}'}\delta_{ab}
\end{equation}
The values of $a$ and the locations of the $\theta$ variables will depend on the theory, but in general there is one value of $a$ per Gauss' Law constraint. We defer further discussion until talking about the individual theories.

The Hamiltonian for every theory will break down into the same basic structure, identical to the structure of the usual lattice $U(1)$ gauge theory, but the form of each term will depend on the theory.
\begin{equation}
H = H_{Maxwell}+H_{Higgs}+H_{Gauss}
\label{eqn:HStructure}
\end{equation}

The Maxwell term provides dynamics for the gauge field $A_{ij}$ and electric field $E_{ij}$ degrees of freedom. Its form is
\begin{align}
H_{Maxwell} = \sum_{\bv{r},i} &\(\tilde{h}_{s} E_{ii}^2 - \frac{1}{g_s^2}\cos(B_{ii})\) \nonumber \\ 
&+ \sum_{\bv{r},i<j} \(\tilde{h}_{f} E_{ij}^2 - \frac{1}{g_{f}^2}\cos(B_{ij})\)
\label{eqn:MaxwellTerm}
\end{align}
where $B_{ij}$ is the simplest (i.e. contains the fewest derivatives) combination of the $A_{ij}$ which is gauge invariant, 
and $\tilde{h}_s$, $\tilde{h}_f$, $g_s$, and $g_f$ are coupling constants. The form and symmetry of $B_{ij}$ will 
depend on the gauge transformation rules of the theory.

The Higgs term contains the dynamics of the matter field(s) $\theta$ and $L$. Most notably, it couples the 
charge-$p$ matter field $\theta$ to the gauge field $A_{ij}$. The form of this coupling depends on the theory 
but is dictated by gauge invariance. This coupling term will be taken strong in order to enter a Higgs phase.

Finally, the Gauss' Law term dynamically identifies $pL$ with the right-hand side of Gauss' Law. This term has the schematic form
\begin{equation}
H_{Gauss} = \tilde{U}\sum_{a,\bv{r}} \(G(E) - pL_a\)^2
\label{eqn:GaussTerm}
\end{equation}
There is one term in the sum for each $L$ variable, that is, one per site in the scalar charge theories and one per link in the vector charge theories. At large $\tilde{U}$, the low-energy subspace is the gauge theory restricted to charge-$p$ particles (since, by hand, we only put charge-$p$ matter into the theory). 
Implementing Gauss' law as an energetic constraint rather than an exact constraint allows the gauge theory to emerge at low energies from a 
local lattice rotor model whose Hilbert space has a local tensor product structure. 

What remains for each theory is to use Gauss' Law to specify the structure of the matter field and to use 
gauge invariance to specify the form of the magnetic field $B_{ij}$ and the form of $H_{Higgs}$.

\subsection{Scalar Charge Theories}

We now define the full set of theories that we consider and the relations between them. These are summarized in Table \ref{tab:theoryList}; there are scalar and vector charge theories. A trace constraint may also be added to these theories\cite{RasmussenFractons}; we leave investigation of such theories to future work.

We begin with the scalar charge theories.

\subsubsection{Set of Theories}
The scalar charge theory with continuous rotation invariance has Gauss' Law
\begin{equation}
\sum_{ij} \Delta_i \Delta_j E_{ij} = \rho
\label{eqn:11ScalarCharge}
\end{equation}
which generates the gauge transformation
\begin{equation}
A_{ij} \rightarrow A_{ij} + \Delta_i \Delta_j \alpha.
\end{equation}
Here $\Delta_i$ is a lattice derivative in the $i$ direction. 

If we demand only the symmetries of a square (cubic) lattice, the diagonal components of $E_{ij}$ are not symmetry-related to the off-diagonal components, and we may modify Gauss' Law to the following
\begin{equation}
m\sum_i \Delta_i^2 E_{ii} + n \sum_{i\neq j}\Delta_i \Delta_j E_{ij} = \rho
\end{equation}
which we call the $(m,n)$ scalar charge theory. Keeping careful track of the factors of 2 in the commutators, this leads to the gauge transformation rule
\begin{equation}
A_{ij} \rightarrow A_{ij} + \begin{cases}
m\Delta_i^2 \alpha & i = j\\
n\Delta_i \Delta_j \alpha & i\neq j
\end{cases}
\label{eqn:ScalarGaugeTransform}
\end{equation}
In all these theories, there is a single matter field which, on the lattice, lives on the sites.

Clearly the scalar charge theory with continuous rotational invariance, Eq. \eqref{eqn:11ScalarCharge}, is the $(1,1)$ scalar charge theory in our nomenclature. How many distinct theories are there? If the gauge group is noncompact, that is, if we do not enforce $A_{ij} \sim A_{ij}+2\pi$, then $E_{ij}$ is not quantized and may be rescaled freely. Therefore all nonzero $m$ and $n$ generate the same theory. However, if the gauge group is compact, $E_{ij}$ is quantized and may only be rescaled by a sign while maintaining the quantization conditions. Therefore, in general different values of $m$ and $n$ are \textit{not} equivalent. We generally require $m$ and $n$ to be relatively prime integers (common factors may be removed by rescaling the matter field $\rho$)\footnote{In fact, exactly one (up to rescalings of the matter field) of $m$ or $n$ could in principle be irrational. However, the compactness conditions of the $A_{ij}$ lead to the magnetic field obeying $B_{ij} \sim B_{ij}+2\pi m$, $B_{ij}+2\pi n$ for $i\neq j$. If one of $m$ or $n$ is irrational, the resulting magnetic field cannot be incorporated into the Hamiltonian in a straightforward way while respecting the compactness condition. We generally expect such theories to be at best highly unstable.}.

\begin{figure}
\subfigure[ ]{
	\includegraphics[width=5cm]{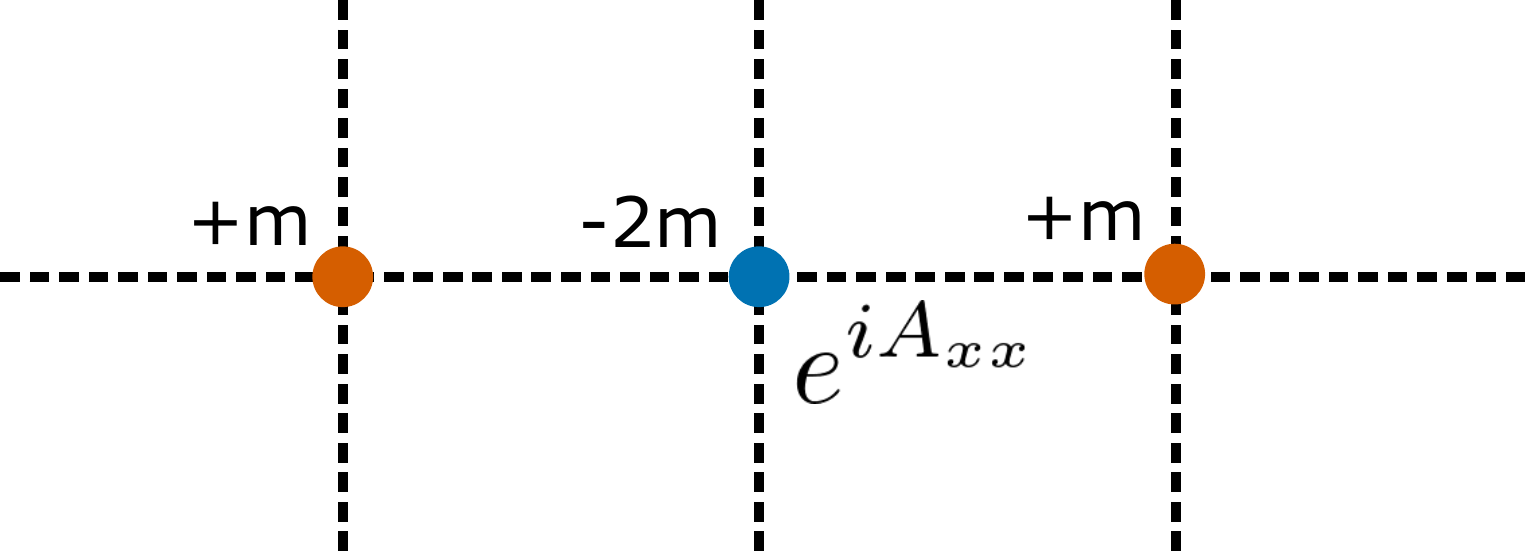}
	\label{fig:scalarAxx}
}
\subfigure[]{
	\includegraphics[width=4cm]{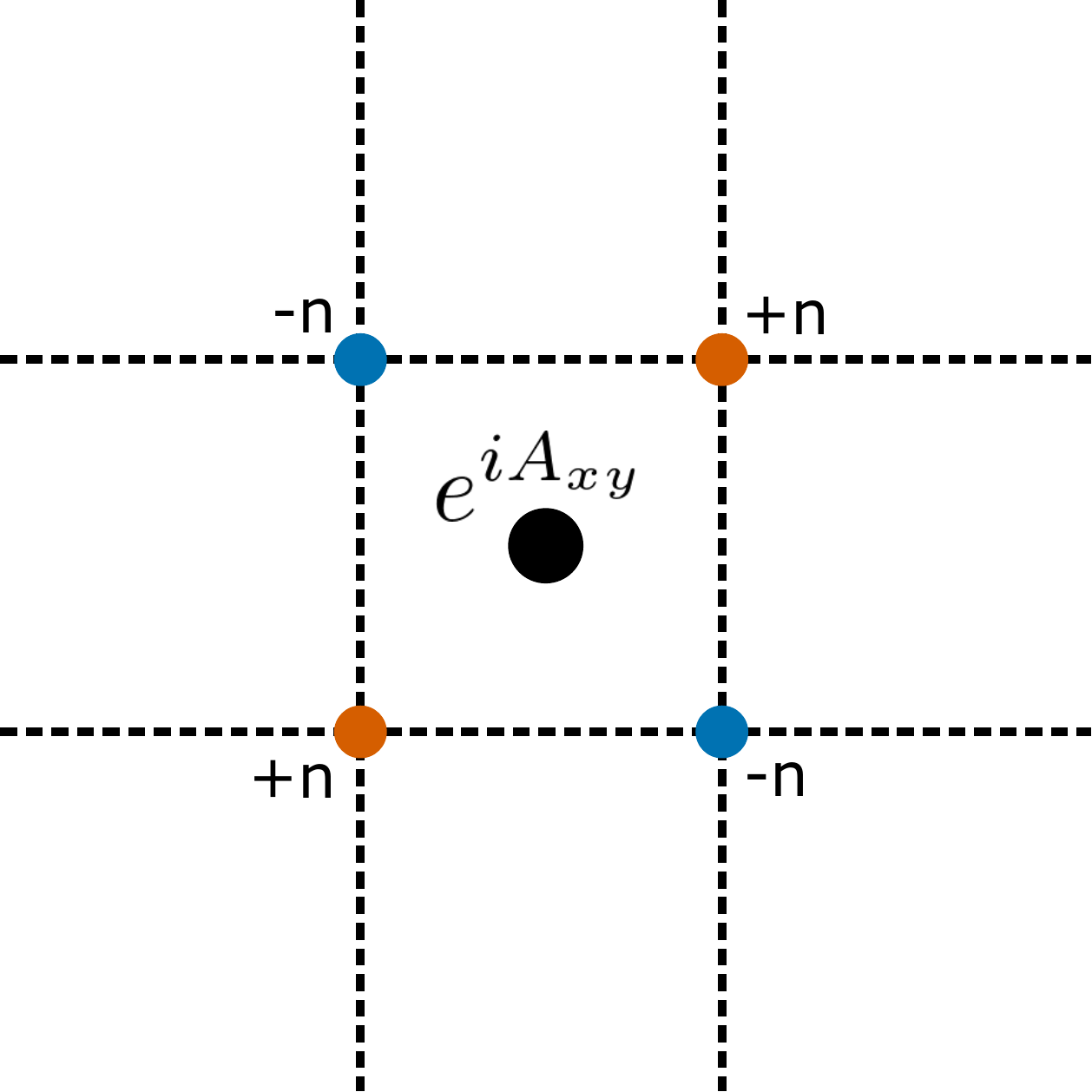}
	\label{fig:scalarAxy}
}
\caption{Charge configurations created in the $(m,n)$ scalar charge theory by (a) $e^{iA_{xx}}$, the raising operator for $E_{xx}$, acting on the center site (b) $e^{iA_{xy}}$, the raising operator for $E_{xy}$, acting on the black plaquette. }
\label{fig:scalarU1Operators}
\end{figure}

In each theory, charges are created by the local operators $e^{\pm iA_{ij}}$, which are raising and lowering operators for the $E_{ij}$. The charge configurations created by these local operators in the $(m,n)$ theories are determined by inspection of Gauss' Law, and are shown in Fig. \ref{fig:scalarU1Operators}.

We now examine different theories individually.

\subsubsection{$(1,0)$ Scalar Charge Theory}
\label{subsubsec:10ScalarSetup}

The $(1,0)$ theory is special in that the off-diagonal components of $A_{ij}$ are gauge-invariant and decouple from the diagonal components. Since the off-diagonal components are decoupled and have no gauge transformation associated with them, we discard them as a trivial sector of the theory.

The magnetic field is
\begin{equation}
B_{i} = \sum_{jk}\epsilon_{ijk} \Delta_j^2 A_{kk}
\end{equation}
for $i=z$ in $d=2$ and $i=x,y,z$ in $d=3$. Here $\epsilon$ is the Levi-Civita symbol. The Higgs term of the Hamiltonian is
\begin{equation}
H_{Higgs} = \sum_{\bv{r}} \frac{L(\bv{r})^2}{2M} - V \sum_i \cos(\Delta_i^2 \theta + p A_{ii})
\end{equation}

In $d=2$ local operators create point-like magnetic excitations, so we expect the theory to be unstable to instanton proliferation. In $d=3$, this theory is self-dual at $V=0$; if we write
\begin{equation}
E_{ii} = \sum_{jk}\epsilon_{ijk}\Delta_j^2 h_{kk}
\end{equation}
where $h$ is a diagonal rank-2 tensor, then $[h_{jj},B_{k}] = i\delta_{jk}$ reproduces the correct commutation relations between $A_{jj}$ and $E_{kk}$. The self-duality of the Hamiltonian can be checked explicitly. The duality and gauge invariance is enough to show, following similar arguments to Refs. \onlinecite{XuFractons2,RasmussenFractons}, that the $(1,0)$ theory in $d=3$ is stable to confinement. Its photon mode has the soft dispersion $\omega \sim k^2$.

Electric charges in this theory are immobile, while dipoles can move only in one dimension, along the direction of their dipole moment. 
To show this, we adapt the arguments of Ref. \onlinecite{PretkoSubdimensional}. In this theory, dipole moments are conserved:
\begin{equation}
\int x_i \rho(\bv{r}) d^d\bv{r}  = \int x_i \sum_j \Delta_j^2 E_{jj} d^d\bv{r} = \int \Delta_i E_{ii} d^d\bv{r} = 0
\end{equation}
where we integrated by parts and used the fact that the $E_{ii}$ are single-valued. Moving an isolated charge would violate this conservation law and is therefore not allowed.

Another way to see this is to examine the local operators which create charge; these operators are $e^{iA_{ii}}$. Using the form of Gauss' Law, it is straightforward to see that $e^{iA_{xx}}$, for example, creates the charge configuration in Fig. \ref{fig:scalarAxx} (with $m=1$). Alternatively, this operator is a hopping operator which moves an $x$-directed dipole by one unit in the $x$ direction; dipoles may move in the direction of their dipole moment. However, because $e^{iA_{xy}}$ is not present in this theory, any operator that moves an $x$-directed dipole in the $y$ direction creates additional charges. Hence, the dipoles are mobile only along their dipole moment. There is no local hopping operator for a single charge. 

\subsubsection{$(0,1)$ Scalar Charge Theory}

In the $(0,1)$ theory the diagonal components of $A_{ij}$ are gauge-invariant and decouple from the off-diagonal components. The diagonal components may then be discarded as a trivial sector of the theory.

In $d=2$, $A_{xy}$ is the only nontrivial degree of freedom remaining. It is not possible to create a gauge-invariant operator solely out of $A_{xy}$; accordingly, no magnetic field can be defined in $d=2$. We therefore expect that the $(0,1)$ theory in $d=2$ is highly degenerate; its Maxwell theory has no $B^2$ term, which corresponds to a speed of light equal to zero. This theory is therefore
highly unstable to perturbations that take the system out of the gauge theory subspace, and thus not well-defined. 

In $d=3$, we can construct a magnetic field
\begin{equation}
B_{ii} = \sum_{ab}\epsilon_{iab}\Delta_a A_{bi}
\end{equation}
The labeling is for later comparison with other theories. Treating $B$ as a (diagonal) rank-2 tensor, we see that it is traceless. The Higgs term of the Hamiltonian is
\begin{equation}
H_{Higgs} = \sum_{\bv{r}} \frac{L(\bv{r})^2}{2M} - V \sum_{i < j} \cos(\Delta_i\Delta_j \theta + p A_{ij})
\end{equation}

Unlike the $(1,0)$ theory, the $d=3$ $(0,1)$ theory is not self-dual. Local operators create point-like magnetic excitations, so we expect that instanton processes induce confinement. Ref. \onlinecite{XuSU4} showed that the theory is indeed unstable to confinement.

The photon mode in this $(0,1)$ theory has dispersion $\omega \sim k$. Following the same arguments as in Sec. \ref{subsubsec:10ScalarSetup}, one can check that isolated electric charges are immobile, and that dipoles can move in any direction perpendicular to their dipole moment. Isolated charges are immobile because, as before, there is a dipole conservation law $\int x_i \rho(\bv{r})d^3{\bv{r}} = 0$. The only local operators which create charge, $e^{iA_{ij}}$ for $i \neq j$, create charges in the set of four shown in Fig. \ref{fig:scalarAxy}. This operator is exactly a transverse hopping operator for a dipole, but no longitudinal hopping operator exists because the diagonal components of the electric field do not exist.

This motion is reminiscent of the restricted mobility of excitations in the X-cube model; we will show later that Higgsing the $d=3$ $(0,1)$ model indeed produces the X-cube model.

\subsubsection{$(m,n)$ Scalar Charge Theory}

We now consider the rest of the $(m,n)$ theories, i.e. $m,n$ relatively prime positive integers.

\begin{widetext}
\begin{equation}
H_{Higgs} = \sum_{\bv{r}} \frac{L(\bv{r})^2}{2M} - V_1 \sum_{\bv{r},i < j} \cos(n\Delta_i\Delta_j \theta + p A_{ij}) - V_2 \sum_{\bv{r},i}\cos(m\Delta_i^2 \theta + p A_{ii})
\label{eqn:mnScalarHiggsTerm}
\end{equation}
\end{widetext}
For the later part of the paper, we will want to take a single strong-coupling limit which Higgses the entire gauge field. We therefore want $V_1$ and $V_2$ to scale to strong coupling at the same rate; for these purposes, it suffices to take them equal.

In $d=2$, the magnetic field has two components
\begin{align}
B_{zx} &= m\Delta_x A_{xy}- n\Delta_y A_{xx}\\
B_{zy} &= m\Delta_y A_{xy}- n\Delta_x A_{yy}
\label{eqn:11ScalarBd2}
\end{align}
The notation is for consistency with $d=3$, where the magnetic field is a traceless, non-symmetric tensor
\begin{equation}
B_{ij} = \begin{cases}
\sum_{ab} \epsilon_{iab}\Delta_a A_{bi} & i = j\\
 \sum_{a\neq i,j} \(m\Delta_j A_{aj} - n \Delta_a A_{jj}\) & i \neq j
\end{cases}
\label{eqn:scalarB}
\end{equation}

In neither $d=2$ nor $d=3$ does this model have a clear self-duality. Local operators create point-like magnetic excitations in $d=2$ but not in $d=3$, so we expect that the theory is unstable to confinement in $d=2$ but is stable in $d=3$, consistent with expectations\cite{PretkoSubdimensional} for the $(1,1)$ model.

\begin{figure}
\includegraphics[width=4cm]{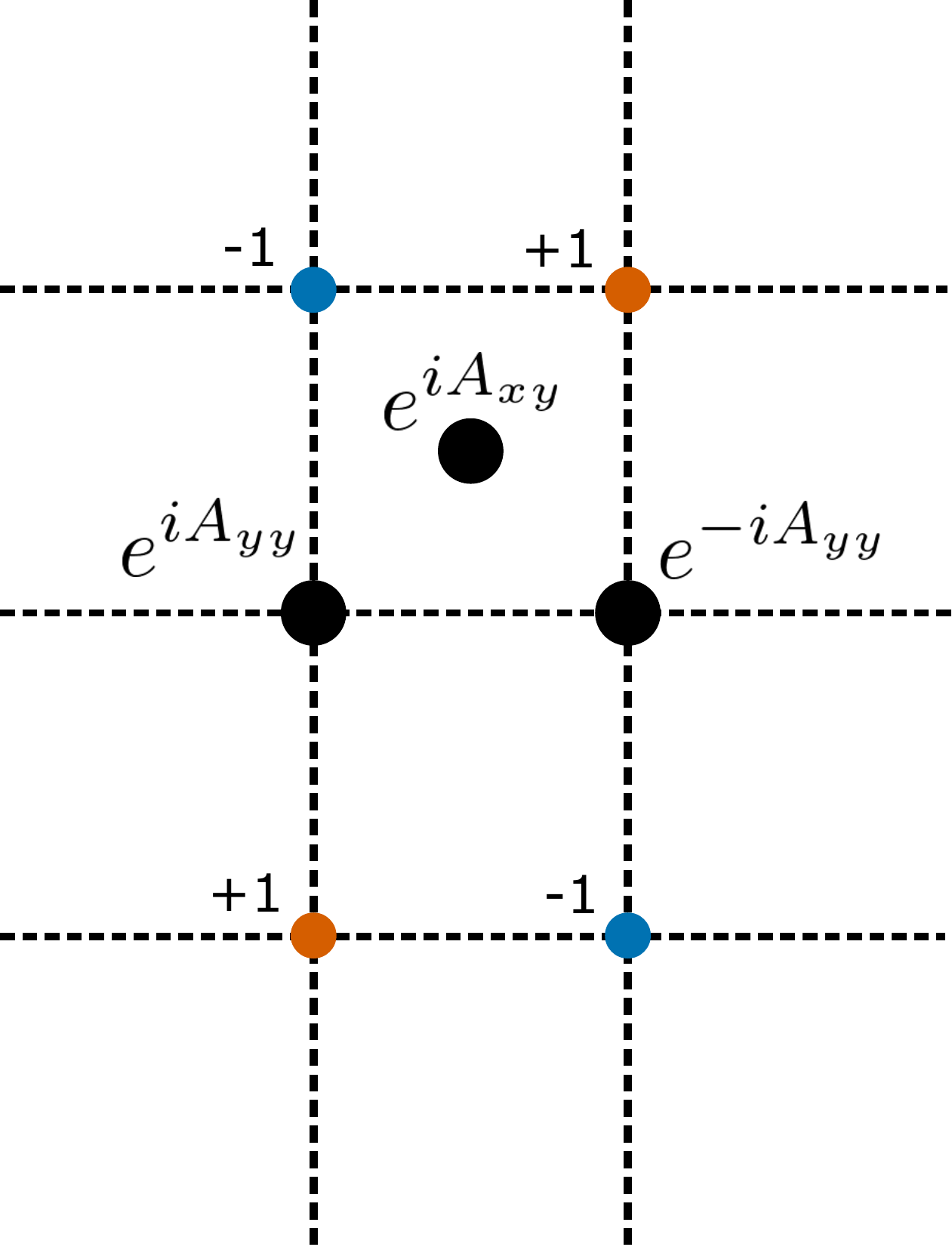}
\caption{Transverse hopping operator for a unit dipole in the $(1,2)$ scalar charge theory. The $e^{iA_{ij}}$ operators act on the black rotors and create the charge configuration shown. Because $e^{iA_{xy}}$ only creates charge-2 objects in the $(1,2)$ theory, it cannot cause a unit dipole to hop, unlike the $(1,1)$ theory.}
\label{fig:dipoleHop}
\end{figure}

The photon mode has dispersion $\omega \sim k$. These theories all have the dipole conservation law 
$\int x_i \rho(\bv{r}) d^d\bv{r}=0$, so electric charges are immobile. Dipoles can propagate in any direction, 
but on the lattice they have the curious property that the distance by which they can move depends on $m$ and $n$. 
For example, in the $(1,1)$ theory, $e^{iA_{ii}}$ is a longitudinal hopping operator for a unit dipole and $e^{iA_{ij}}$ for $i\neq j$ is a transverse hopping operator, as can be deduced from Fig. \ref{fig:scalarU1Operators}. In the $(1,2)$ theory, $e^{iA_{ij}}$ for $i \neq j$ is a transverse hopping operator for dipoles with moment 2 (see Fig. \ref{fig:scalarAxy}); the simplest transverse hopping operator for a unit dipole is shown in Fig. \ref{fig:dipoleHop} and moves the unit dipole by \textit{two} lattice units. 

As a further note, it can be checked explicitly that the $(m,n)$ theory for $m,n \neq 0$ can be produced by condensing a bound state of charge $n$ in the $(1,0)$ theory and charge $-m$ in the $(0,1)$ theory. This means that in some sense one may think of the $(1,0)$ and $(0,1)$ theories as the fundamental scalar charge theories.

\subsection{Vector Charge Theories}

We next examine the possible vector charge theories, proceeding similarly to the scalar charge theories.
Here, the matter field consists of rotor variables $\theta_a(\bv{r} )$ that are defined on links that are 
oriented in the $a$ direction, and which transform as a vector under rotations. Consequently, there are
$d$ different types of charges, which are subject to $d$ Gauss' Laws. 

\subsubsection{Set of Theories}

The vector charge theory with continuous rotational invariance has the following $d$ Gauss' Laws:
\begin{equation}
2\sum_{i} \Delta_i E_{ij} = \rho_j .
\label{eqn:21VectorCharge}
\end{equation}
The factor of $2$ is present to cause $\rho_j$ to take on integer values; this is needed because the off-diagonal components of $E_{ij}$ may be half-integers. This generates the $d$ gauge transformations
\begin{equation}
A_{ij} \rightarrow A_{ij} + \Delta_i \alpha_j + \Delta_j \alpha_i
\end{equation}
where the $\alpha_i$ are $d$ independent gauge transformations.

As before, demanding only the symmetries of a square (cubic) lattice allows modifications 
\begin{equation}
m \Delta_i E_{ii} + 2n \sum_{i\neq j}\Delta_i E_{ij} = \rho_j
\end{equation}
which we call the $(m, n)$ vector charge theory. The factor of $2$ is again present to make $\rho_j$ integer-valued when $m$ and $n$ are integers. This leads to the gauge transformation rule
\begin{equation}
A_{ij} \rightarrow A_{ij} + \begin{cases}
m\Delta_i \alpha_i & i = j\\
n\(\Delta_i \alpha_j + \Delta_j \alpha_i\) & i\neq j
\end{cases}
\label{eqn:VectorGaugeTransform}
\end{equation}
In these theories, there are $d$ matter fields $\theta_i$ which, on the lattice, live on the $i$-directed links.

Clearly the rotationally invariant scalar charge theory Eq. \eqref{eqn:21VectorCharge} is the $(2,1)$ vector charge theory in our nomenclature. We proceed as before to classify the set of possible theories. Again, if the gauge field is noncompact, all nonzero $m$ and $n$ generate the same theory because $E_{ij}$ can be rescaled. For the same reasons as before, with a compact gauge group, relatively prime integers $m$ and $n$ generate distinct theories.

\begin{figure}
\subfigure[]{
	\includegraphics[width=5cm]{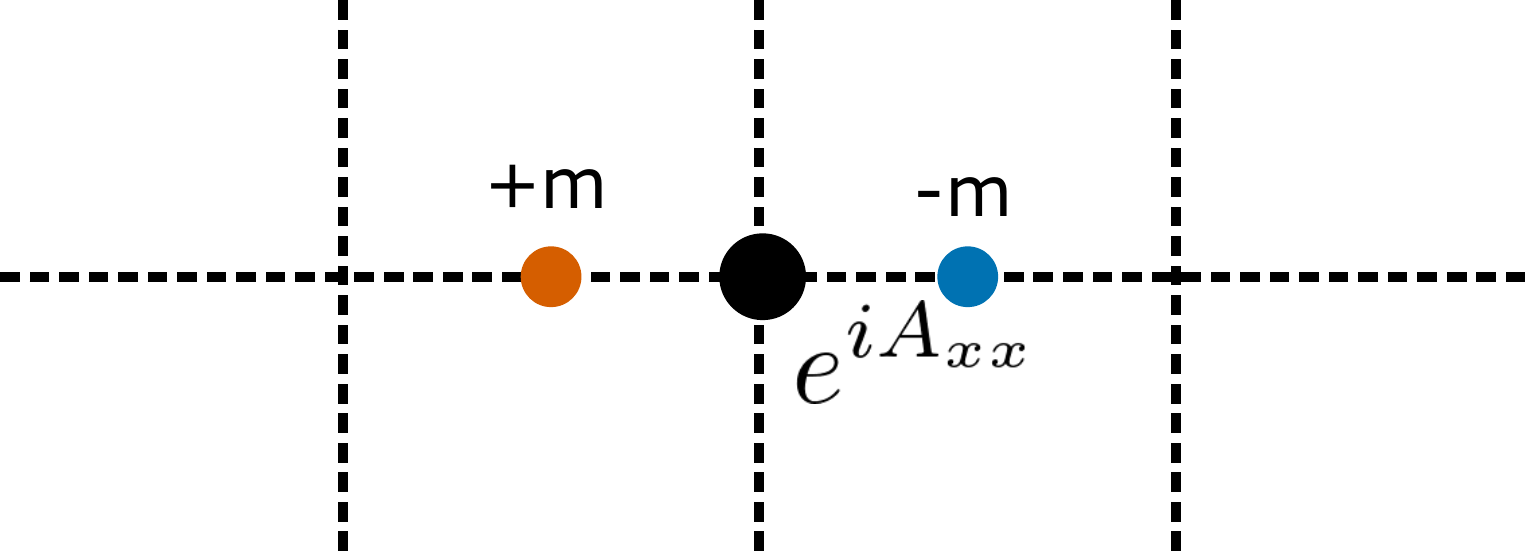}
	\label{fig:vectorAxx}
}
\subfigure[]{
	\includegraphics[width=4cm]{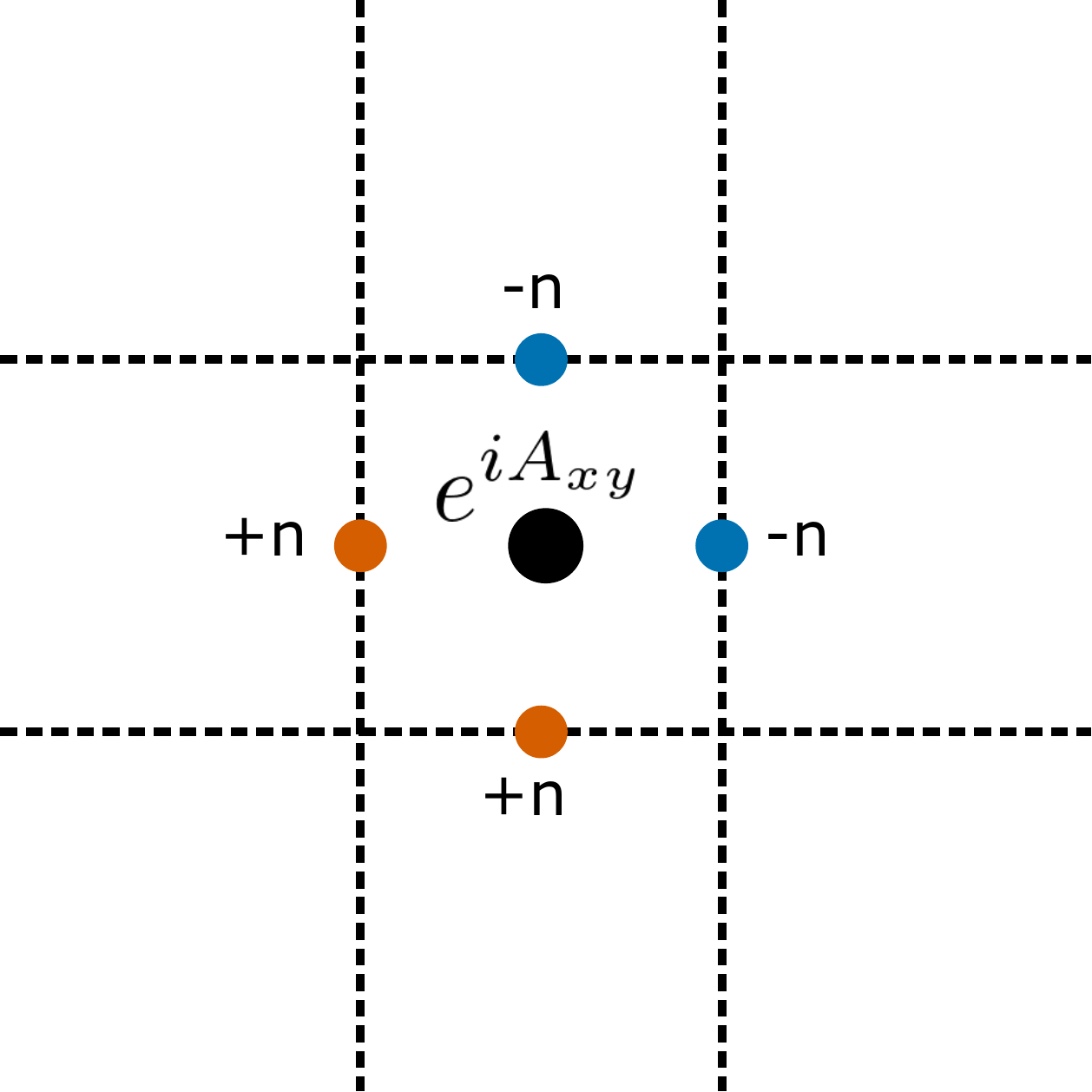}
	\label{fig:vectorAxy}
}
\caption{Charge configurations created in the $(m,n)$ vector charge theory by (a) $e^{iA_{xx}}$, the raising operator for $E_{xx}$, acting on the black site (b) $e^{iA_{xy}}$, the raising operator for $E_{xy}$, acting on the black plaquette.}
\label{fig:vectorU1Operators}
\end{figure}

The charge configurations created by local operators in the $(m,n)$ theories are again determined by inspection of Gauss' Law, and are shown in Fig. \ref{fig:vectorU1Operators}.

We now examine different theories individually.

\subsubsection{$(1,0)$ and $(0,1)$ Vector Charge Theories}

As in the scalar charge case, the $(1,0)$ vector charge theory has decoupled, gauge-invariant off-diagonal components of $A_{ij}$ which can 
be discarded as trivial. Likewise, the diagonal components of $A_{ij}$ may be discarded in the $(0,1)$ theory. However, it can be checked that 
in neither of these theories can a magnetic field be defined; no gauge-invariant linear combination of the remaining $A_{ij}$ exist. These 
theories effectively have zero ``speed of light" and are thus highly degenerate and unstable to perturbations that take the theory out of 
the gauge invariant subspace.  Although a Higgs mechanism can still be defined, it will turn out to produce trivial Higgs phases in these theories.

\subsubsection{$(m,n)$ Vector Charge Theory}

Although the $(1,0)$ and $(0,1)$ vector charge theories are unstable, the theories we now consider, i.e. $(m,n)$ theories with $m,n$ relatively prime positive integers, are much better behaved.

The Higgs term in the Hamiltonian is of the form
\begin{align}
H_{Higgs} = &\sum_{\bv{r},i} \frac{L_i(\bv{r})^2}{2M} - V_1 \sum_{\bv{r},i} \cos(m\Delta_i \theta_i + p A_{ii}) \nonumber \\
&- V_2 \sum_{\bv{r},i<j}\cos(n(\Delta_i \theta_j + \Delta_j \theta_i) + p A_{ij})
\end{align}
For the same reasons as in the scalar charge theories, we will choose to take $V_1=V_2$ for simplicity.

In $d=2$, the magnetic field has one component
\begin{align}
B_{zz} = n \sum_{a \neq b} \Delta_a^2 A_{bb} - m\Delta_x\Delta_yA_{xy} 
\end{align}
The notation is for consistency with $d=3$, where the magnetic field is a symmetric tensor
\begin{widetext}
\begin{equation}
B_{ij} = \begin{cases}
\frac{1}{2}\sum_{a \neq b \neq i} \(2n\Delta_a^2 A_{bb}-m\Delta_a\Delta_b A_{ab}\) & i=j\\
\frac{2}{3-(-1)^m}\sum_{k \neq i,j}\left[m\(\Delta_i \Delta_k A_{jk} + \Delta_j \Delta_k A_{ik} - \Delta_k^2 A_{ij} \) - 2n \Delta_i \Delta_j A_{kk}\right] & i \neq j
\end{cases}
\label{eqn:vectorB}
\end{equation}
\end{widetext}

The peculiar factor in front of the off-diagonal terms, which is $1$ when $m$ is odd and $1/2$ when $m$ is even, merits explanation. In the expression for $B_{ij}$ for $i \neq j$, $A_{kk}$ appears with an even coefficient $2n$. Therefore, under $A_{kk}({\bf r}) \rightarrow A_{kk}({\bf r}) + 2\pi$ at a specific single site ${\bf r}$, we have $B_{ij}({\bf r}) \rightarrow B_{ij}({\bf r})+4n\pi$. 
Under $A_{ij}({\bf r}) \sim A_{ij}({\bf r}) +2\pi$ at a single specific site ${\bf r}$, we have $B_{ij}({\bf r}) \rightarrow B_{ij}({\bf r})+2m\pi$. If $m$ is odd, the fact that $m$ and $2n$ are relatively prime implies that $B_{ij} \sim B_{ij} + 2\pi$, and $\cos(B_{ij})$ is the minimal term in the Hamiltonian which respects this identification. If $m$ is even, then the factor of $1/2$ maintains $B_{ij} \sim B_{ij} + 2\pi$; were the factor of $1/2$ not present, a term $\cos(B_{ij}/2)$ would be allowed in the Hamiltonian. The prefactor just absorbs that factor of $1/2$ into the definition of the magnetic field. 

The $(2,1)$ model at $V=0$ has been shown\cite{XuFractons2} to confine in $d=2$ and to be self-dual and stable in $d=3$. These arguments carry through with minimal modification for the general $(m,n)$ theories ($m,n\neq 0$): these theories are also confining in $d=2$ and self-dual and stable in $d=3$.

The photon mode has dispersion $\omega \sim k^2$. The motion of charges depends on the values of $m$ and $n$. The operator $e^{iA_{ii}}$ is a longitudinal hopping operator for particles with vector charge $m\hat{x}_i$, (here $\hat{x}_i$ is the elementary vector charge on an $i$-directed link) as shown in Fig. \ref{fig:vectorAxx}, while $e^{iA_{ij}}$ for $i \neq j$ creates a loop of charges of magnitude $n$, as shown in Fig. \ref{fig:vectorAxy}. Charges whose charge components are all multiples of $m$ may only move in one dimension, along the direction of the charge, while other charges are confined. 

Similarly to the scalar charge theories, the $(m,n)$ theory for $m,n \neq 0$ can be produced starting from the decoupled $(1,0)$ and $(0,1)$ theories. One does this by condensing the bound states of charge $n\hat{x}_i$ in the $(1,0)$ theory with charge $-m\hat{x}_i$ in the $(0,1)$ theory.

\section{General Comments on Higgs Phases}
\label{sec:intuition}

Before explicitly solving the models that arise upon spontaneously breaking the higher rank $U(1)$ gauge symmetry, we will provide
in this section some general comments and intuition for what to expect for the properties of the resulting Higgs phases. 

\subsection{Scalar Charge Theories}

Let us first discuss the $(1,1)$ scalar charge theory. Recall that the operators that create electric charges 
in the $U(1)$ theories are $e^{iA_{ij}}$, which are raising operators for the $E_{ij}$ and accordingly modify 
the eigenvalues of the charge density on various sites. In particular, $e^{iA_{xx}}$ creates a line of charges 
of value $1, -2,$ and $1$ on neighboring lattice sites as shown in Fig. \ref{fig:scalarAxx}, and $e^{iA_{xy}}$ 
creates a square of charge-$1$ particles as shown in Fig. \ref{fig:scalarAxy}.

\begin{figure}
\includegraphics[width=5cm]{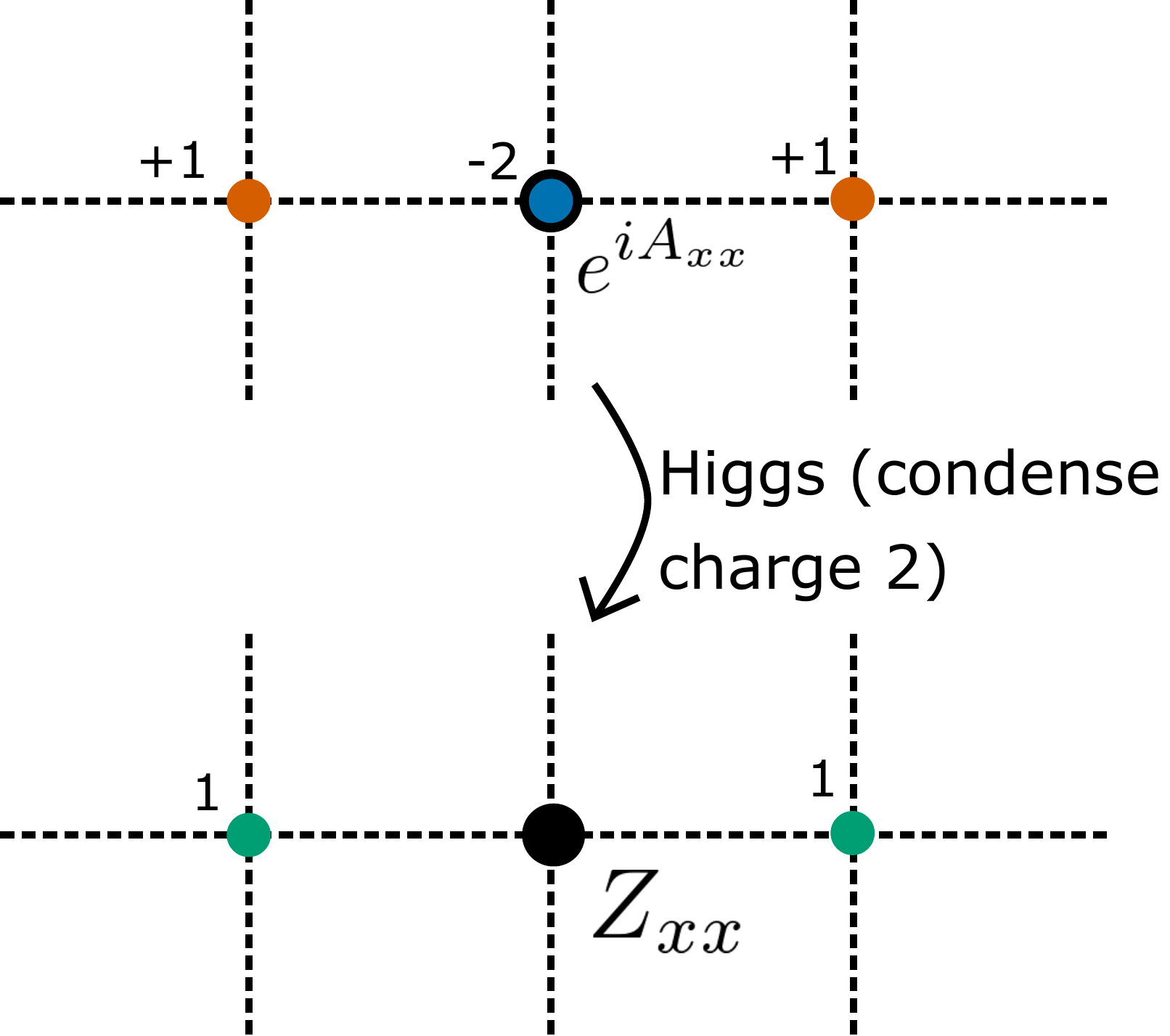}
\caption{Effect of the Higgs mechanism on the local operator $e^{iA_{xx}}$ in the $(1,1)$ scalar charge theory. The original charge configuration (top) is modified when charge-2 particles are condensed; the charge-2 particle is absorbed into the condensate, and charges $+1$ and $-1$ become equivalent.}
\label{fig:dipoleHopHiggs}
\end{figure}

The Higgs procedure condenses charge $p$ particles; for the moment, we specialize to $p=2$. Hence only the 
parity of charges is well-defined after Higgsing. In particular, the charge $-2$ particle created by $e^{iA_{xx}}$ 
may be absorbed into the condensate, and $e^{iA_{xx}}$ becomes the distance-2 hopping operator 
$Z_{xx}$ for $\mathbb{Z}_2$ charges (the reason for the notation will be made clear later). This process 
is shown in Fig. \ref{fig:dipoleHopHiggs}. The only change to the action of $e^{iA_{xy}}$ (see Fig. \ref{fig:scalarAxy}) is 
that $+1$ and $-1$ charges are now equivalent since they differ by a condensed charge. Therefore, we see that 
in the Higgs phase, individual charges are now free to propagate and are no longer immobile. We expect, then, that the
resulting Higgs phase will possess some form of conventional topological order. 

The above line of reasoning extends to general $(m,n)$ scalar charge theories. By examining 
Fig. \ref{fig:scalarU1Operators}, it is clear that if $m$ is odd, $e^{iA_{ii}}$ becomes a hopping 
operator for $\mathbb{Z}_2$ charge by 2 units in the $i$ direction, and charges become mobile. If $m$ is even, $e^{iA_{xx}}$ 
acts trivially because all the charges that it creates can be absorbed into the condensate. Likewise, 
if $n$ is odd, the Higgsed version of $e^{iA_{xy}}$ simply creates four (identical) $\mathbb{Z}_2$ charges. 
If $n$ is even, then $e^{iA_{xy}}$ acts trivially.

The behavior of the local operators in the electric sector of the scalar charge theory after Higgsing 
therefore depend entirely on the parities of $m$ and $n$ (which are relatively prime). We label the 
classes of Higgsed theories using representations for $m$ and $n$ which make their parities clear. The distinct classes of Higgsed scalar charge theories  are those arising from the $(1,0)$, $(0,1)$, $(2r+1,2s+1)$, $(2r+2,2x+1)$, $(2r+1,2s+2)$ scalar charge theories with $r,s$ nonnegative integers. ($(0,1)$ and $(1,0)$ are distinguished separately because their magnetic fields behave differently from other values of $(m,n)$.)

In the Higgsed ($2r+1,2s$) theory, electric excitations can hop two lattice sites in any direction. Since 
$n$ is even, there is no one-site hopping operator, so we expect $2^d$ decoupled copies of the resulting theory, 
each living on a different sublattice of lattice constant 2. This will turn out to be $2^d$ copies of the $\mathbb{Z}_2$ 
toric code. 

The difference between the Higgsed $(2r+1,2s)$ and $(2r+1,2s+1)$ scalar charge theories is that in the latter 
the operator $e^{iA_{xy}}$ (considered after Higgsing) acts nontrivially. In $d = 2$, it couples the four ``copies" by locally 
creating or annihilating a bound state of the charges on all four sublattices. That is, the $d= 2$ Higgsed $(2r+1,2s+1)$ scalar charge
theory should be produced from the $(2r+1,2s)$ theory by condensing the four-charge bound state. 
The topological order turns out to be three copies of the $\mathbb{Z}_2$ toric code. The analogous consideration in $d = 3$ leads to four
copies of the $\mathbb{Z}_2$ toric code, as explained in detail in the subsequent sections. 

Finally, the Higgsed $(2r,2s+1)$ theory only allows electric particles to be created in sets of four, which is reminiscent
of fracton phases. In $d = 3$, we find that indeed these $(2r, 2s+1)$ scalar charge theories yield the X-cube model upon breaking the 
$U(1)$ higher rank gauge symmetry to its $\mathbb{Z}_2$ subgroup. In $d = 2$ however, careful examination reveals that 
we obtain the trivial gapped phase. 

\begin{figure}
\includegraphics[width=6cm]{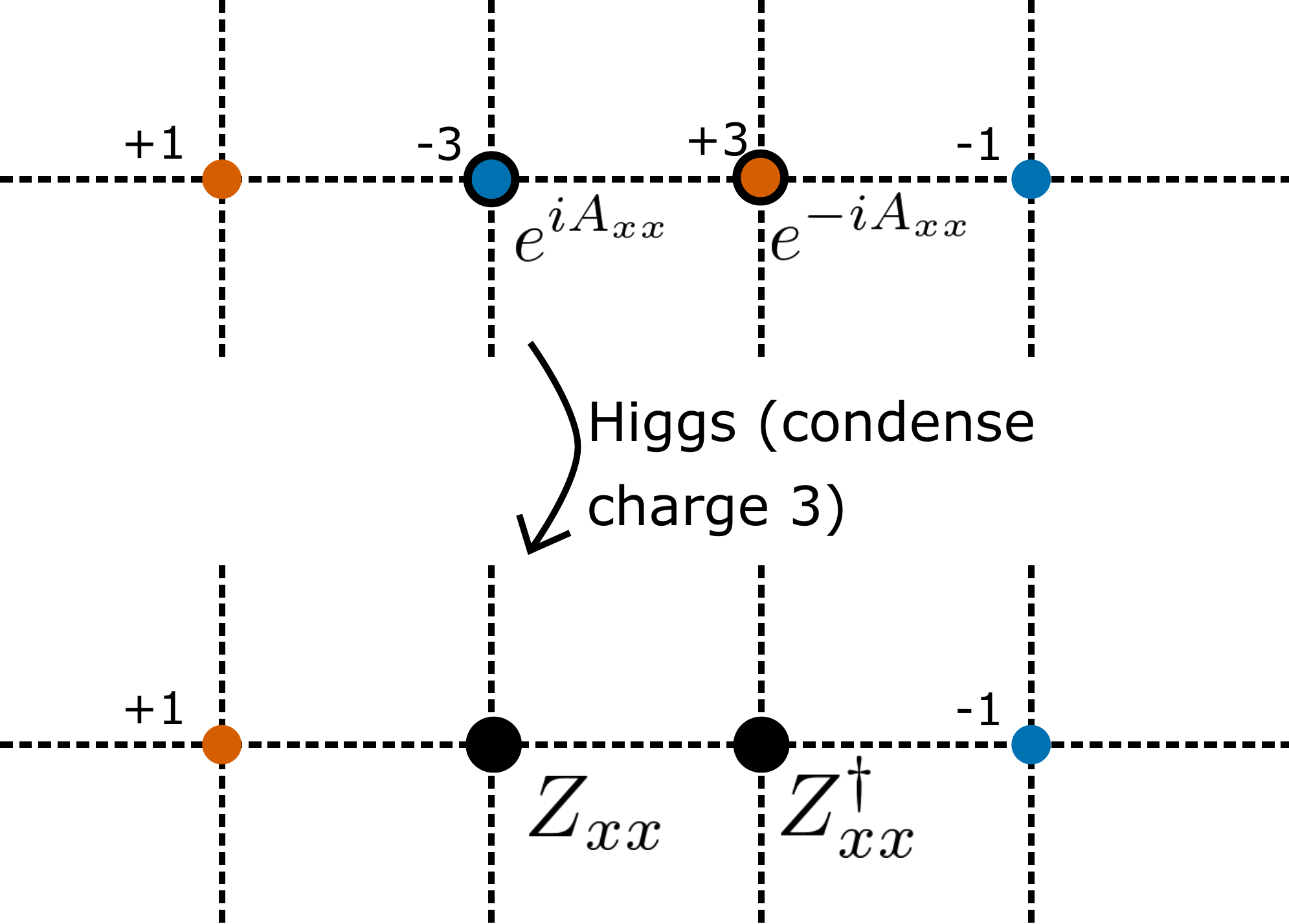}
\caption{Operator generating a $\mathbb{Z}_3$ charge hopping operator in the $p=3$ Higgsed $(1,1)$ scalar charge theory.}
\label{fig:dipoleHopHiggsp3}
\end{figure}

The above results can be readily generalized to the condensation of charge $p$, with $p>2$, although the analysis is slightly more complicated. 
For example, in a $U(1)$ theory with $m=1$, consider the operator shown in Fig. \ref{fig:dipoleHopHiggsp3}. Before Higgsing,
it creates four charges, $+1$, $-3$, $+3$, and $-1$ in a line. If charge $p=3$ condenses, then the $\pm 3$ charges can be 
absorbed into the condensate and this operator becomes a hopping operator for $\mathbb{Z}_3$ charges. It is a straightforward 
generalization to show if charge $p$ is condensed, then there is a distance-$p$ hopping operator in the Higgsed theory when 
$m$ and $p$ are relatively prime. Accordingly, if $m$ and $p$ are relatively prime, then the Higgsed $(m,0)$ theory decouples 
into sublattices of lattice constant $p$, and the topological order turns out to be $p^2$ copies of the $\mathbb{Z}_p$ toric code. 
More generally, the resulting theory depends only on the values of $m$ and $n$ modulo $p$. We will not consider $p>2$ in much further detail.

\subsection{Vector Charge Theories}

\begin{figure}
	\includegraphics[width=8cm]{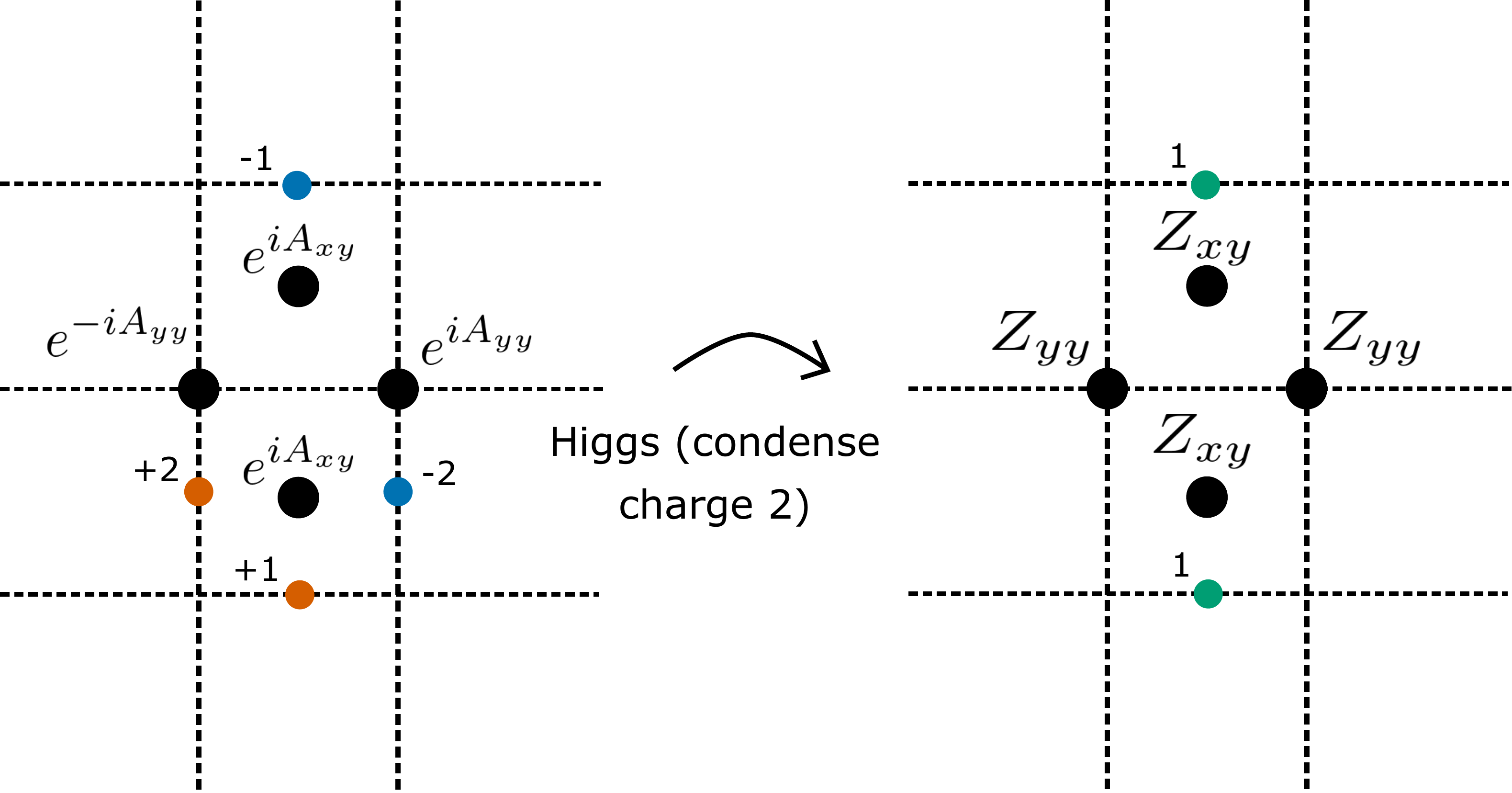}
	\caption{Operator in the $(1,1)$ vector charge theory that, after $p=2$ Higgsing, allows $\mathbb{Z}_2$ charges on $x$-directed links to hop in the $y$ direction.}
	\label{fig:vectorTransverseHopHiggs}
\end{figure}

Similar logic may be used on the vector charge theories. When $m$ is odd, in the $\mathbb{Z}_2$ Higgs phase, $e^{iA_{ii}}$ becomes a hopping operator for charges on the $i$ links (see Fig. \ref{fig:vectorAxx}); therefore such $i$-directed charges are mobile in the $i$ direction. If $m$ and $n$ are both odd, then in the $\mathbb{Z}_2$ Higgs phase charges become mobile in all $d$ dimensions. The operator which moves a particle in a direction transverse to the link it lives on has a non-obvious form, shown in Fig. \ref{fig:vectorTransverseHopHiggs}. If $m$ is even, then charges are confined because $e^{iA_{ij}}$ ($i \neq j$) can only create closed strings of charge and $e^{iA_{ii}}$ acts trivially. 

As such, the behavior depends mostly on the parities of $m$ and $n$. However, there are some subtleties in the magnetic sector in $d=3$. The $\mathbb{Z}_2$ Higgsed $(m,n)$ vector charge models in $d=3$ are actually labeled not just by the parity of $m$ and $n$. Specifically, if $m$ is even and nonzero, then there is a distinction between the $m \equiv 0$ mod 4 and $m \equiv 2$ mod 4 theories. We will discuss this further in Sec. \ref{subsec:vectorHiggsChanges}.

The distinct classes of $\mathbb{Z}_2$ Higgsed vector charge theories in $d=3$ thus arise from the $(1,0)$, $(0,1)$, $(2r+1,2s+1)$, $(2r+1, 2s+2)$, $(4r+2,2s+1)$, and $(4r+4,2s+1)$ vector charge theories, where $r$ and $s$ are nonnegative integers.

\section{Scalar Charge Higgs in $d=2$}
\label{sec:Scalar2DHiggs}

Here we will discuss the Higgs phases of the scalar charge theory in $d= 2$. 
For a review of relevant aspects of the Higgs mechanism in standard (rank-1) 
compact $U(1)$ gauge theory, see Appendix \ref{app:HiggsReview}.

Our focus here is on explaining the details of theories for which the $\mathbb{Z}_2$ Higgs 
phases are non-trivial. This occurs for the $(2r+1,2s+1)$ and $(1,0)$ scalar charge theories. 
Other cases, where the $\mathbb{Z}_2$ Higgs phases are trivial theories, are discussed in Appendix \ref{app:trivialHiggs}.  All the Higgs phases are summarized in Table \ref{tab:HiggsPhases}.

\subsection{Higgsing Procedure}
\label{subsec:HiggsingProcedure}

We begin by describing the models that we obtain by taking the gauge-matter coupling in the Hamiltonian Eq. \eqref{eqn:HStructure}
to be large. These induce condensation of the charge $p$ matter fields, inducing a $\mathbb{Z}_p$ Higgs transition. 
We first explain the example of the $(1,1)$ scalar charge theory, and subsequently discuss the generalization to $(m,n)$. 

We recall the general form of the Hamiltonian Eq. \eqref{eqn:HStructure}, and take 
$V$ (see Eq. \eqref{eqn:mnScalarHiggsTerm}) much larger than all other scales in the problem, which freezes
\begin{equation}
\Delta_i \Delta_j \theta + p A_{ij} = 2\pi n
\end{equation}
Given any initial gauge choice, $\theta$ may be set uniformly to zero by choosing the gauge transformation $\alpha(\bv{r}) = -\theta(\bv{r})/p$ where the $\theta$ on the right-hand side is defined using the initial gauge choice. In this gauge, 
\begin{equation}
A_{ij} = \frac{2\pi}{p} n
\label{eqn:PinningA}
\end{equation}
for $n \in \mathbb{Z}$. For simplicity, we specialize to $p=2$. Then $e^{iA_{ij}} = \pm 1$ on each site or plaquette. 
Furthermore, since $e^{iA_{ij}}$ is a raising operator for $E_{ij}$, its action flips the sign of $(-1)^{(2-\delta_{ij})E_{ij}} = \pm 1$. 
The factor of 2 for the off-diagonal piece is present because of the factor of 1/2 in its commutation relations, see Eq. (\ref{eqn:comms}). 
Therefore, the spectrum and the commutation relations of $e^{iA_{ij}}$ and $(-1)^{(2-\delta_{ij})E_{ij}}$ in the 
low-energy subspace are reproduced by the identification $e^{iA_{ij}} = Z_{ij}$ and $(-1)^{E_{ij}} = X_{ij}$ where $X_{ij} = X_{ji}$ and $Z_{ij}=Z_{ji}$ are Pauli matrices and have
\begin{align}
[X_{ab},Z_{cd}] = 0 & \text{ if } (a,b) \neq (c,d) \nonumber \\
\lbrace X_{ab},Z_{cd} \rbrace = 0 & \text { if } (a,b)=(c,d)
\end{align} 
We emphasize that the indices $a,b,c,d$ label which spin the operator is associated with. That is, $X_{xx}$ is a $2\times 2$ matrix, not a matrix element. The arrangement and labeling of the spin degrees of freedom is inherited from the parent rotor variables shown in Fig. \ref{fig:2DLatticeSetup}. The generalization to $p>2$ simply replaces the spin-1/2 particles by $p$-state clock variables and $X_{ij}$ and $Z_{ij}$ by generalized Pauli matrices. We specialize to $p=2$ for the rest of the paper; the generalizations are mostly straightforward.

\begin{figure}
\subfigure[]{\includegraphics[width=7cm]{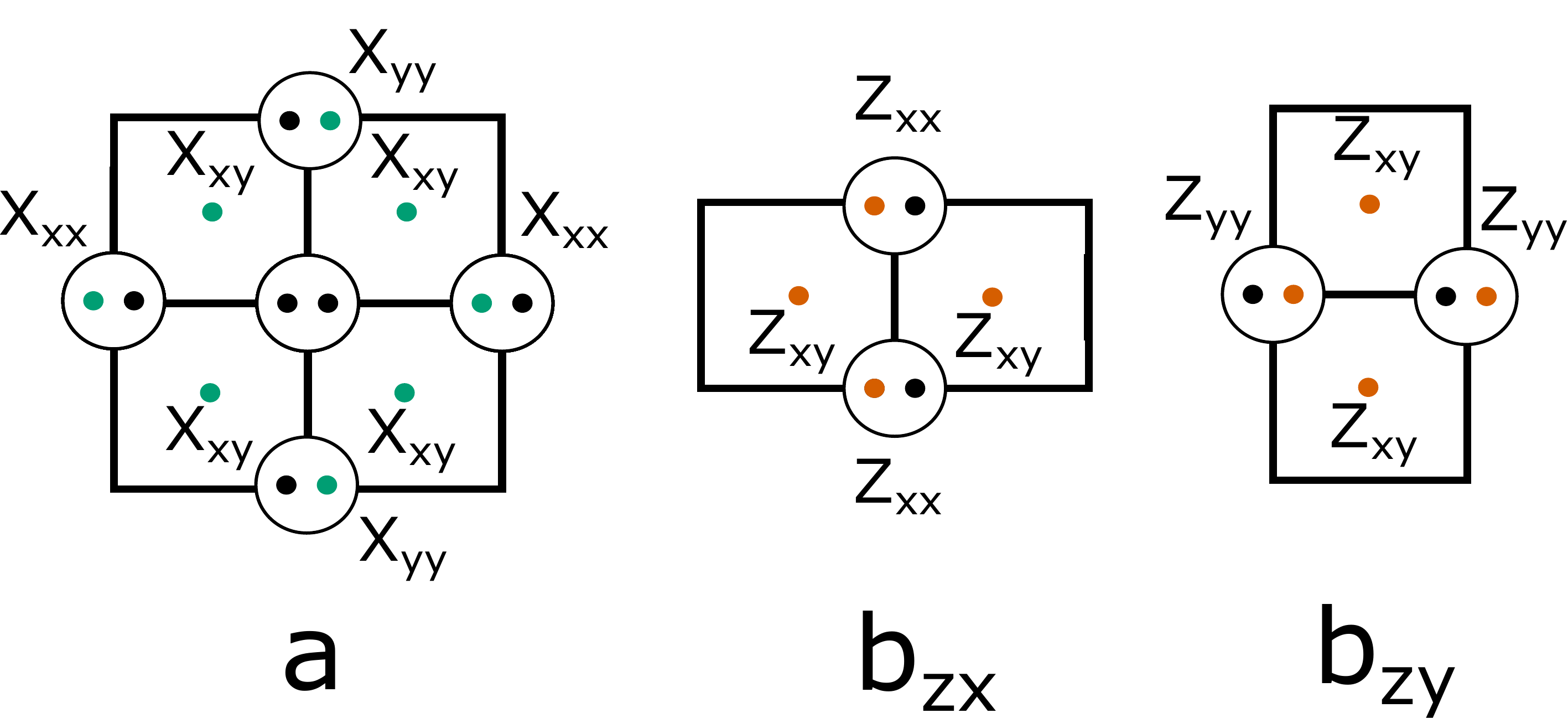}}
\subfigure[]{\includegraphics[width=4.5cm]{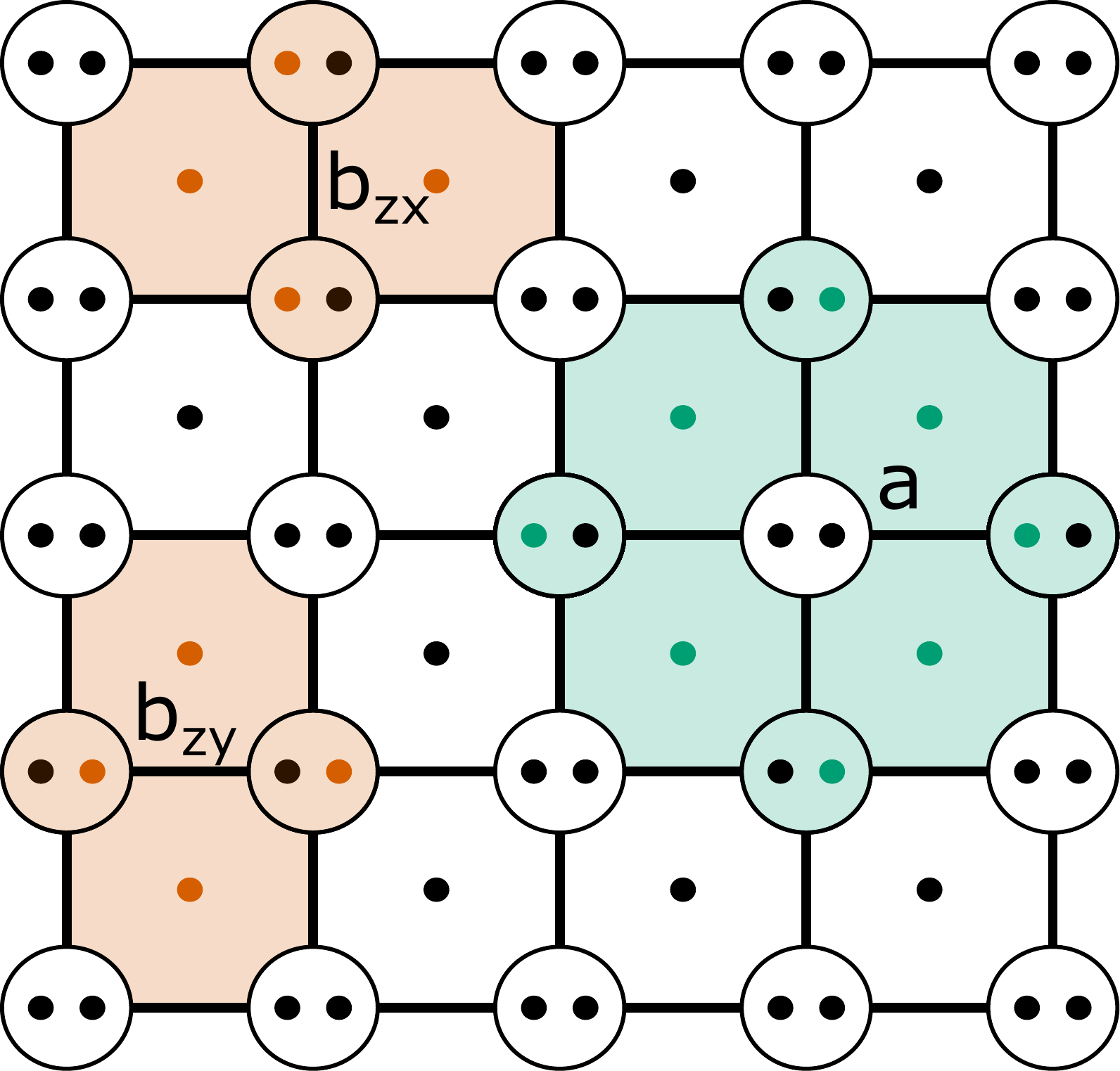}}
\caption{Terms in the $d=2$ Higgsed $(1,1)$ scalar charge theory Hamiltonian (a) in isolation and (b) in the context of the larger lattice. $a$ is associated with the center site and is the product of the eight indicated Pauli $X$ operators, which act on the green spins. The $b_{zi}$ operators are are associated with the center links and are products of four Pauli $Z$ operators acting on the orange spins. The sites/plaquettes on which operators act (in addition to the spins themselves) have been shaded for easier visibility. }
\label{fig:2DScalarHam}
\end{figure}

The magnetic terms $\cos{B_{ij}} \equiv b_{ij}$ in the Hamiltonian \eqref{eqn:MaxwellTerm} are precisely products of $Z$ operators after Higgsing. Their forms, which can be straightforwardly deduced from the form of $B_{ij}$ in Eq. \eqref{eqn:11ScalarBd2}, are shown in Fig. \ref{fig:2DScalarHam}.

The operator in the Gauss' Law term Eq. \eqref{eqn:GaussTerm} of the Hamiltonian is strongly fluctuating because charge 
is condensed. However, since only charge-2 matter fields are condensed, charge parity is still a good quantum number. 
The $U(1)$ version of $H_{Gauss}$ should then be replaced by a mod 2 Gauss' law, which is enforced by the term
\begin{equation}
-U\sum_{\bv{r}} a(\bv{r}) \equiv  -U\sum_{\bv{r}} (-1)^{\sum_{i, j} \Delta_i \Delta_j E_{ij}},
\label{eqn:aterm}
\end{equation}
with $U \propto \tilde{U}$. Above we have suppressed the site labels for $E_{ij}$. In the gauge theory ($U \rightarrow \infty$) language, 
$a$ is constrained to equal 1, which just says that $\sum_{i,j} \Delta_i \Delta_j E_{ij}$ is even, in accordance with Gauss' Law.
Note that charge 1 excitations in this model cost an energy on the order of $U \propto \tilde{U}$, which is due to the fact only charge $p$ excitations
exist below the energy scale $\tilde{U}$ in the $U(1)$ theory. 

Note that several diagonal terms drop out of the expression of Gauss' Law:
\begin{align}
(-1)^{\Delta_i^2 E_{ii}(\bv{r})} &= X_{ii}(\bv{r}+\hat{x}_i )(X_{xx}(\bv{r}))^{-2}X_{xx}(\bv{r}-\hat{x}_i) 
\nonumber \\
&= X_{xx}(\bv{r}+\hat{x}_i )X_{xx}(\bv{r}-\hat{x}_i)
\end{align}
This is precisely the manifestation of the intuition we saw in Sec. \ref{sec:intuition}. The $E_{xx}(\bv{r})$ terms 
were responsible for the charge-2 particles created by $e^{iA_{xx}}$; these terms drop out of the expression for 
$a$ because charge-2 particles are condensed. 

From this expression, we find that the operator $a(\bv{r})$ (which lives on sites) involves 8 spins in $d=2$. Its form is shown in Fig. \ref{fig:2DScalarHam}. The final Hamiltonian is
\begin{align}
H_{2D} = -\frac{1}{g^2}&\sum_{\text{links}}(b_{zx}+b_{zy}) - h_s \sum_{\text{sites},i}X_{ii} \nonumber \\
&- h_p \sum_{\text{plaquettes}} X_{xy} - U \sum_{\text{sites}} a \label{eqn:2DScalarHiggs}
\end{align}
Compared to Eq. \eqref{eqn:MaxwellTerm}, we have set $g_f = g_s = g$ for simplicity and renamed 
$h_f$ ($f$ for ``face") to $h_p$ ($p$ for ``plaquette") to be more appropriate for $d=2$. The $h_s$ and $h_p$
terms induce fluctuations in the gauge field, analogous to the $\tilde{h}_s$ and $\tilde{h}_f$ terms in the parent $U(1)$ theory. 

The Higgs procedure for general $(m,n)$ scalar charge theories is exactly the same as for the $(1,1)$ theory. The large-$V$ limit is still well-defined, $\theta$ can be gauged away, and the condition Eq. \eqref{eqn:PinningA} still results, independent of $m$ and $n$. The operator identifications are therefore identical. The only differences come in the form of Gauss' Law and the magnetic field operators. In particular, only the $m$th power of site operators $X_{ii}$ and the $n$th power of the plaquette operator $X_{xy}$ appear in the $(m,n)$ version of Eq. \eqref{eqn:aterm}. Therefore, the form of $a$ is determined only by the parity of $m$ and $n$. In particular, if $m$ ($n$) is even, the site (plaquette) operators are absent in $a$.

Likewise, the form of the magnetic field term depends on $m$ and $n$. For $m,n \neq 0$, only the $m$th power of $Z_{xy}$ and the $n$th power of $Z_{ii}$ appear in the expressions for $b_{zi}$, and again the form of the $b_{zi}$ depend only on the parity of $m$ and $n$ if both are nonzero. The $(0,1)$ theory has no magnetic field, while the $(1,0)$ theory has a somewhat different form of the magnetic field; these cases must be treated separately but analogously.

In accordance with the intuition from Sec. \ref{sec:intuition}, we have found that the Higgsed theory depends entirely on the parity of $m$ and $n$ (for $m,n \neq 0$). There are therefore five scalar charge theories to consider: the $(1,0)$, $(0,1)$, $(2r+1,2s+1)$, $(2r+1,2s)$, and $(2r,2s+1)$ theories. (Recall that $m$ and $n$ can always be defined to be relatively prime, so there is no $(2r,2s)$ theory.) More generally, for $p > 2$, the Higgsed theory depends on $m$ and $n$ modulo $p$. 

\subsection{$(2r+1,2s+1)$ Scalar Charge Theory}

\begin{figure}
\includegraphics[width=5cm]{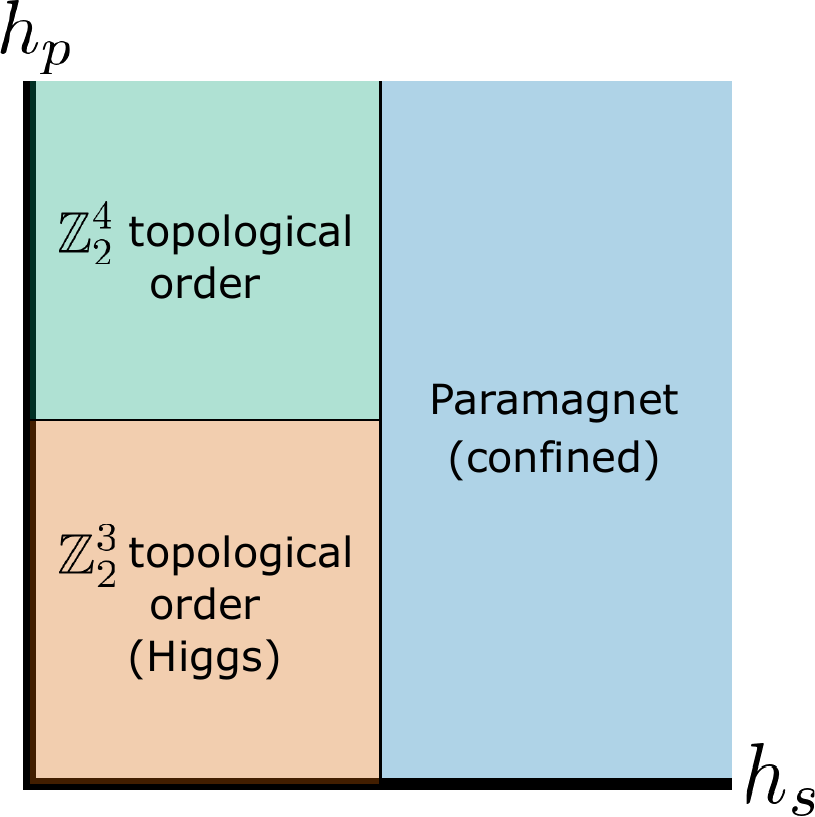}
\caption{Schematic phase diagram at $V = \infty$ for the $d=2$ $(1,1)$ scalar charge theory. The phases at large and small $h$ are accurate, but the phase boundaries and behavior at intermediate coupling are schematic. The direct transition from $\mathbb{Z}_2^3$ to paramagnetism is protected by $C_4$ rotation symmetry.}
\label{fig:PhaseDiagram2DScalarA}
\end{figure}

The central claim of this section is that the Higgsed model Eq. \eqref{eqn:2DScalarHiggs}, which describes the $\mathbb{Z}_2$ Higgs phase of the 
$(2r+1, 2s+1)$ scalar charge theories, has the schematic phase diagram shown in Fig. \ref{fig:PhaseDiagram2DScalarA}. We will show that the Higgs phase has $\mathbb{Z}_2^3$ topological order, which can be driven either into $\mathbb{Z}_2^4$ topological order or to a confined (paramagnetic) phase by tuning the
``Zeeman fields'' $h_s$ and $h_p$. 

\subsubsection{Explicit Solution at $h_s = h_p = 0$}

We first show explicitly that  at $h_s=h_p=0$, the Higgsed $(1,1)$ scalar charge model in $d=2$ has $\mathbb{Z}_2^3$ topological order (three copies of the toric code). This will be shown by computing the ground state degeneracy, the excitations, and their fusion and braiding rules.

At $h_s = h_p = 0$, Eq. \eqref{eqn:2DScalarHiggs} is a commuting projector model and thus is exactly soluble. Ground states $\ket{G}$ must obey the simultaneous but not necessarily independent constraints $(a-1)\ket{G} = 0$ and $(b_{zi}-1)\ket{G} = 0$. The number of ground states is simply $2^{N-C}$ where $N$ is the number of spins and $C$ is the number of independent constraints. For commuting projector models of spin-1/2 particles, each constraint can be encoded as a binary vector such that independent constraints produce linearly independent (over $\mathbb{Z}_2$) vectors; see Ref. \onlinecite{HsiehFractonsPartons} for the details. Therefore $C$ is equal to the rank over $\mathbb{Z}_2$ of the matrix consisting of all these binary vectors. Using this method we checked numerically (for even $L\leq 50$) that the ground state degeneracy of Eq. \eqref{eqn:2DScalarHiggs} on an $L \times L$ torus is $2^{6}$, which is the correct degeneracy for $\mathbb{Z}_2^3$ topological order.

One such ground state is constructed in the string-net picture by starting from the state $\ket{\lbrace X = +1 \rbrace}$ in which all spins are in the $X=+1$ eigenstate. Obviously this satisfies all the $a=1$ constraints but not the $b_{zi}=1$ constraints. One ground state $\ket{0}$ is formed as the superposition
\begin{equation}
\ket{0} = \sum_{\lbrace n_i(\bv{r}) \rbrace \in \lbrace 0, 1 \rbrace^{2L^2}}\prod_{\bv{r},i} (b_{zi}(\bv{r}))^{n_i(\bv{r})}\ket{\lbrace X = +1 \rbrace}
\end{equation}
That is, $\ket{0}$ is a superposition of all possible products of $b_{zi}$ applied to the spin-polarized state $\ket{\lbrace X = +1 \rbrace}$.

The other ground states are, as usual, created by acting on $\ket{0}$ with Wilson loop operators wrapping around the handles of the torus. To understand the ground states, it suffices to understand the excitations of the model, as the Wilson loops can be constructed from the string operators which create pairs of anyons.

\begin{figure}
\subfigure[]{
	\includegraphics[width=8cm]{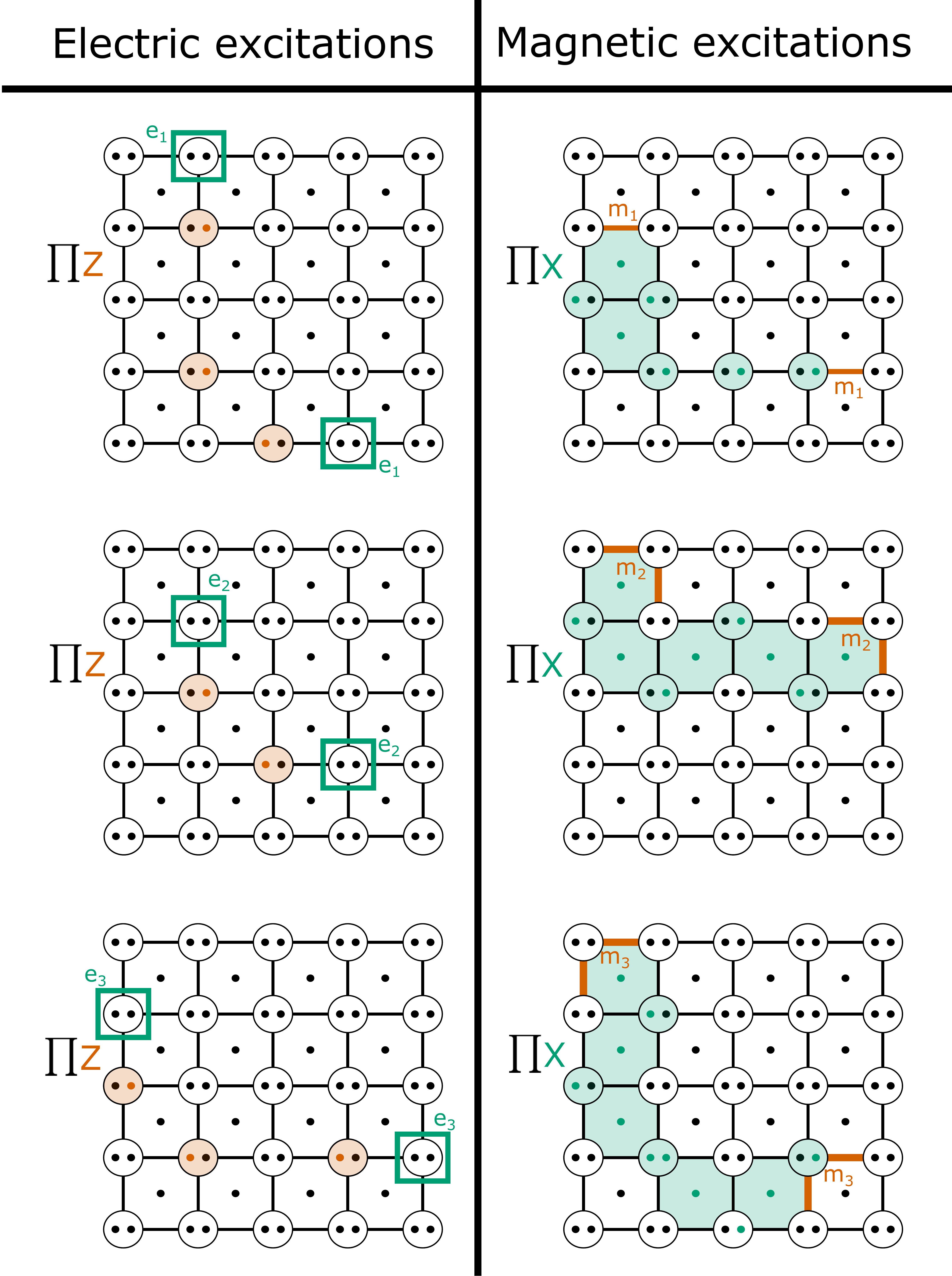}
	\label{fig:2DExcitationPairs}
}
\subfigure[]{
	\includegraphics[width=5cm]{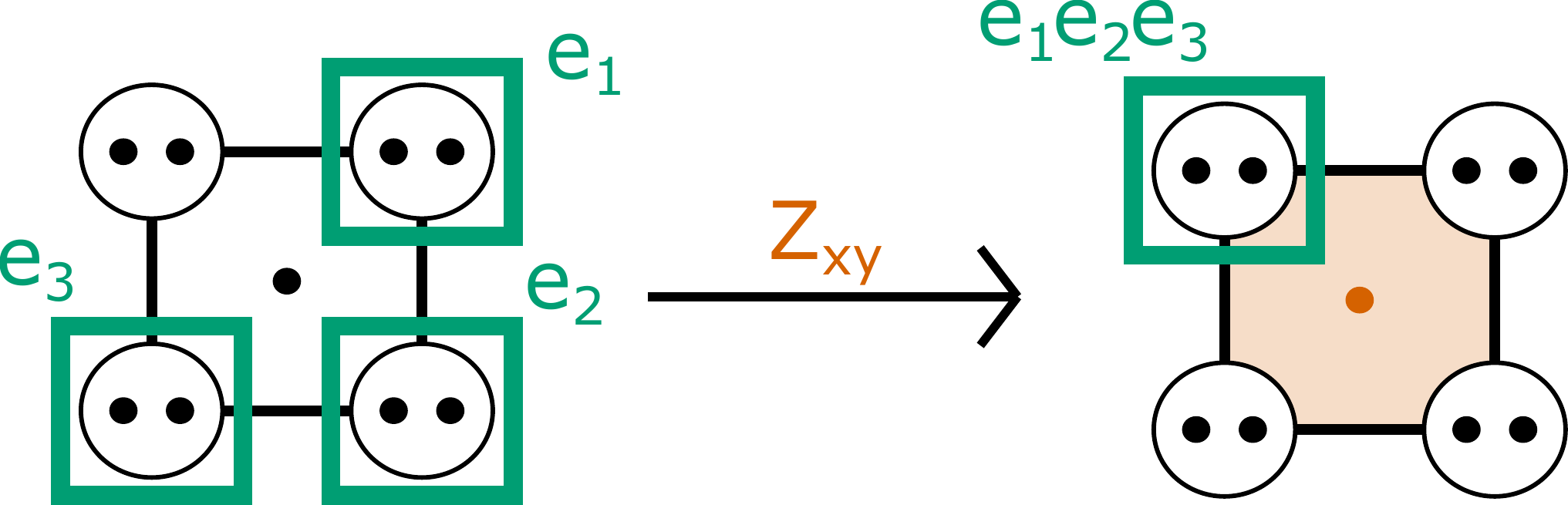}
	\label{fig:ZxyLocalAction}
}
\caption{(a): String operators and the anyons they create in the Higgsed $(1,1)$ scalar charge model. Electric excitations, shown on the left, consist of (green-boxed) sites where the eigenvalue of $a$ is $-1$ and are created by a string of $Z$ operators acting on the orange spins. Magnetic excitations, shown on the right, consist of a set of (heavy orange) links where the eigenvalue of $b_{zi}$ is $-1$ and are created by a string of $X$ operators acting on the green spins.  For the purposes of checking braiding, all the pictures should be regarded as representing the same portion of the square lattice. The sites/plaquettes on which an operator acts have been shaded (light orange/green) for easier visibility. (b) The local action of $Z_{xy}$ on a bound state of $e_1$, $e_2$, and $e_3$ converts the bound state to a single-charge excitation on the top-left site.}
\end{figure}

Consider the state $Z_{ii}(\bv{r})\ket{0}$; it is still an eigenstate of all the terms in the Hamiltonian, but since $Z_{ii}$ anticommutes with $a$, the eigenvalue of $a(\bv{r}\pm \hat{i})$ is $-1$. That is, $Z_{ii}$ creates a pair of electric excitations separated by two sites in the $i$ direction.  If, for the moment, we disregard the action of $Z_{xy}$, this motivates a guess that there are up to four topologically distinct single-electric-charge excitations in the model, one on each of the four sublattices of lattice constant 2. A pair of such excitations is created by a string of $Z_{ii}$ operators separated by two sites; three types of excitations are shown on the left-hand side of Fig. \ref{fig:2DExcitationPairs} . 

Considering only the action of the site operators $Z_{ii}$, it would seem that there are four topologically distinct electric excitations. However, $Z_{xy}$ anticommutes with the four $a$ operators which touch its plaquette. That is, the local action of $Z_{xy}$ converts the bound state of three charges on a single plaquette into a single-charge state on the fourth sublattice, shown in Fig. \ref{fig:ZxyLocalAction}.
Therefore there are only three independent electric anyons, but each one carries a local degree of freedom which is modified by $Z_{xy}$.

The story is similar in the magnetic sector; the independent excitations are slightly more complicated but produced similarly, and are shown on the right-hand side of Fig. \ref{fig:2DExcitationPairs}. Again there are three distinct excitations, and it can be easily checked that $X_{xy}$ applied to the end of a string modifies local degrees of freedom.

It is straightforward to check from the string operators that, as labeled in Fig. \ref{fig:2DExcitationPairs}, $e_i$ and $m_i$ braid as $e$ and $m$ in the toric code and $e_i$ and $m_j$ braid trivially for $i \neq j$. Therefore, this model indeed has $\mathbb{Z}_2^3$ topological order.

\subsubsection{Condensation Transition From Large $h_p$}

The local degree of freedom associated with the action of $X_{xy}$ or $Z_{xy}$ is important in that it allows for transitions to other nontrivial phases. To illustrate the point, we will show that at large $h_p$, two excitations which differ only by a local operator become topologically inequivalent, driving the model to $\mathbb{Z}_2^4$ order. Equivalently, starting at large $h_p$ and reducing it condenses the four-electric-charge bound state of $\mathbb{Z}_2^4$ topological order. A related mechanism will occur in several other models that we study.

\begin{figure}
\subfigure[ ]{
	\includegraphics[width=6cm]{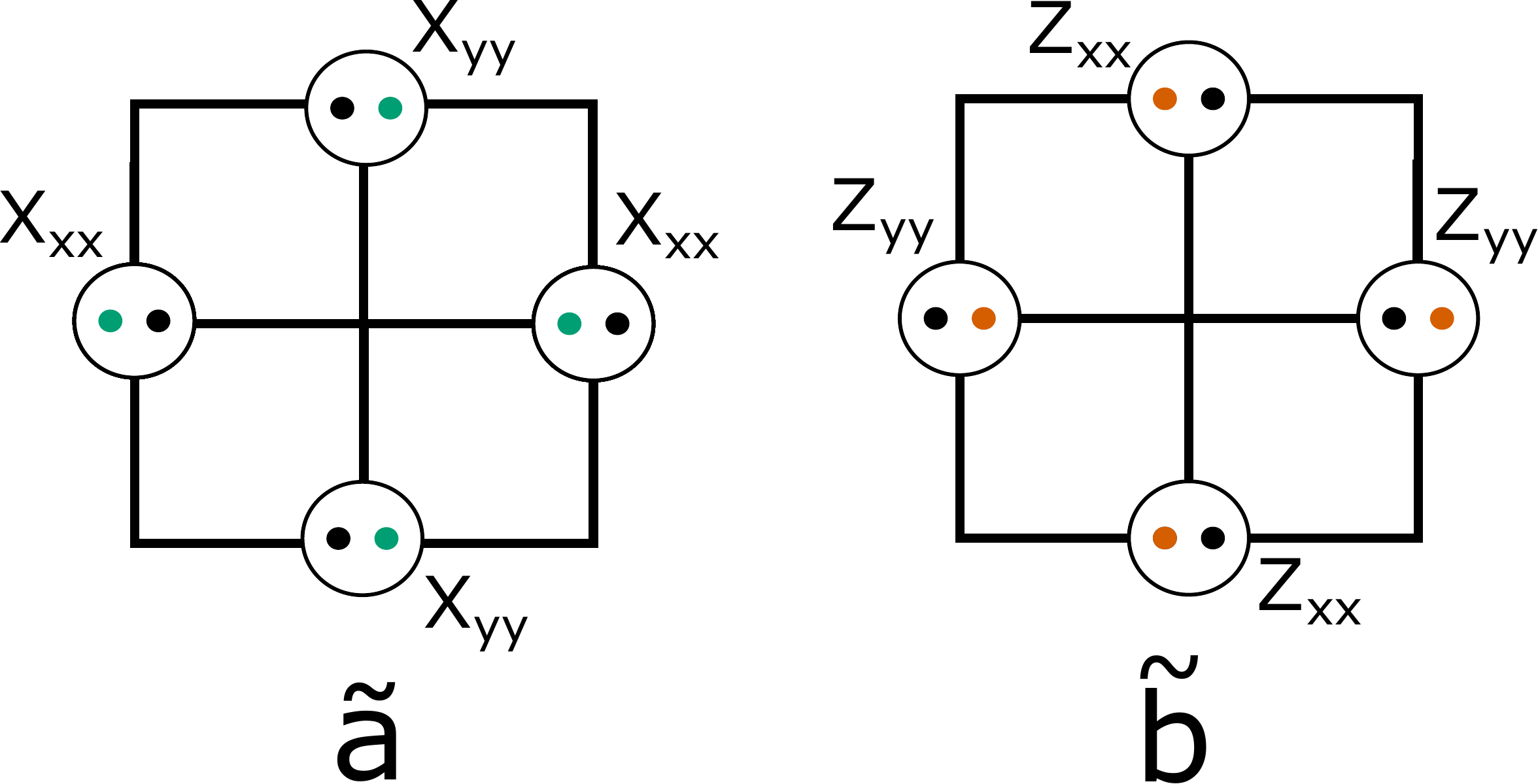}
	\label{fig:2DDecoupledTC}
}

\subfigure[ ]{
	\includegraphics[width=3.5cm]{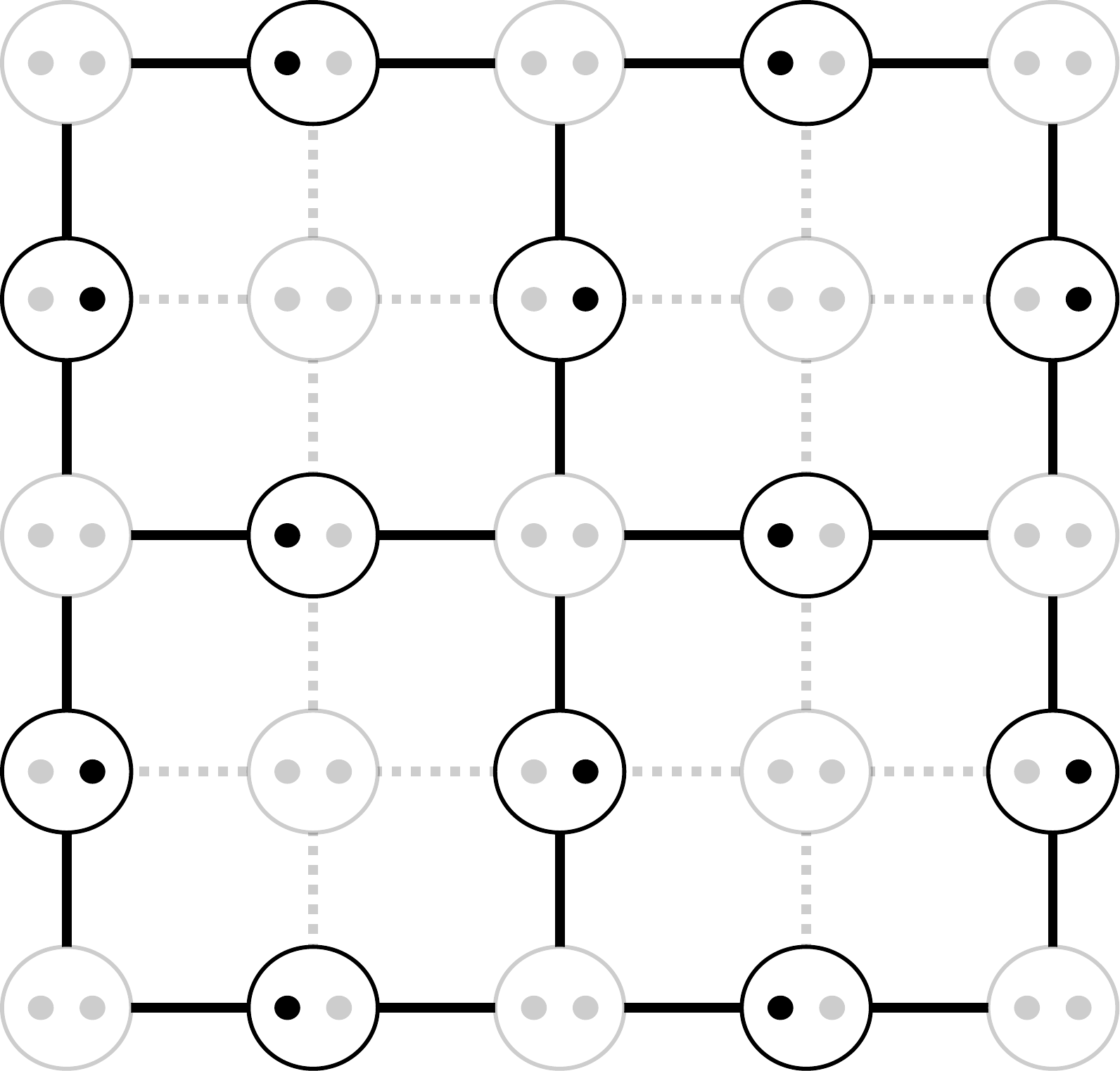}
	\label{fig:2DSublattices}
}
\caption{(a) Terms in the Hamiltonian at large $h_p$. The operators are products of the indicated spins; the plaquette spins are frozen out. (b) Sublattice of spins (dark) which form one of four decoupled copies of the toric code at large $h_p$. The large-$h_p$ effective Hamiltonian Eq. \eqref{eqn:2DScalarLargeHpHamiltonian} does not couple the dark spins to any others. Some links of the lattice have been lightened to make it easier to see the relation to the toric code.}
\end{figure}

We begin by simply finding the effective Hamiltonian at large $h_p$. At zeroth order, the low-energy subspace consists of any spin configuration on the sites and all the plaquette spins pinned to the eigenstate $X_{xy} = +1$. This leads to extensive ground state degeneracy, with low-energy states labeled entirely by the site spin configuration. This degeneracy is split in degenerate perturbation theory by $U, 1/g^2  >0 $. The lowest-order contributions are first-order in $U/h_p$ and fourth-order in $1/g^2h_p$; the effective Hamiltonian is
\begin{equation}
H_{eff} = -\sum_{\text{sites}}\(U\tilde{a} + K \tilde{b} \)
\label{eqn:2DScalarLargeHpHamiltonian}
\end{equation}
where $\tilde{a}$ and $\tilde{b}$ are the operators shown in Fig. \ref{fig:2DDecoupledTC} and $K \sim 1/(g^8h_p^3)$

We claim that this effective Hamiltonian is precisely \textit{four} copies of the toric code. To see this, consider the sublattice of spins shown in Fig. \ref{fig:2DSublattices}. If we interpret these spins as living on the links of a lattice of length 2, as indicated by the set of darkened bonds in Fig. \ref{fig:2DSublattices}, by inspection each $\tilde{a}$ acts on exactly one such sublattice. On that sublattice, it acts exactly as a toric code star operator. Likewise, each $\tilde{b}$ acts on a single sublattice as a toric code plaquette operator. Hence each of the four distinct sublattices is one copy of the toric code.

Notice that all the operators that create electric excitations at large $h_p$ also create electric excitations on the same sites at $h_p=0$. However, as we saw previously, at $h_p=0$ the bound state of three of those excitations is equivalent to the fourth \textit{up to a local application of} $Z_{xy}$. By contrast, at large $h_p$, $Z_{xy}$ takes the system out of the low-energy subspace; accordingly, the process in which a bound state of three types of charge is fused into the fourth type of charge is no longer allowed. That is, at small $h_p$, this fourth type of electric charge is equivalent to the fusion of the other three types, but at large $h_p$ it is a topologically distinct excitation.

\subsubsection{Large-$h_s$ limit}
Finally, we may ask about the topological order of the large $h_s$ limit. In this case, the electric sector is confined because electric strings have finite tension, and one may expect a trivial paramagnetic limit. We demonstrate this using degenerate perturbation theory.

The low-energy subspace consists of all states with the site spins in the $X_{ii} = +1$ eigenstate. The $a$ and $h_p$ terms both contribute at first order in perturbation theory; $a$ is a product of $X_{xy}$ around a plaquette of the dual lattice, and $h_p$ is a longitudinal magnetic field. To first order, the model is classical and fully gapped. Importantly, the $b_{zi}$ terms contribute in degenerate perturbation theory only at $L$th order in $1/(g^2h_s)$, with $L$ the linear system size (At this order, the string of $b_{zi}$ operators consists only of face spins and thus commutes with the $h_s$ terms). In this limit, $1/g^2$ needs to be larger than $U$ and $h_p$ by an amount exponentially large in the system size in order to have an effect comparable to the gap in the first-order model. Hence, in the thermodynamic limit the system is indeed a trivial paramagnet.

As can be seen from Fig. \ref{fig:2DExcitationPairs}, $C_4$ rotation symmetry rotates electric particles into each other. If this symmetry is preserved, then all of the electric particles should condense at the same time and $\mathbb{Z}_2^3$ transitions directly to a paramagnetic phase. Breaking rotational symmetry generally leads to intermediate phases.

\subsection{$(1,0)$ Scalar Charge Theory}

Recall that the $(1,0)$ scalar charge theory has no (nontrivial) plaquette degrees of freedom. The Higgsed Hamiltonian can be checked straightforwardly to be equal to the effective Hamiltonian Eq. \eqref{eqn:2DScalarLargeHpHamiltonian} of the large-$h_p$ limit of the Higgsed $(2r+1,2s+1)$ scalar charge theory, with $K$ replaced by $1/g^2$ (see Fig. \ref{fig:2DDecoupledTC} for operator definitions). This Higgsed theory is therefore exactly four decoupled copies of the $\mathbb{Z}_2$ toric code. It confines at large $h_s$. 

\section{Vector Charge Higgs in $d=2$}
\label{sec:HiggsOtherd2}

The Higgsing procedure for the vector charge theories has only minor differences from the scalar charge theories. We will comment on those differences, then discuss the theories that produce nontrivial Higgs phases; see Appendix \ref{app:trivialHiggs} for discussion of trivial Higgs phases. All the Higgs phases are summarized in Table \ref{tab:HiggsPhases}.

\subsection{Changes to the Higgsing Procedure}
\label{subsec:vectorHiggsChanges}

As in the scalar charge theories, the matter field can still be gauged away, and the operator identifications all go through; 
if $m$ ($n$) is even, then the site (plaquette) spins drop out of Gauss' Laws and the plaquette (site) spins drop out of the magnetic field. 
In $d=2$, by the same arguments as for the scalar charge theories, the Higgsed $(m,n)$ vector charge theory for nonzero $m$ and $n$ 
is determined entirely by the parity of $m$ and $n$ (or, for $p>2$, the values of $m$ and $n$ modulo $p$). There are therefore 
five total theories to consider. This changes slightly in $d=3$, as will be discussed in Sec. \ref{sec:oddOddVector_d3}.

\subsection{$(2r,2s+1)$ Vector Charge}

Note that this case includes the $(2,1)$ theory, which is the theory with continuous rotational invariance.

\begin{figure}
\subfigure[]{
	\includegraphics[width=7cm]{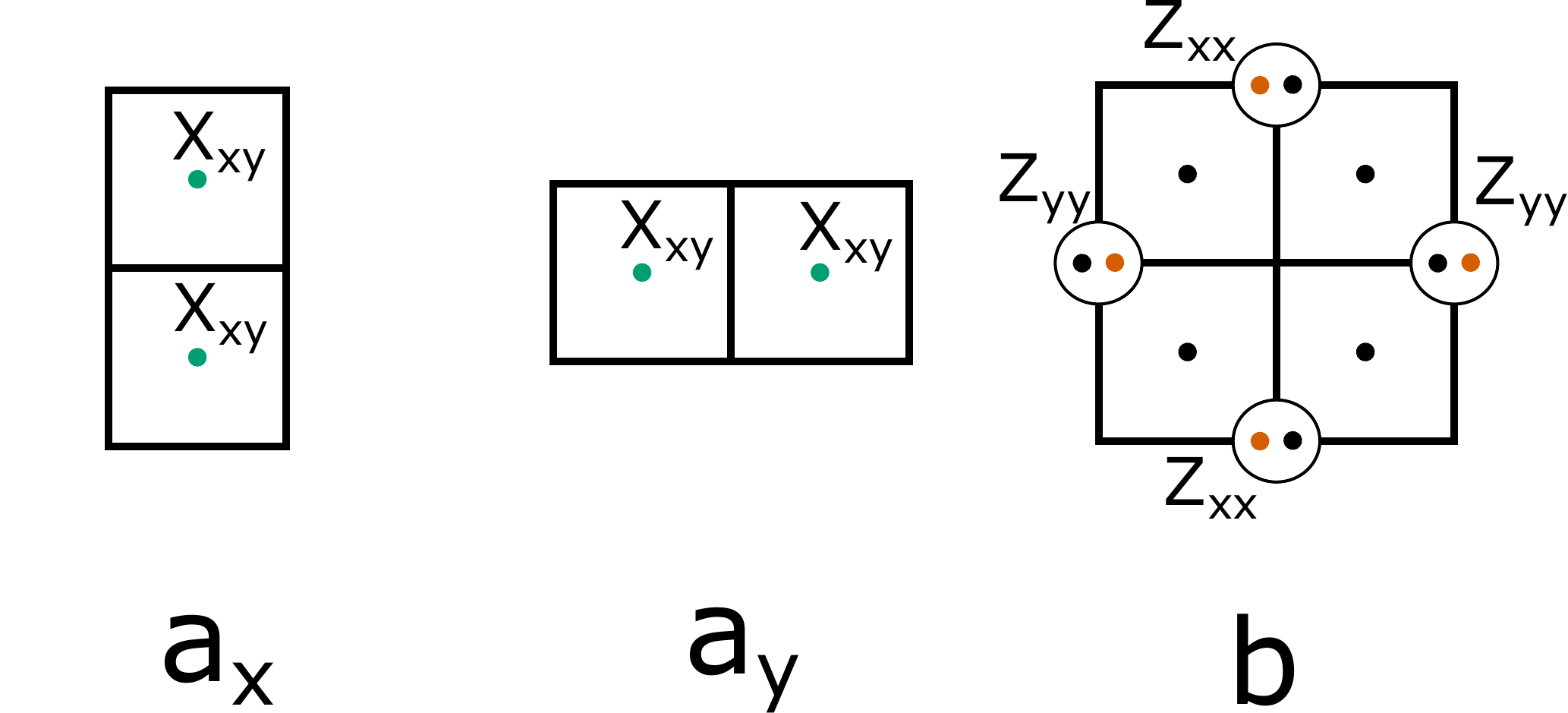}
	\label{fig:2DVectorAHam}
}
\subfigure[]{
	\includegraphics[width=4cm]{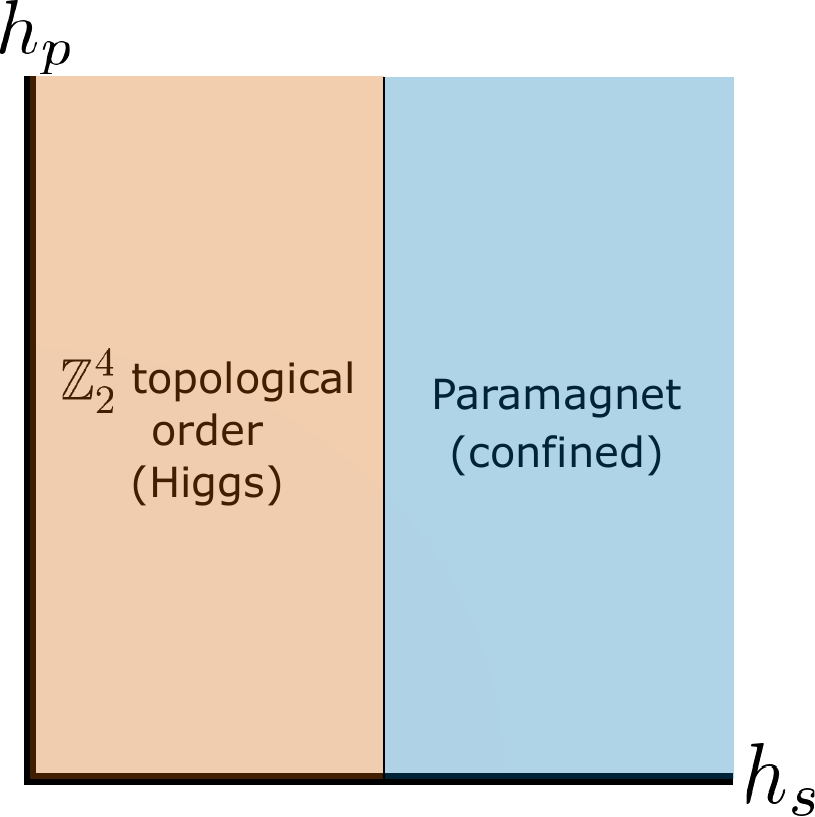}
	\label{fig:PhaseDiagram2DVectorA}
}
\caption{(a) Terms in the Hamiltonian and (b) phase diagram of the $d=2$ Higgsed $(2r,2s+1)$ vector charge Hamiltonian. $a_i$ are associated with the center links and is the product of the two indicated Pauli $X$ operators, which act on the green spins. The $b$ operator is associated with the center site and is a product of the indicated Pauli $Z$ operators on the four orange spins. The direct transition in (b) is protected by $C_4$ rotation symmetry.}
\end{figure}

The Higgsed model has Hamiltonian
\begin{align}
H = -\frac{1}{g^2}&\sum_{\text{sites}} b  - U \sum_i \sum_{i-\text{links}}a_i  \nonumber \\
&- h_s \sum_{\text{sites},i}X_{ii} - h_p \sum_{\text{plaquettes}}X_{xy}
\label{eqn:vectorA2DHam}
\end{align}
where the forms of the operators $a_i$ and $b$ are shown in Fig. \ref{fig:2DVectorAHam}. The site and plaquette spins are decoupled.

By inspection the plaquette sector of this model is the classical Ising model in a magnetic field, which has a unique ground state for all $h_p \neq 0$.

As for the site sector, for $h_s \ll 1/g^2$, i.e. in the Higgs phase, we can perform degenerate perturbation theory. The $h_s$ term contributes at fourth order in $1/g^2h_s$ and produces the same model as the large-$h_p$ limit of the $(2r+1,2s+1)$ scalar charge model Eq. \eqref{eqn:2DScalarLargeHpHamiltonian} (compare $b$ in Fig. \ref{fig:2DVectorAHam} to $\tilde{b}$ in Fig. \ref{fig:2DDecoupledTC}). Its topological order is four copies of the toric code, i.e. $\mathbb{Z}_2^4$ lattice gauge theory. At large $h_s$, we simply have a trivial paramagnet. The phase diagram is shown in Fig. \ref{fig:PhaseDiagram2DVectorA}; the direct transition to a paramagnet is protected by $C_4$ rotation symmetry, which permutes the decoupled copies of toric code.

\subsection{$(2r+1,2s+1)$ Vector Charge}

\begin{figure}
\subfigure[]{
	\includegraphics[width=7cm]{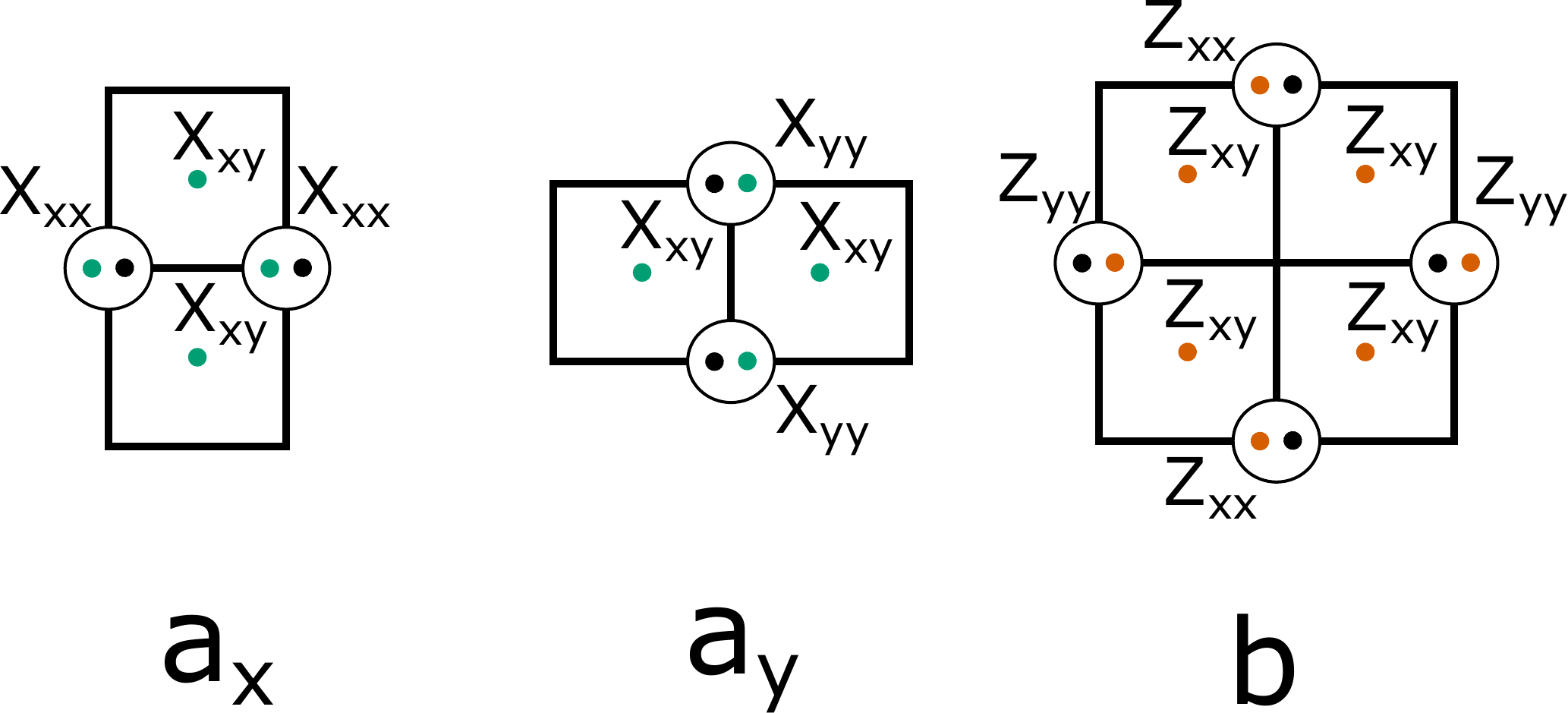}
	\label{fig:VectorEHam2D}
}
\subfigure[]{
	\includegraphics[width=4cm]{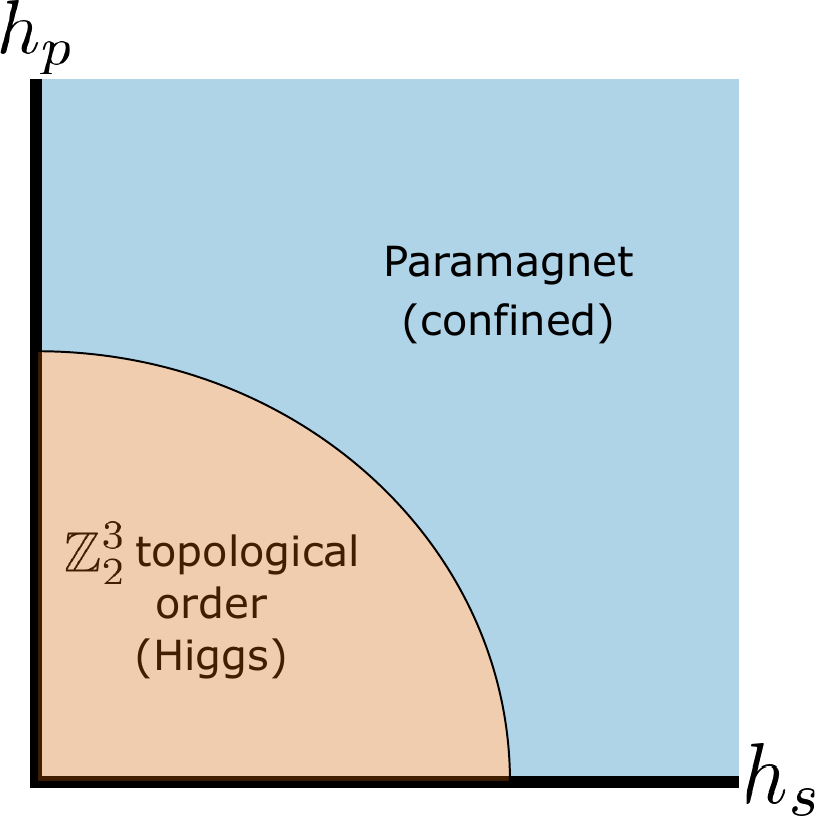}
	\label{fig:PhaseDiagram2DVectorE}
}
\caption{(a) Terms in the Hamiltonian and (b) schematic phase diagram of the $d=2$ Higgsed $(2r+1,2s+1)$ vector charge theory Hamiltonian. $a_i$ are associated with the center bonds and is the product of the four indicated Pauli $X$ operators, which act on the green spins. The $b$ operator is associated with the center site and is a product of the indicated Pauli $Z$ operators on the eight orange spins. The direct transition to a paramagnet in (b) is protected by $C_4$ rotational symmetry.}
\end{figure}

The Hamiltonian for the $(2r+1,2s+1)$ vector charge theory takes the same form as the Hamiltonian Eq. \eqref{eqn:vectorA2DHam} for the $(2r,2s+1)$ theory, but $a_i$ and $b$ take different forms because of the difference in factors of 2 in Gauss' Law and the magnetic field. Their forms are shown in Fig. \ref{fig:VectorEHam2D}.

At $h_s = h_p=0$, the Higgsed $(2r+1,2s+1)$ vector charge theory is equivalent to the model obtained in the Higgsed $(2r+1,2s+1)$ scalar charge theory, as can be 
seen by comparing Figs. \ref{fig:VectorEHam2D} and \ref{fig:2DScalarHam} and performing a global spin rotation. The Higgs phase of 
this model therefore also has $\mathbb{Z}_2^3$ topological order.

However, the ``electric field" terms $hX_{ij}$ are not dual to those in the $(2r+1,2s+1)$ scalar charge theory, so their effects should be studied as well. Since each magnetic excitation is created by an $X_{ii}$ operator, large $h_s$ should condense the entire magnetic sector, confining the electric sector and leading to a trivial paramagnet. In degenerate perturbation theory, this arises from the fact that $X_{xy}$ and the $a_i$ contribute at first order (the site spins simply drop out of the $a_i$) and $b$ only contributes at order $L^2$ in $1/g^2h_s$, where $L$ is the linear system size. Therefore, up to exponentially small corrections in the system size, the first-order effective Hamiltonian, which describes a classical paramagnet, describes the system accurately.

At large $h_p$, the story is similar; the $X_{ii}$ and $a_i$ contribute at first order in perturbation theory and describe a classical paramagnet of the site spins, while $b$ contributes at $L$th order and can be neglected.

The phase diagram is summarized in Fig. \ref{fig:PhaseDiagram2DVectorE}. Much like the Higgsed $(2r+1,2s+1)$ scalar charge theory, the direct transition to a paramagnet is protected by the $C_4$ rotational symmetry of the square lattice, which permutes magnetic particles.

\section{$(2r,2s+1)$ Scalar Charge Higgs in $d = 3$ and Fracton Order}

We now turn to the $d=3$ models. For the scalar charge models, the Higgsing procedures are all identical to the $d=2$ case, as is the argument that the resulting model depends only on the parity of $m$ and $n$.  Accordingly, we will simply state the resulting Hamiltonian and then analyze the phase diagram. 

We begin by studying the models which have fractonic Higgs phases.

\subsection{$(0,1)$ Scalar Charge}

\begin{figure}
	\includegraphics[width=7cm]{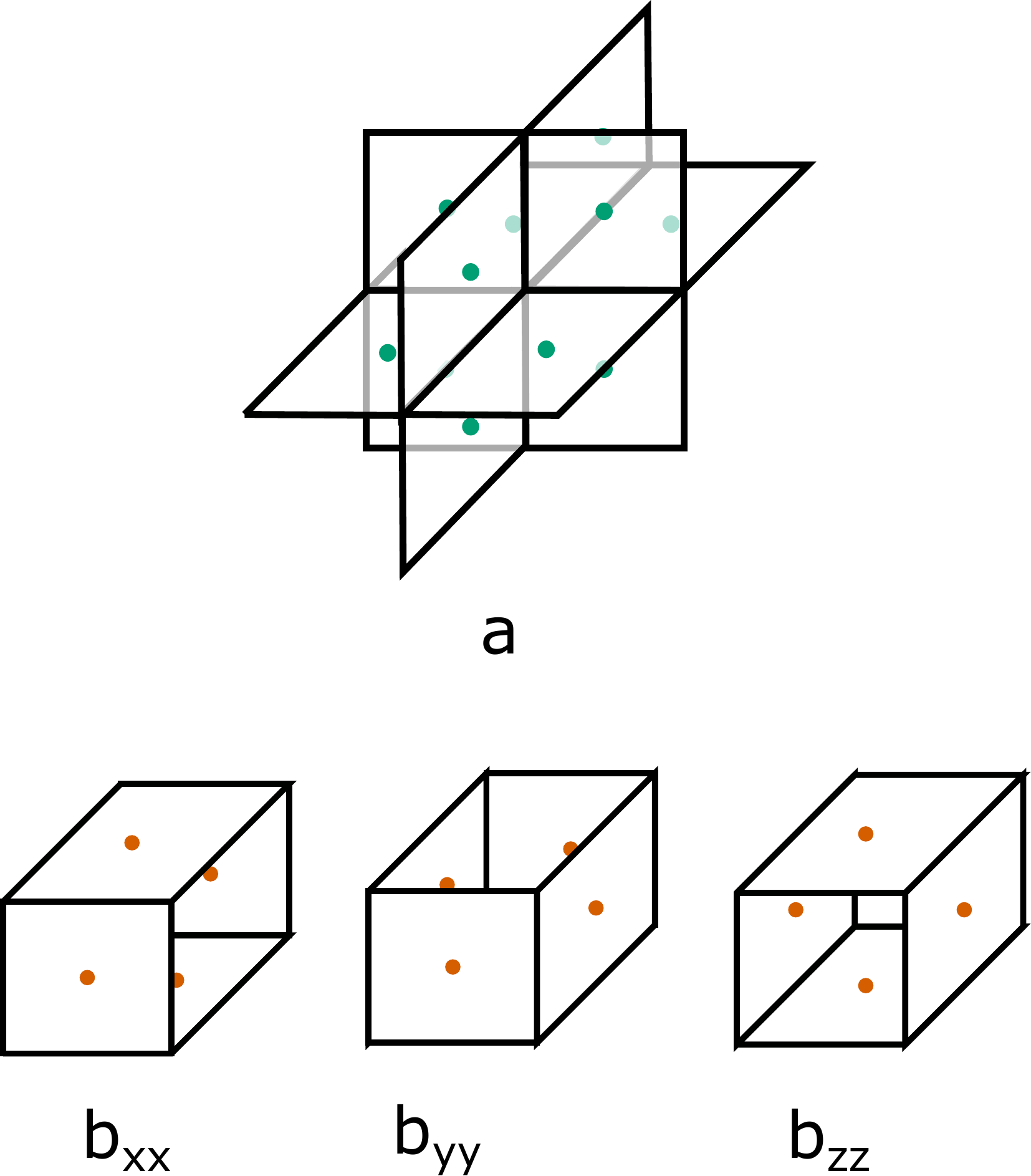}
	\caption{Terms in the Hamiltonian Eq. \eqref{eqn:XCube} for the Higgsed $(0,1)$ scalar charge model in $d=3$. $a$ is a product of Pauli $X$ operators on the twelve green spins, while the $b_{ii}$ are products of four Pauli $Z$ operators on the orange spins.}
	\label{fig:XCube}
\end{figure}

The Hamiltonian of the $d=3$ Higgsed $(0,1)$ scalar charge model is
\begin{equation}
H = -U\sum_{\text{sites}} a - \frac{1}{g^2}\sum_{\text{cubes},i} b_{ii} - h_f \sum_{\text{faces},i<j} X_{ij}
\label{eqn:XCube}
\end{equation}
where $\tilde{a}$ and $\tilde{b}_{ii}$, shown in Fig. \ref{fig:XCube}, are just $a$ and $b_{ii}$ with the site operators frozen out. This is precisely the X-cube model\cite{VijayGaugedSubsystem} on the dual cubic lattice. Accordingly, this phase has fracton order at small $h_f$. At large $h_f$, the system becomes paramagnetic.

The $(0,1)$ scalar charge model is the natural $U(1)$ generalization of the X-cube model, so the relationship between them is not surprising. However, it can be checked that the $(0,1)$ scalar charge model has pointlike magnetic monopoles and may therefore suffer from confinement; the relevance of the monopole creation operators must be studied in more detail. The X-cube phase, on the other hand, is known to be stable.

\subsection{$(2r+2,2s+1)$ Scalar Charge}

The $(2r+2,2s+1)$ scalar charge model (for $r,s \geq 0$) must be considered separately from the $(0,1)$ theory because of the presence of the $A_{ii}$ degrees of freedom in the $U(1)$ theory. 

\begin{figure}
	\includegraphics[width=6cm]{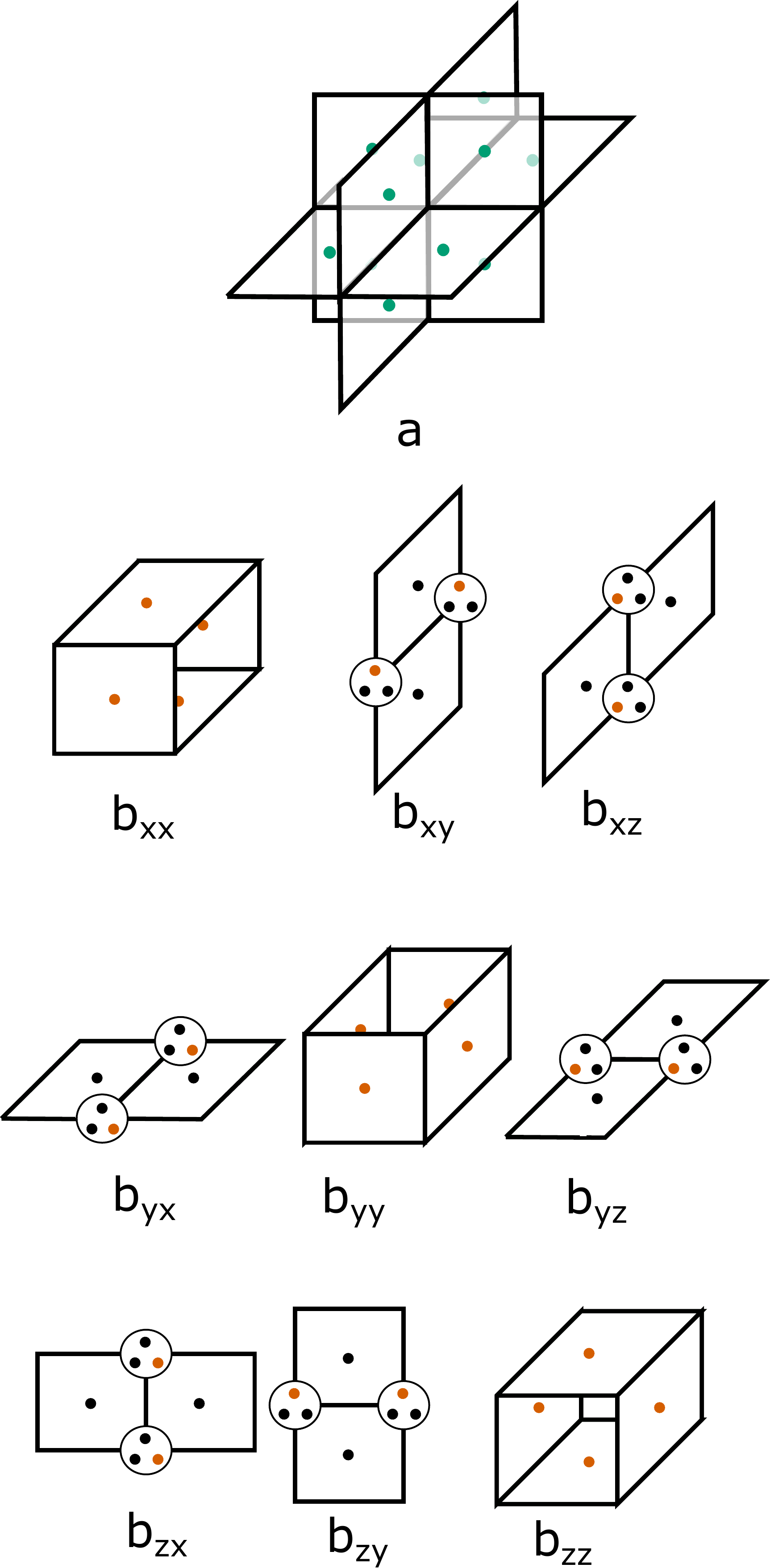}
\caption{Terms in the Hamiltonian for the Higgsed $(2r,2s+1)$ scalar charge theory in $d=3$. The $a$ operators are associated with the sites and are products of twelve $X_{ij}$ operators acting on the green spins. The $b_{ij}$ operators are products of two (off-diagonal terms, associated with links) or four (diagonal terms, associated with cubes) Pauli $Z_{ij}$ operators acting on the orange spins.}
	\label{fig:ScalarHamEvenOdd3D}
\end{figure}

The Hamiltonian for the $d=3$ Higgsed $(2r+2,2s+1)$ scalar charge theory is
\begin{align}
H = -U &\sum_{\text{sites}} a - \frac{1}{g^2}\sum_{\text{cubes}, i}b_{ii}  - \frac{1}{g^2}\sum_{\text{links},i,j}b_{ij}\nonumber \\
& - h_s \sum_{\text{sites},i}X_{ii} - h_f \sum_{\text{faces},i<j}X_{ij}
\label{eqn:3DScalarHam}
\end{align}
with the operators $a$ and $b_{ij}$ defined in Fig. \ref{fig:ScalarHamEvenOdd3D}. 

By inspection, the face spins form the X-cube model Eq. \eqref{eqn:XCube} (c.f. Fig. \ref{fig:XCube}), and the site spins form decoupled, interpenetrating planes of transverse field 2+1-dimensional Ising models. The various decouplings in this model are, of course, fine-tuned. In particular, there are operators in the $U(1)$ theory which are irrelevant (in the renormalization group sense) in the photon phase but which will couple the site spins to the face spins (as well coupling the planes of Ising model to each other). However, the topological stability of the X-cube model ensures that these operators do not destroy the fracton order.

The $2+1$-dimensional Ising order of the site degrees of freedom, on the other hand, is fine-tuned, with $3L$ degenerate ground states at $h_s=0$. The fate of this degeneracy is non-universal and depends on what operators are added to the theory.

We have found an infinite class of $U(1)$ scalar charge models with a fractonic Higgs phase. Note that among the scalar charge theories, the only theory with continuous rotational invariance is the $(1,1)$ theory, which is not in the class we consider here. We will find that this is a general feature of the models we consider - X-cube order is never obtained by Higgsing a model with continuous rotational symmetry.

\section{$(2r+1,2s+1)$ Scalar Charge Higgs in $d=3$}
\label{sec:oddOddScalar_d3}

For the next two sections, we will focus on the $(2r+1,2s+1)$ scalar and vector charge models. Their Higgs phases are described by interesting models for multiple copies of $d=3$ toric code topological order, and they have rich phase diagrams.

\begin{figure}
\subfigure[]{
	\includegraphics[width=5.5cm]{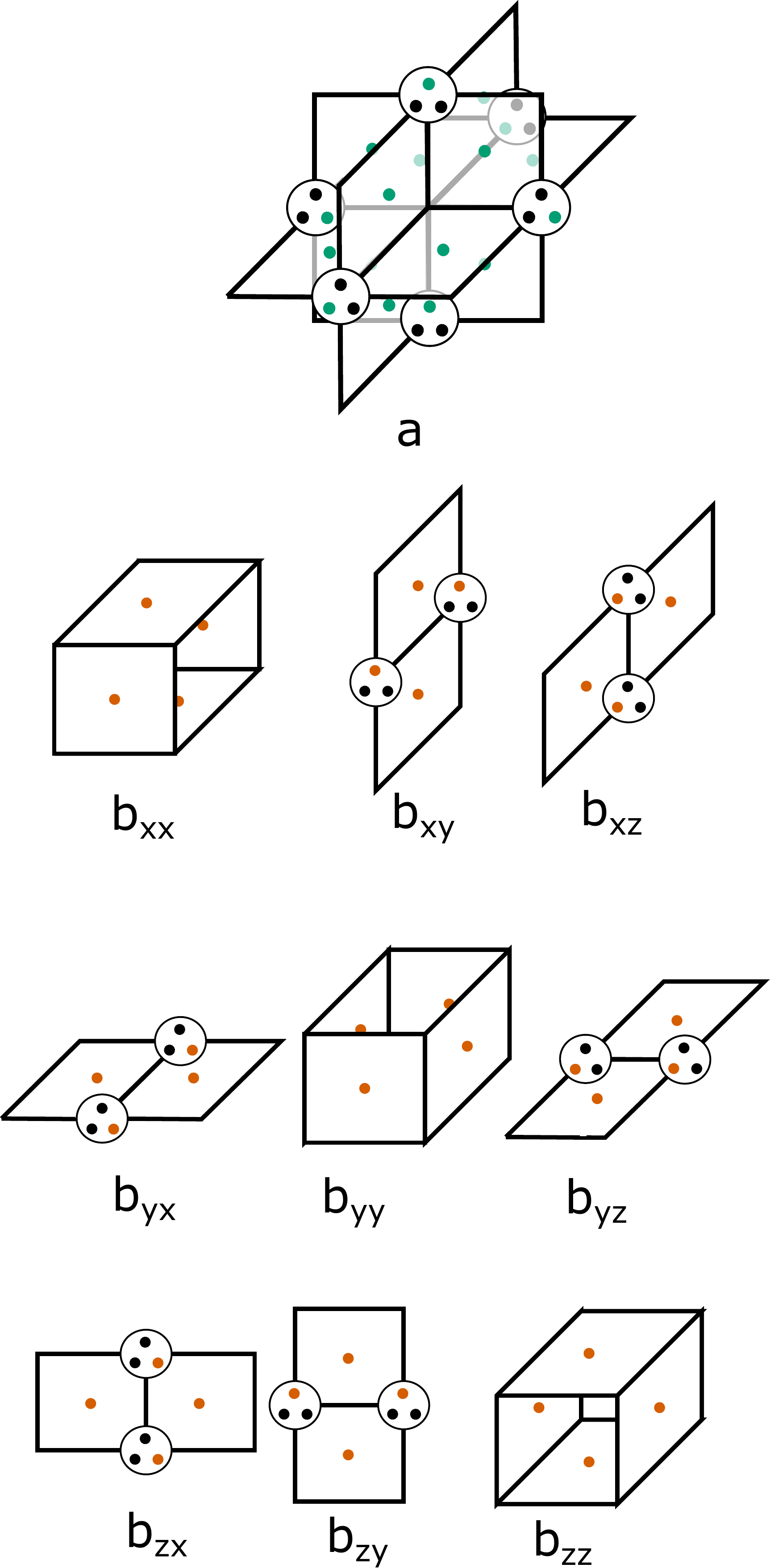}
	\label{fig:3DScalarHam}
}
\subfigure[]{
	\includegraphics[width=4cm]{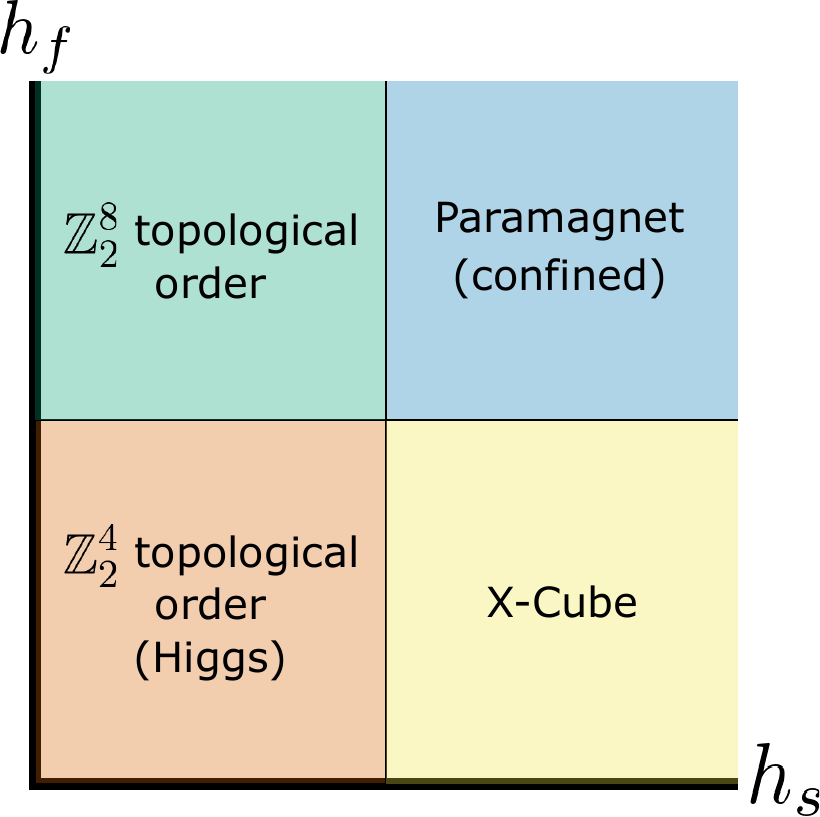}
	\label{fig:PhaseDiagram3DScalarA}
}
\caption{(a) Terms in the Hamiltonian and (b) schematic phase diagram of the $d=3$ Higgsed $(2r+1,2s+1)$ scalar charge theory Hamiltonian. $a$ are associated with the sites and is the product of eighteen Pauli $X$ operators, which act on the green spins. The $b_{ij}$ operators are products of four Pauli Z operators acting on the orange spins. The diagonal components $b_{ii}$ are associated with cubes and the off-diagonal components are associated with links. The direct transitions in (b) are protected by cubic rotation symmetry.}
\end{figure}

The $\mathbb{Z}_2$ Higgs phase of the $(2r+1,2s+1)$ scalar charge model produces an interesting model with $\mathbb{Z}_2^4$ topological order, and 
upon breaking rotational invariance it can be driven to a fracton phase -- the X-cube model. Its phase diagram is shown schematically in Fig. \ref{fig:PhaseDiagram3DScalarA}.

The Hamiltonian for the $d=3$ Higgsed $(2r+1,2s+1)$ scalar charge theory is the same form Eq. \eqref{eqn:3DScalarHam} as for the $(2r+2,2s+1)$ models we considered in the previous section, but the operators $a$ and $b_{ij}$ are defined differently - they are shown in Fig. \ref{fig:3DScalarHam}. The operator $a$ is an 18-spin operator, while all the $b_{ij}$ are 4-spin operators. We now analyze this model in various limits.

\subsection{Higgs Phase}

We show explicitly that at $h_f=h_s=0$, this model realizes $\mathbb{Z}_2^4$ topological order, i.e. it is equivalent to 
four copies of the $\mathbb{Z}_2$ toric code.

At $h_f=h_s=0$ we again have a commuting projector model and the ground state degeneracy can be 
computed as in the $d=2$ models; we find that it is $2^{12}$ on the $L\times L \times L$ $3-$torus for $L$ even. 
This is consistent with $\mathbb{Z}_2^4$ topological order.

As usual, one ground state is obtained by starting from the state with all spins in the $X_{ij}=+1$ eigenstate 
and superposing over all applications of $b_{ij}$ operators on this reference state. The other ground states 
are obtained by applying Wilson loops around handles of the 3-torus, so we turn to the excitations.

\begin{figure*}
\includegraphics[width=9cm]{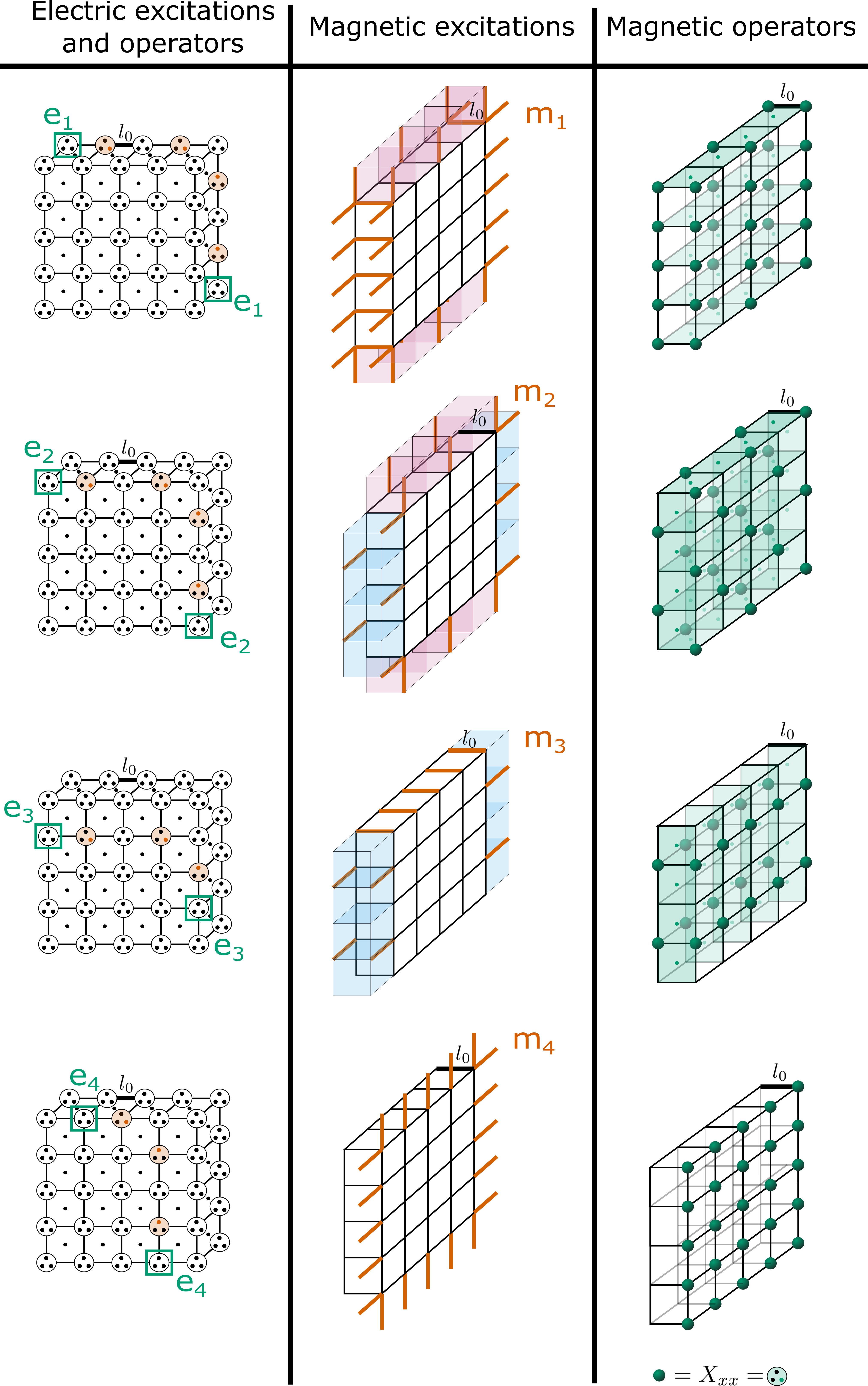}
\caption{String and membrane operators and excitations in the Higgsed $d=3$ $(2r+1,2s+1)$ scalar charge model. The bold link labeled $l_0$ is the same link in all images so that braiding statistics can be checked; $e_i$ excitations braid nontrivially with $m_i$ strings but no others. Left: pairs of electric particles (green boxes) are created by a string of $Z$ operators acting on the orange spins (relevant sites shaded in orange), and have $a=-1$. Middle: excitation patterns of magnetic strings. The orange links have either $b_{yx}$ or $b_{zx} = -1$, as appropriate for the link orientation. Pink cubes have $b_{xx}=b_{zz}=-1$, and blue cubes have $b_{xx}=b_{yy}=-1$. Right: membrane operators, consisting of products of $X_{ij}$ on the green spins, for producing the magnetic strings shown in the middle. The membrane operators for $m_1$ and $m_2$ include all $X_{xz}$ operators within the membrane, while those for $m_2$ and $m_3$ include all $X_{xy}$ operators within the membrane.}
\label{fig:ExcitationPairs3D}
\end{figure*}

The behavior of this model in the electric sector is similar to the $d=2$ $(2r+1,2s+1)$ scalar charge theory; 
$Z_{ii}$ creates a pair of excitations which can hop by two sites in any direction. Temporarily ignoring the 
face spins, there are eight obvious electric excitations, created by acting with $Z_{ii}$ on any of the eight 
sublattices with lattice constant 2. Four are shown on the left side of Fig. \ref{fig:ExcitationPairs3D}.

As in the $d=2$ $(2r+1,2s+1)$ scalar charge theory, these excitations are not all independent. The local 
action of $Z_{ij}$ is the same as in $d=2$ (see Fig. \ref{fig:ZxyLocalAction}); it locally turns any three 
excitations on the sites of a single face into a single excitation on the fourth site of that face. Straightforward 
counting shows that this leaves four independent electric excitations, shown in Fig. \ref{fig:ExcitationPairs3D}, 
and that $Z_{ij}$ toggles local degrees of freedom for these excitations.

As expected, there are four independent magnetic string excitations, created by membrane operators, shown in Fig. \ref{fig:ExcitationPairs3D}. Using the labeling in Fig. \ref{fig:ExcitationPairs3D}, it can be checked by inspection that a Wilson loop for $e_i$ anticommutes with the membrane operator creating an $m_i$ string and commutes with the operators creating an $m_j$ string for $i \neq j$. Accordingly the topological order is $\mathbb{Z}_2^4$.

\subsection{Large $h_f$ Limit}
\label{subsec:scalarToEightCopies}

The large $h_f$ limit is treated analogously to the $d=2$ case; the face spins are pinned to the $X_{ij} = +1$ eigenstate. The operator $Z_{ij}$ which transmutes a three-charge bound state of the $h_f=0$ model to the fourth charge on the face now leaves the low-energy subspace; accordingly, the fourth charge on a face is now an independent excitation from the other three charges on that face. Therefore, all the eight electric excitations discussed previously become topologically distinct, and the system should have $\mathbb{Z}_2^8$ topological order.

\begin{figure}
\subfigure[]{
	\includegraphics[width=6cm]{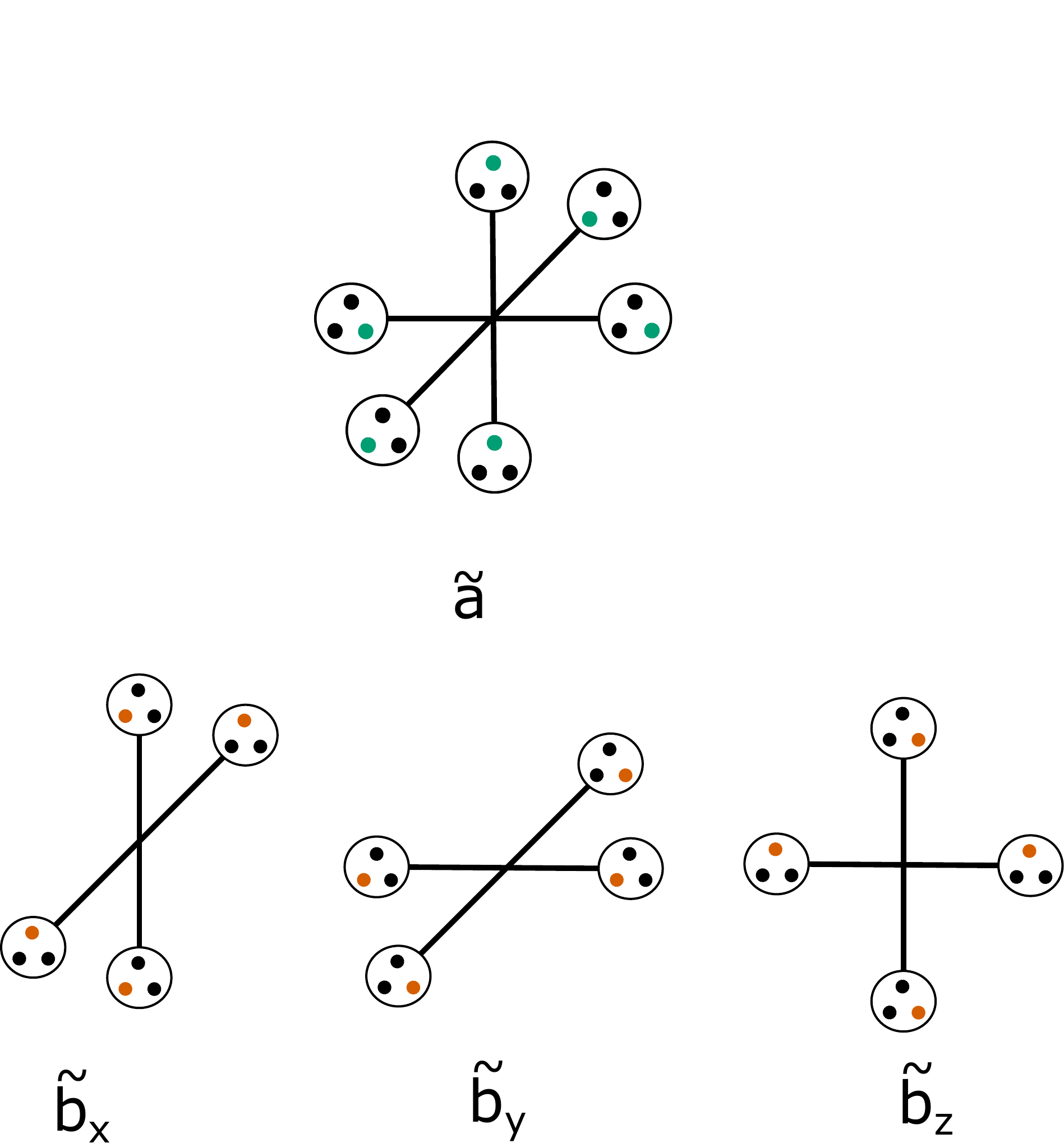}
	\label{fig:3DDecoupledTC}
}
\subfigure[]{
	\includegraphics[width=5cm]{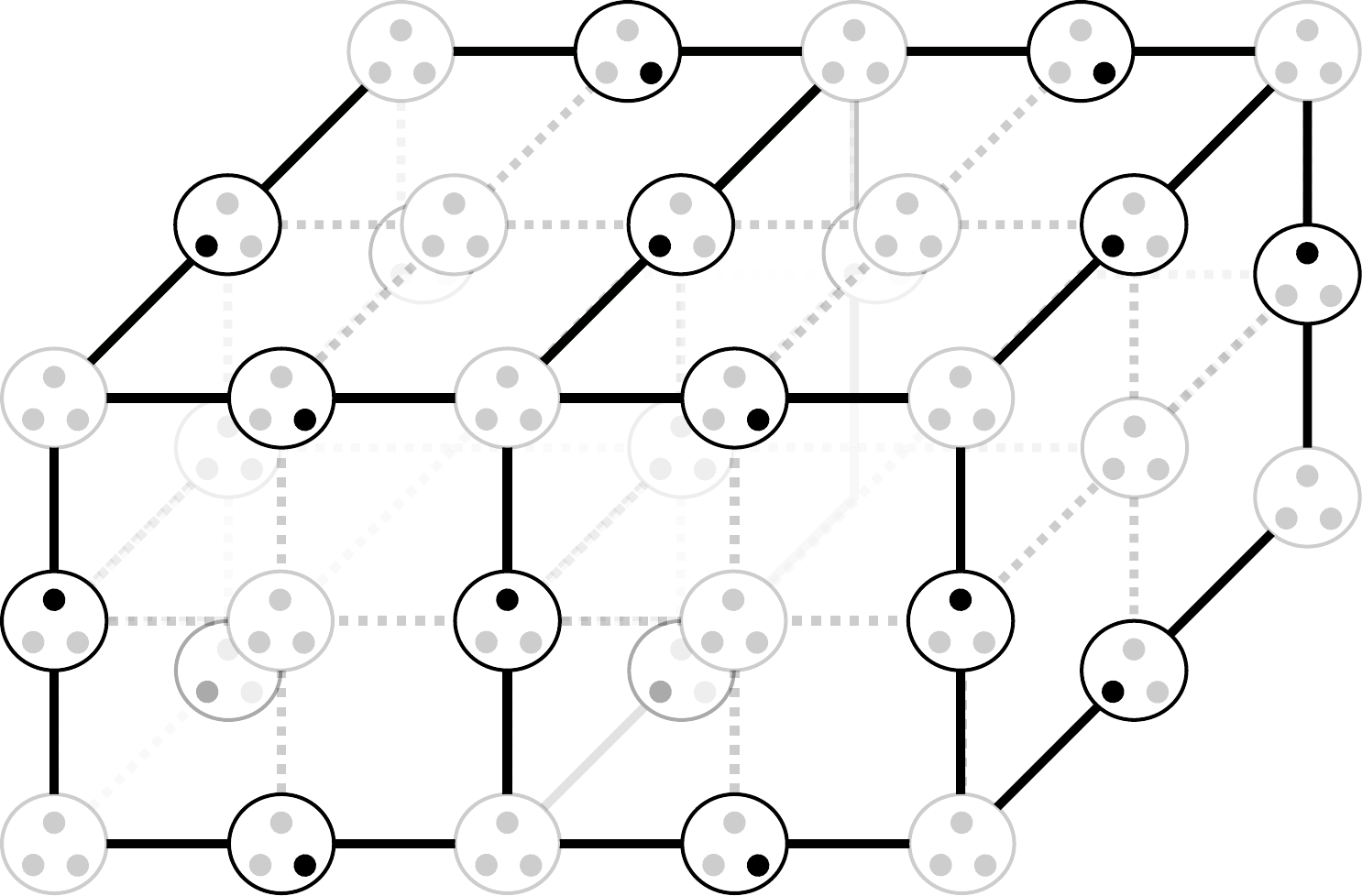}
	\label{fig:3DSublattice}
}
\caption{(a) Terms in the effective Hamiltonian Eq. \eqref{eqn:EightToricCodesHam} for the $d=3$ $(2r+1,2s+1)$ scalar charge model at large $h_f$. (b) One of eight sublattices (dark) of site spins; the terms in (a) do not couple different sublattices. Some bonds of the lattice have been lightened to make the relation to the $d=3$ toric code clearer.}

\end{figure}

This conclusion can be checked explicitly by computing the effective Hamiltonian with degenerate perturbation theory. The electric terms contribute at first order in $U/h_f$ and the off-diagonal magnetic terms contribute at fourth order in $1/g^2h_f$; the effective Hamiltonian is
\begin{equation}
H_{eff} = - \sum_{\text{sites}}\(U \tilde{a} + K \sum_i \tilde{b}_i\)
\label{eqn:EightToricCodesHam}
\end{equation}
where $\tilde{a}$ and $\tilde{b}_i$ are shown in Fig. \ref{fig:3DDecoupledTC} and $K \sim 1/g^8h_f^3$. (The diagonal magnetic terms only contribute at higher order in $1/g^2h_f$.)

Similarly to the $d=2$ case, this model is precisely eight copies of the $d=3$ toric code. The proof is similar to $d=2$; consider only the spins shown in Fig. \ref{fig:3DSublattice}. There are eight such sublattices of spins (translate the sublattice by one unit in any set of lattice basis directions), and each term in the Hamiltonian acts on only a single sublattice; the $a$ terms as a toric code star operator and the $b_i$ terms as a toric code plaquette operator. Collecting all the terms which act on each of the eight possible sublattices produces exactly one copy of the toric code for each of the eight sublattices.

The transition from large $h_f$ to small $h_f$ is therefore a condensation transition of four-charge bound states, where the four charges live on a single face of the lattice. Equivalently, as $h_f$ increases, a local degree of freedom gets frozen out, causing excitations which differed only by that local degree of freedom to become topologically distinct.

In the absence of any symmetries, a direct transition from $\mathbb{Z}_2^8$ to $\mathbb{Z}_2^4$ would either be 
first order, or it would correspond to a multicritical point. Here the lattice symmetries, which permute 
the different electric charges, can protect the critical point, leaving only one relevant perturbation that tunes between
the $\mathbb{Z}_2^4$ and $\mathbb{Z}_2^8$ phases. Analogous statements hold also for the transition to the trivial 
paramagnetic phase. 

As an aside, the Hamiltonian for the Higgsed $(1,0)$ scalar charge model happens to be exactly Eq. \eqref{eqn:EightToricCodesHam} (with operator definitions in Fig. \ref{fig:3DDecoupledTC}). This model therefore also has $\mathbb{Z}_2^8$ topological order and confines at large $h_s$. 

\subsection{Large $h_s$ Limit - X-Cube Phase}

The large $h_s$ limit can also be treated in degenerate perturbation theory, much like the large $h_f$ limit. In $d=2$, this limit produced a trivial paramagnet. In $d=3$, however, the resulting model is very different. The $a$ operator again contributes at first order (as does $h_f$), but so do $b_{ii}$; these operators did not exist in $d=2$. The remaining spins live on the faces of the lattice, and the effective Hamiltonian is the X-cube model Eq. \eqref{eqn:XCube}.

Remarkably, then, as $h_s$ increases this model may have a direct transition from $\mathbb{Z}_2^4$ conventional topological 
order to X-cube order. Such a transition is of considerable interest; we defer it to future work.

Recall that the $(1,1)$ model, which Higgses to the model under consideration, has continuous rotational invariance. In taking $h_s$ large with $h_f$ small, we have broken the continuous 
rotational invariance down to the symmetry of the cubic lattice. Thus we see that in this model, 
the appearance of the X-cube model required terms whose continuum limit breaks continuous 
rotational symmetry down to discrete rotational symmetry of the cubic lattice.

\section{$(2r+1,2s+1)$ Vector Charge Higgs in $d=3$}
\label{sec:oddOddVector_d3}

Before discussing the $(2r+1,2s+1)$ vector charge theory, we briefly comment on a small modification of the results of the Higgs procedure in $d=3$ for vector charge theories. The factor of $2/(3-(-1)^m)$ in the off-diagonal components of $B_{ij}$ in Eq. \eqref{eqn:vectorB} leads to a difference between $m \equiv 0$ mod 4 and $m \equiv 2$ mod 4. Specifically, if $m = 2m_0$ with $m_0$ an integer, then for $i \neq j$
\begin{widetext}
\begin{equation}
B_{ij} = \sum_{k \neq i,j} \left[ m_0 \(\Delta_i \Delta_k A_{jk} + \Delta_j \Delta_k A_{ik} - \Delta_k^2 A_{ij}\) - n \Delta_i \Delta_j A_{kk}\right]
\end{equation}
\end{widetext}
Upon Higgsing, this operator becomes, in the low-energy subspace, a product of $Z_{ij} = e^{iA_{ij}}$. In particular, the face spins $Z_{ij}$ for $i \neq j$ always appear raised to the power of $m_0$. If $m_0$ is even, then the face spins will all drop out. If $m_0$ is odd, then the face spins are present. Therefore the Higgsed theories with different parities of $m_0 = m/2$ are distinct.

\begin{figure}
\subfigure[]{
	\includegraphics[width=6cm]{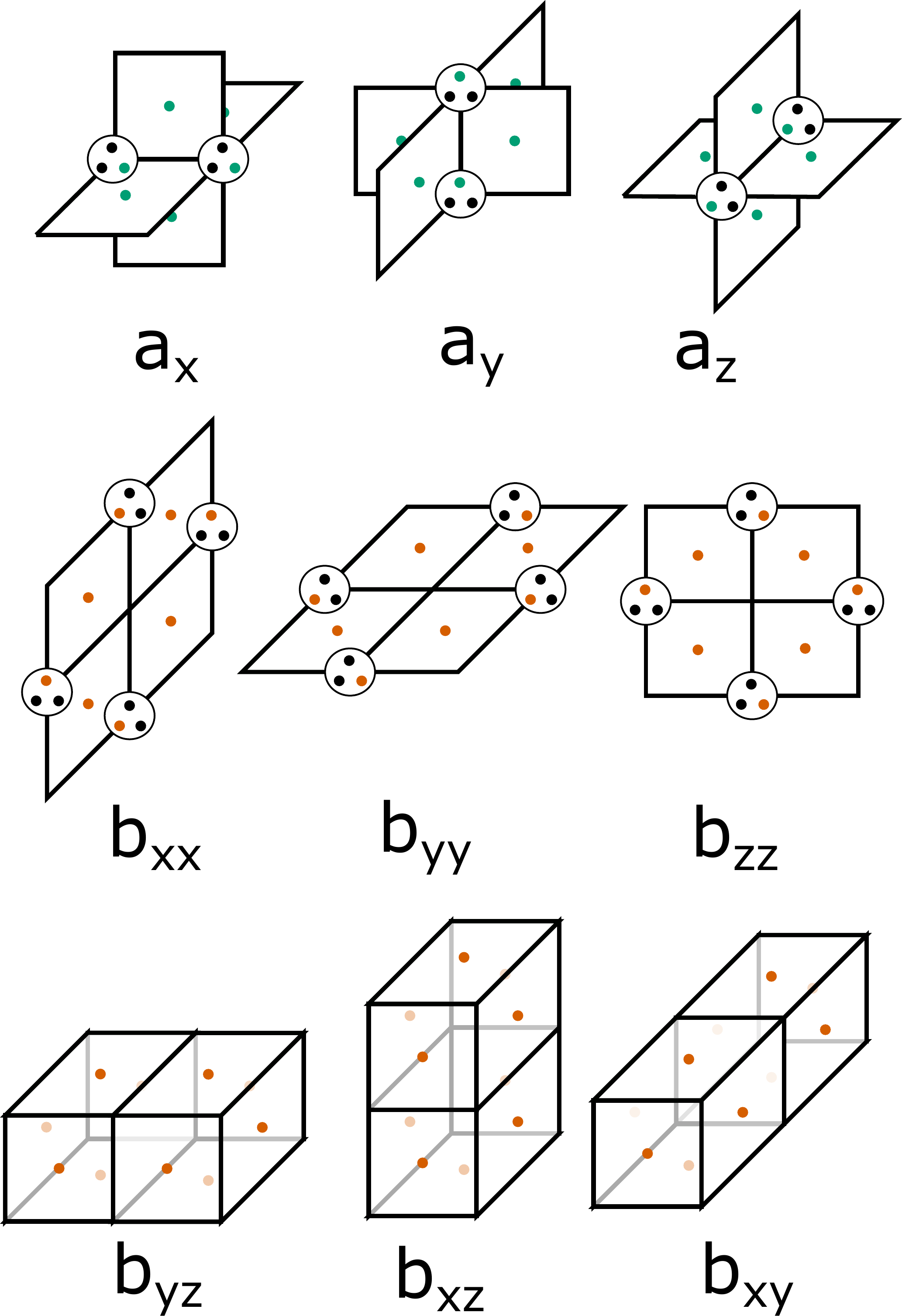}
	\label{fig:3DVectorOddOddHam}
}
\subfigure[]{
	\includegraphics[width=4cm]{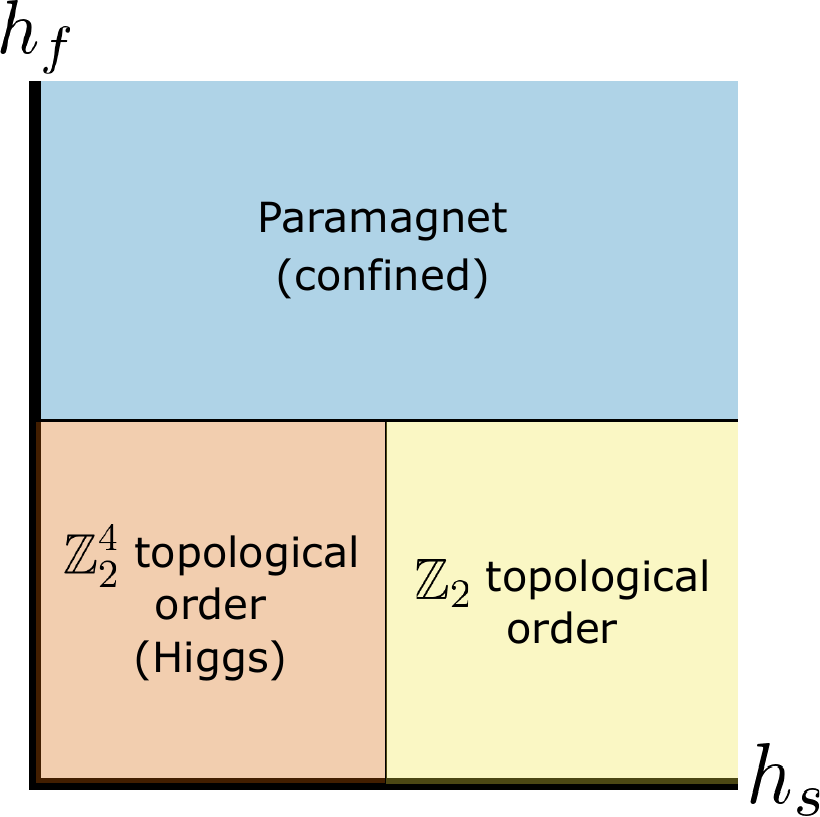}
	\label{fig:PhaseDiagram3DVectorOddOdd}
}
\caption{(a) Terms in the Hamiltonian and (b) phase diagram for the $d=3$ Higgsed $(2r+1,2s+1)$ vector charge theory. The labels for the Pauli operators are omitted to avoid clutter. The $a_i$ are associated with the center bonds and are the product of Pauli $X$ operators on the six green spins. The diagonal (off-diagonal) $b_{ij}$ operators are associated with the center site (face) and are products of Pauli $Z$ operators on the eight (ten) orange spins. The direct transitions from the $\mathbb{Z}_2^7$ phase in (b) are protected by the rotational symmetries of the cubic lattice.}
\end{figure}

The Hamiltonian for the $d=3$ $(2r+1,2s+1)$ vector charge model is 
\begin{widetext}
\begin{equation}
H = -U \sum_i \sum_{i-\text{links}} a_i - \frac{1}{g^2}\sum_{\text{sites}, i}b_{ii} - \frac{1}{g^2}\sum_{\text{faces},i<j}b_{ij} - h_s \sum_{\text{sites},i}X_{ii} - h_f \sum_{\text{faces},i<j}X_{ij}
\label{eqn:OddOddVectorHam3D}
\end{equation}
\end{widetext}
The forms of $a_i$ and $b_{ij}$ are shown in Fig. \ref{fig:3DVectorOddOddHam}, and its phase diagram (a summary of this section) is given in Fig. \ref{fig:PhaseDiagram3DVectorOddOdd}.

\subsection{Higgs Phase}

The Higgs phase is, as usual, understood from the $h_f=h_s=0$ commuting projector model. Its ground state degeneracy is computed to be $2^{22}$. The local operator $c = \prod_{\text{cube}} Z_{ij}$, for any elementary cube, commutes with the Hamiltonian. There is only one such independent operator (the others are generated by multiplying $c$ by various $b_{ij}$ for $i \neq j$). The simplest operator which commutes with $H$ but anticommutes with $c$ is the product $C$ of all $X_{xy}$ in planes with even values of $z$, all $X_{xz}$ in planes with even values of $y$, and all $X_{yz}$ in planes with even values of $x$. We can therefore think of $C$ as generating an Ising ($\mathbb{Z}_2$) global symmetry, but such a symmetry is spontaneously broken in 3+1D. After accounting for the symmetry enrichment, the topological degeneracy is at most $2^{21}$; we claim that the remaining degeneracy is topological, and that the topological order is $\mathbb{Z}_2^7$.

\begin{figure}
\subfigure[]{
	\includegraphics[width=3cm]{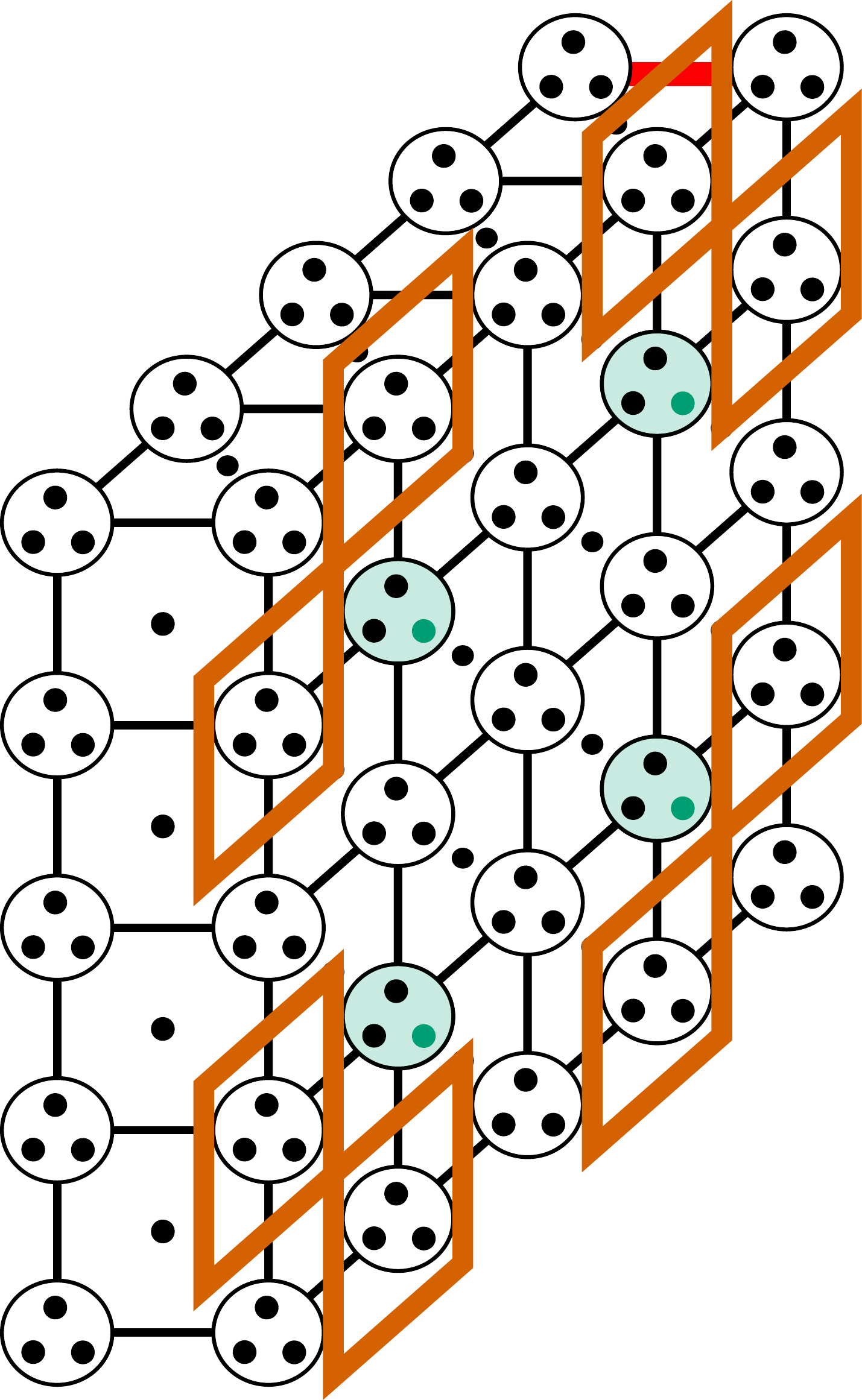}
	\label{fig:VectorESingleMagnetic3D}
}
\subfigure[]{
	\includegraphics[width=3cm]{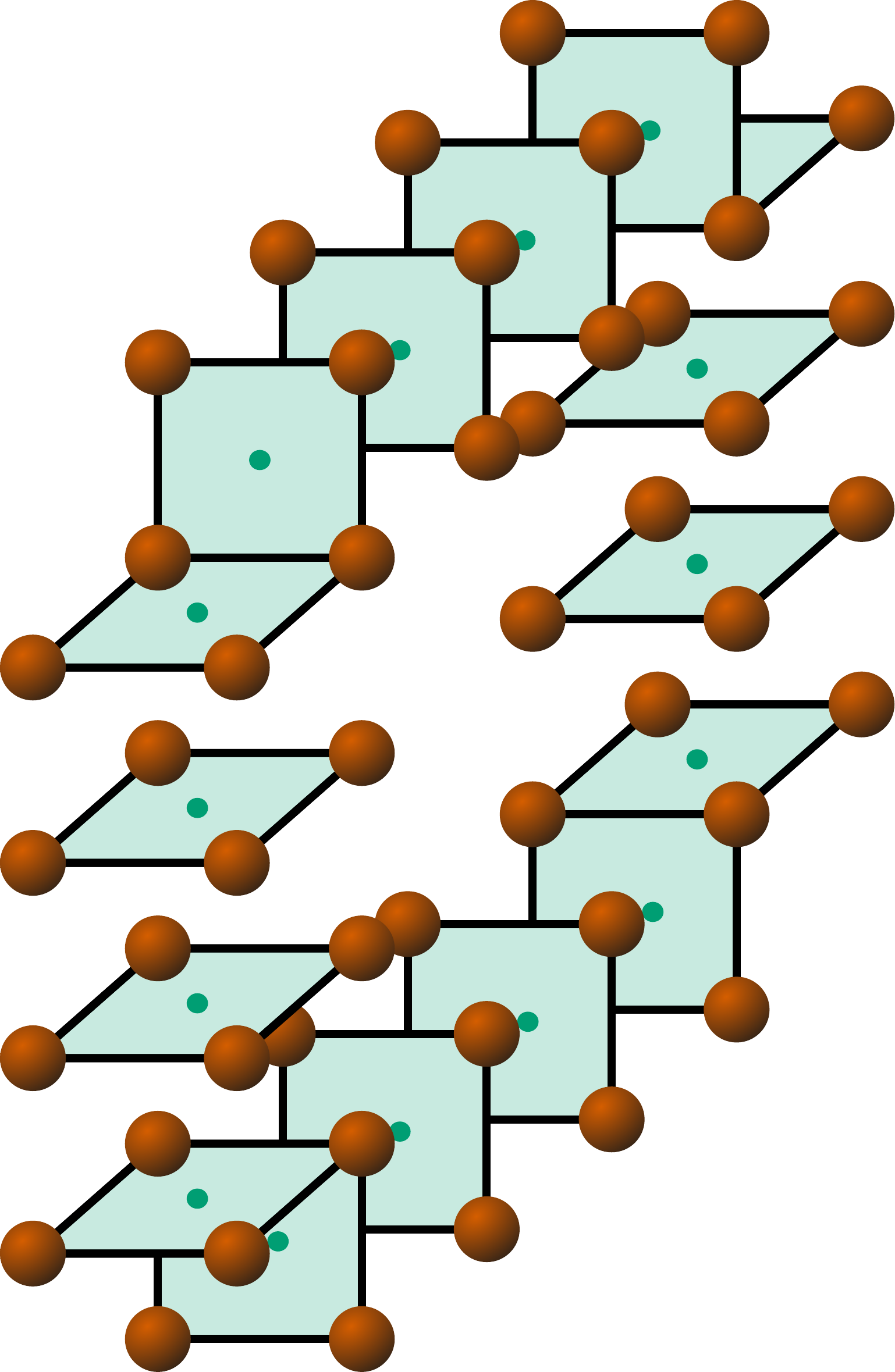}
	\label{fig:VectorEEightMagnetic3D}
}
\caption{(a) Elementary membrane operator ($X_{xx}$ acting on green site spins) and associated string of excitations of $b_{yy}$ and $b_{zz}$ terms (orange squares) for the Higgsed $(2r+1,2s+1)$ vector charge model in $d=3$. This membrane can live on eight possible sublattices of lattice constant 2. (b) Action of a string of $X_{ij}$ operators on green face spins, creating excitations of $b_{yy}$ and $b_{zz}$ on sites (orange spheres). This collection of excitations is equal to a bound state of all eight of the strings discussed in part (a); such a bound state is topologically trivial since it can be destroyed by a string operator.}
\end{figure}

\begin{figure}
	\includegraphics[width=3.5cm]{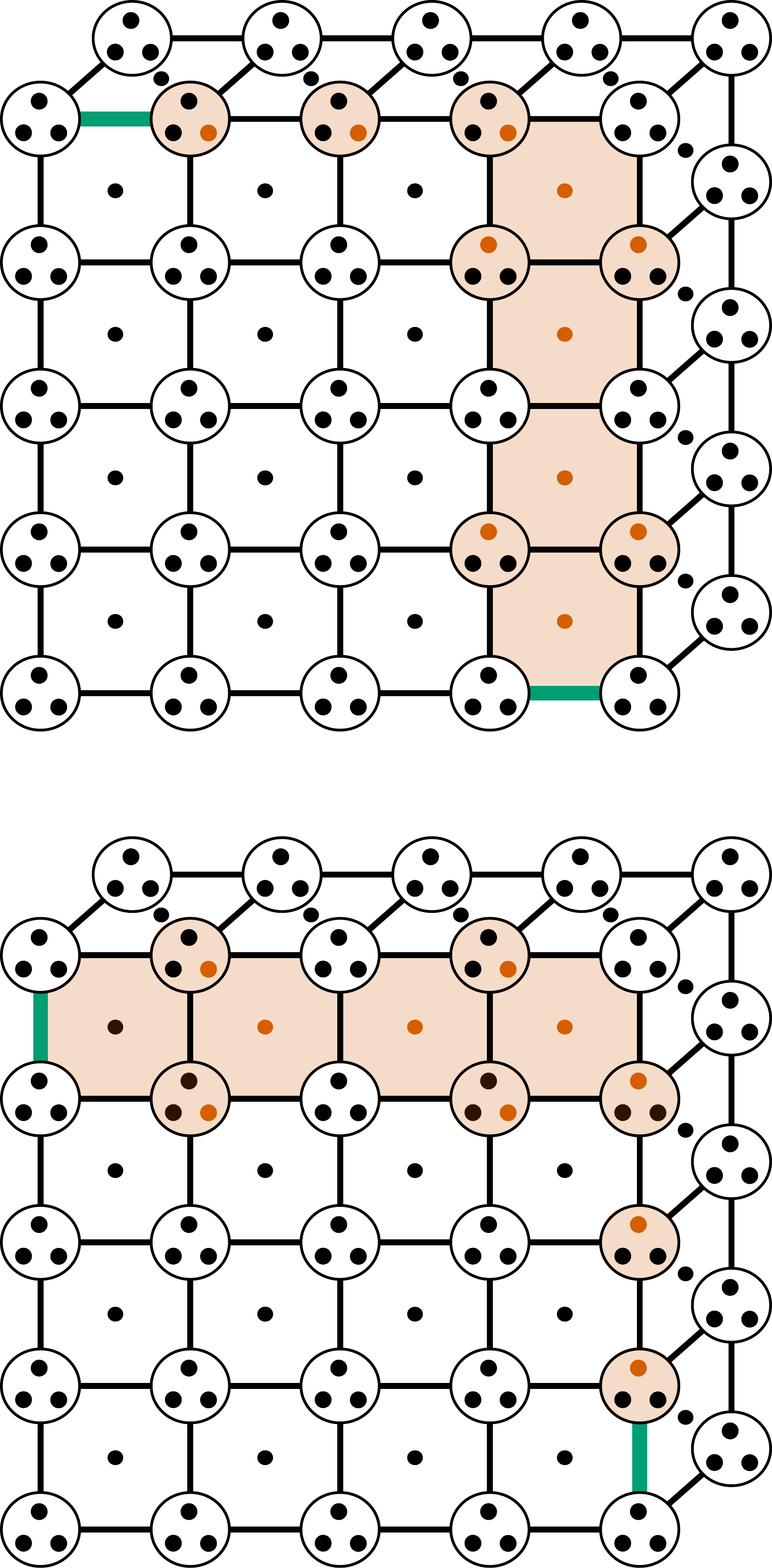}
\caption{Two types of string operators ($Z$ operators acting on orange spins) and their associated electric particles (green links) in the Higgsed $d=3$ $(2r+1,2s+1)$ vector charge model. Eight strings total are obtained by shifting these operators by zero or one unit in one or both lattice directions perpendicular to the green links, but only seven of are topologically distinct because bound states of eight excitations can be destroyed with a local operator.}
	\label{fig:ElectricStringsVectorE3D}
\end{figure}

It is simplest to analyze the magnetic sector. Breaking the site degrees of freedom into eight sublattices as usual, the action of the $b_{ii}$ on the site degrees of freedom looks like toric code plaquette operators on lattice constant 2 sublattices. Accordingly, membranes of $X_{ii}$ operators living on the sublattices create eight types of simple magnetic strings, an example of which is shown in Fig. \ref{fig:VectorESingleMagnetic3D}. As usual, though, we must account for action of $X_{ij}$.  Applying a string of $X_{ij}$ creates the set of excitations shown in Fig. \ref{fig:VectorEEightMagnetic3D}. It is straightforward to check that this pattern of excitations is precisely the bound state of all eight simple strings. Therefore, there are only seven independent magnetic strings, as a bound state of seven simple strings may be converted into the eighth simple string using the aforementioned string operator. It is straightforward but laborious to show that there are also seven independent electric excitations which braid appropriately for $\mathbb{Z}_2^7$ topological order; two examples are shown in Fig. \ref{fig:ElectricStringsVectorE3D}.

\subsection{Large-field limits}

\subsubsection{Large $h_s$}

At $h_s=0$, every magnetic string excitation in the $\mathbb{Z}_2^7$ topological order can be created exclusively with $X_{ii}$. When $h_s$ is taken large, one expects all the magnetic strings to condense and the electric particles to confine. Remarkably, this process simultaneously causes a point-like magnetic excitation to \textit{deconfine}, and the large-$h_s$ model has $\mathbb{Z}_2$ topological order.

\begin{figure}
\includegraphics[width=4cm]{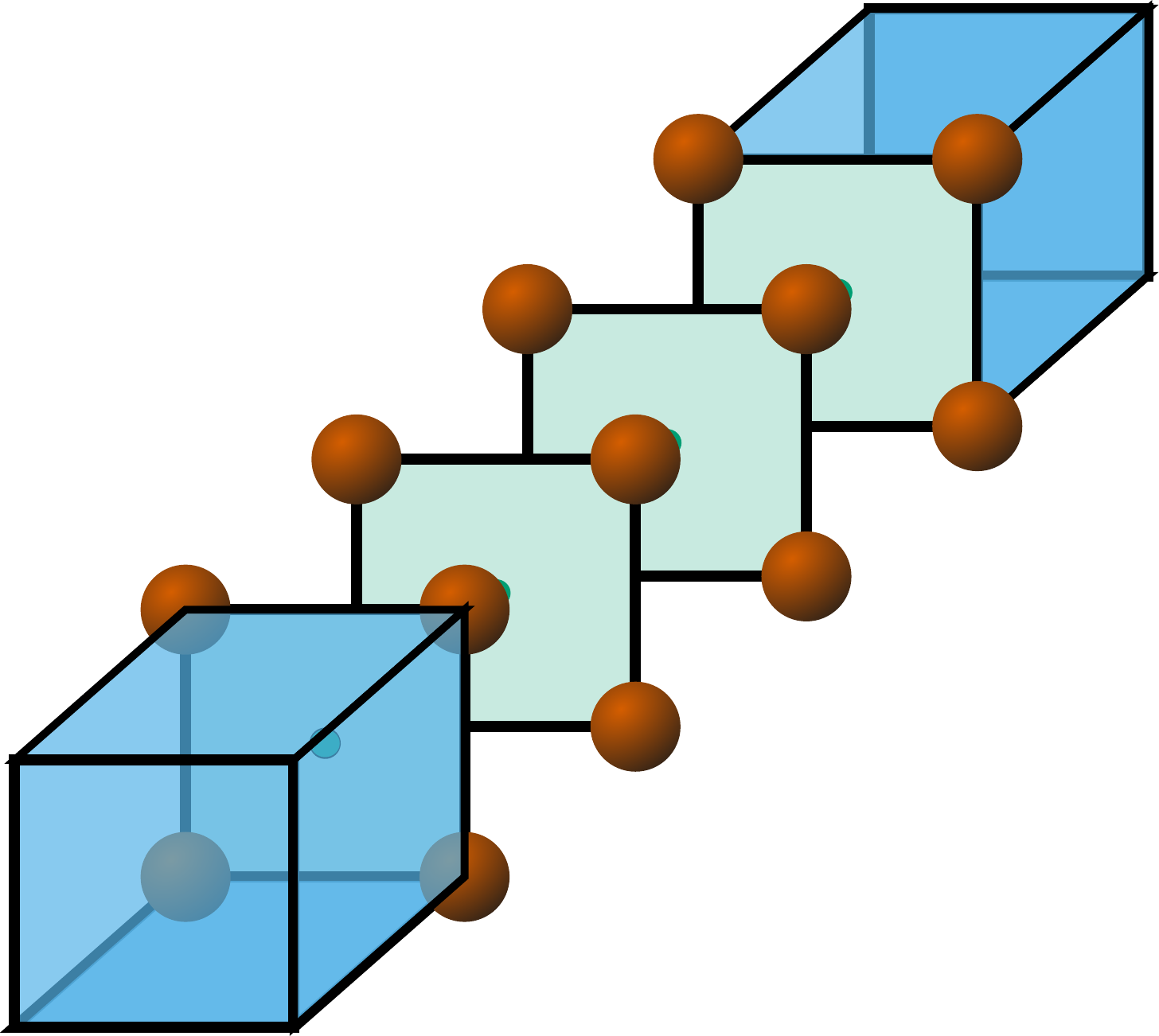}
\caption{Open string of $X_{xy}$ acting on the green faces and its associated excitations in the Higgs phase of the $d=3$ vector charge $(2r+1,2s+1)$ model. The blue faces have magnetic excitations; off-diagonal $b_{ij}$ terms have eigenvalue $-1$ on those phases. The orange spheres are site excitations ($b_{ii}$ which have eigenvalue $-1$) and give the string tension; at large $h_s$, they condense, deconfining the collection of blue face excitations.}
\label{fig:OpenFaceStringVectorE3D}
\end{figure}

To understand what has happened from an intuitive confinement picture, consider the action of a string of $X_{ij}$ operators on the ground state at $h_s=h_f=0$, shown in Fig. \ref{fig:OpenFaceStringVectorE3D}. This string has tension because it anticommutes with magnetic site terms $b_{ii}$ along the string, shown as orange spheres in the figure, and also anticommutes with a set of off-diagonal magnetic terms at each end of the string, shown as the blue faces in the figure. At large $h_s$, the site excitations (rather, strings of them) are condensed; this causes the string to lose tension and the collection of magnetic face excitations at each end of the string becomes a deconfined anyon. 

More explicitly, we can perform the usual degenerate perturbation theory. In this limit, the site spins freeze out of the $a_i$ and the off-diagonal magnetic terms contribute at first order. The diagonal magnetic terms contribute at sixth order, but they generate operators equal to products of four off-diagonal terms and thus do not affect the analysis. The effective Hamiltonian is 
\begin{equation}
H_{eff} = -U\sum_{i \text{ links}}\tilde{a}_i - \frac{1}{g^2}\sum_{\text{faces},i < j}b_{ij} - h_f\sum_{\text{faces}}X_{ij}
\label{eqn:effective3DTC}
\end{equation}
where $\tilde{a}_i$ are $a_i$ with the site operators removed. This is a commuting projector model at $h_f=0$.

To understand the $h_f=0$ model, note that the $\mathbb{Z}_2$ symmetry generator $C$ still commutes with the Hamiltonian. To make the discussion clearer, let us imagine breaking this symmetry explicitly by adding $-\lambda \sum_{\text{cubes}} c$ to the Hamiltonian for $\lambda$ small. Noting that $a_i$ and $c$ are the plaquette and star operators of the $d=3$ toric code on the dual lattice, we see that this symmetry-broken model is exactly the $d=3$ toric code. At $\lambda = 0$, then, the system is in a $\mathbb{Z}_2$ toric code phase with additional spontaneous breaking of an Ising symmetry. The topological order must therefore be stable to a small but nonzero $h_f$.

\subsubsection{Large $h_f$}

At large $h_f$, the system has a unique, approximately classical ground state. As usual, we check this by degenerate perturbation theory. The $a_i$ and $h_s$ contribute at first order. If $1/g^2=0$, then this first-order effective model is immediately seen to be classical and have a unique ground state.

The lowest-order contribution of the $b_{ij}$ is at tenth order in $1/g^2h_f$ (six $b_{ii}$ form a closed 2x2x2 cube and four $b_{ij}$ for $i \neq j$ remove the rest of the face spins), but it just contributes an overall constant. The lowest-order nontrivial contribution occurs at $L$th order in degenerate perturbation theory. Therefore, in the thermodynamic limit, with any $1/g^2 \ll h_f$, the classical effective model remains valid.

\section{$(4r+2, 2s+1)$ Vector Charge Higgs in $d=3$}
\label{sec:evenOddVector_d3}

Recall that the $(2,1)$ vector charge model has continuous rotational invariance. In this section we analyze the Higgs mechanism for the class of theories that include said $(2, 1)$ vector charge model. 

\begin{figure}
\includegraphics[width=7cm]{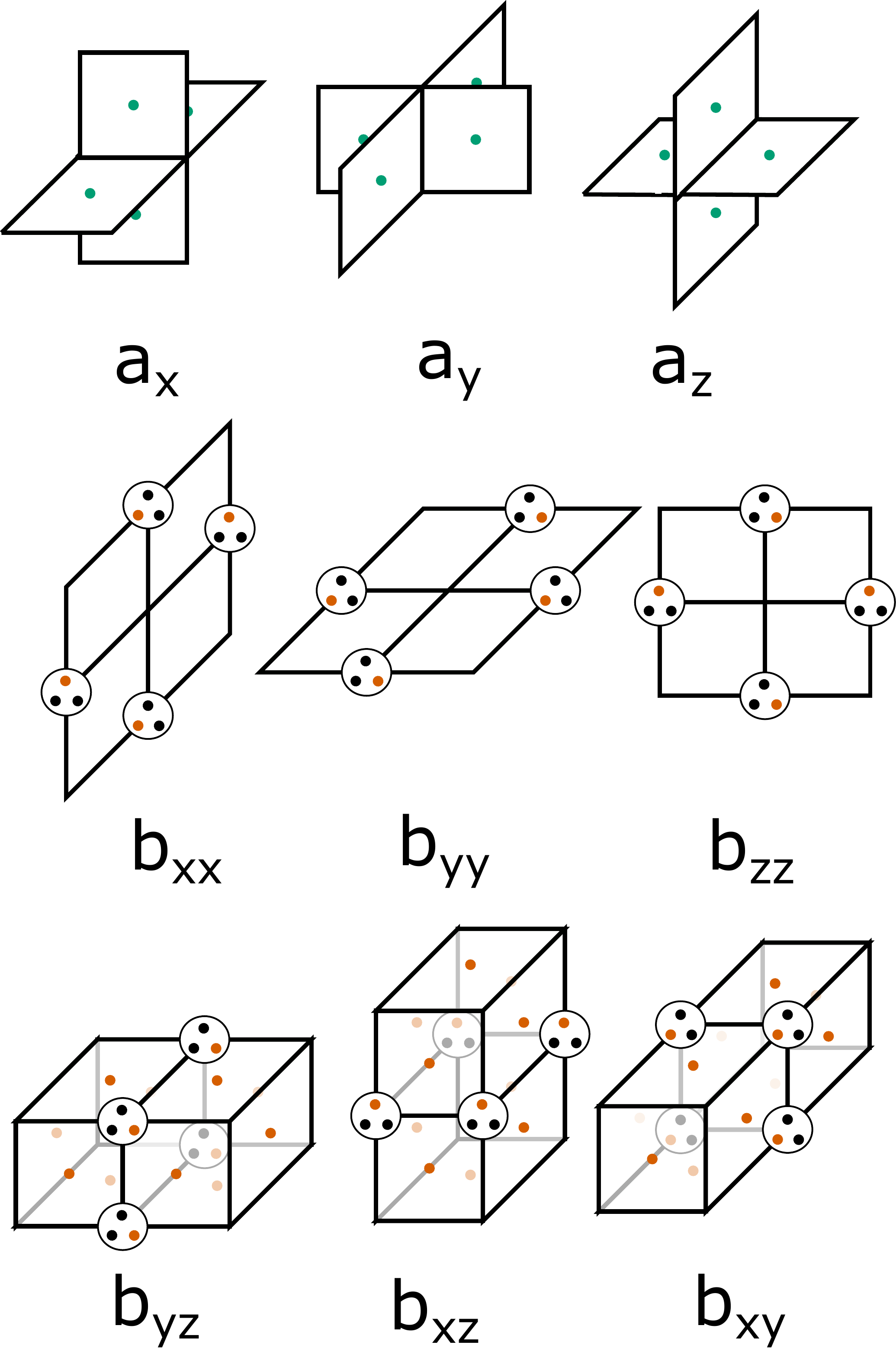}
\caption{Terms in the $d=3$ Higgsed $(4r+2,2s+1)$ vector charge theory Hamiltonian. The $a_i$ are associated with the center bonds and are the product of Pauli $X$ operators on the four green spins. The diagonal (off-diagonal) $b_{ij}$ operators are associated with the center site (face) and are products of Pauli $Z$ operators on the four (fourteen) orange spins.}
\label{fig:3DVectorAHam}
\end{figure}

The Hamiltonian for the $d=3$ Higgsed $(4r+2,2s+1)$ vector charge theory is
\begin{widetext}
\begin{equation}
H = -U \sum_i \sum_{i-\text{links}} a_i - \frac{1}{g^2}\sum_{\text{sites}, i}b_{ii} - \frac{1}{g^2}\sum_{\text{faces},i<j}b_{ij} - h_s \sum_{\text{sites},i}X_{ii} - h_f \sum_{\text{faces},i<j}X_{ij}
\label{eqn:3DVectorAHam}
\end{equation}
\end{widetext}
with the operators shown in Fig. \ref{fig:3DVectorAHam}.

We analyze the Higgs phase by taking $h_s=h_f=0$. In this limit, it is obvious that $Z_{ii}$ on every site commutes with the Hamiltonian. The simplest product of site $X$ operators which commutes with the Hamiltonian is the product of all $X_{ii}$ operators in a plane perpendicular to the $i$ direction; these membrane operators anticommute with the $Z_{ii}$. Hence this model has a non-topological degeneracy which scales as the linear system size $L$. Adding a small $h_f$ obviously does not split this degeneracy, and a small $h_s$ only contributes at order $L^2$ in degenerate perturbation theory, splitting the ground state degeneracy in an exponentially small fashion. 

This limit of the model is therefore fine-tuned, as adding $-\gamma \sum_i Z_{ii}$ to the Hamiltonian with $\gamma$ arbitrarily small splits the degeneracy associated with the site degrees of freedom. Although $\sum_i Z_{ii}$ does not obviously appear from Higgsing a gauge-invariant operator, it commutes with the residual $\mathbb{Z}_2$ gauge symmetry generated by the $a_i$. On general field-theoretic grounds, a nonzero $\gamma$ should therefore be generated. 

More generally, any operator involving $X_{ii}$ splits the non-topological degeneracy by a energy exponentially small in the system size so we can ignore all such operators.
Any operator involving only the $Z_{ii}$ that splits the non-topological degeneracy simply freezes the site spins to some particular product state configuration, although the particular configuration depends on what operators we add. As such, we are justified in calling the topological order of the resulting phase the ``Higgs phase" so long as the phase associated with the plaquette sector is gapped.

If the site spins are frozen to $Z_{ii}$ eigenstates, then it is straightforward to see that the effective model becomes that of the large-$h_s$ limit of the Higgsed $(2r+1,2s+1)$ vector charge model Eq. \eqref{eqn:effective3DTC}. That model has $\mathbb{Z}_2$ topological order, so the Higgs phase of the present $(4r+2,2s+1)$ model also has $\mathbb{Z}_2$ topological order.

Several other theories behave similarly to the $(4r+2,2s+1)$ vector charge model discussed here, except in those cases splitting the non-topological degeneracy leads to topologically trivial phases. We discuss them in Appendix \ref{subsec:illDefined}.

\section{Discussion}

We have defined a large class of rank-2 $U(1)$ gauge theories, which we refer to as the $(m,n)$ scalar and vector charge theories. These
are invariant under the discrete rotational symmetries of the square (cubic) lattices in $d = 2$ ($d = 3$). The previously studied rank-2 theories 
whose continuum limit possesses continuous rotational symmetry correspond to the $(1,1)$ scalar and $(2,1)$ 
vector charge theories. Remarkably, we find most of the $(m,n)$ scalar and vector theories correspond to stable gapless 
phases of matter in $d = 3$, and to critical or multi-critical points in $d = 2$, and thus form an interesting class of field theories
that are worthy of further study. The matter field in these theories is constrained to move along subdimensional manifolds, 
in a manner dictated by $(m,n)$ and whether it is a scalar or vector charge theory. 

Breaking the $U(1)$ gauge symmetry down to a discrete subgroup, such as $\mathbb{Z}_2$, gives rise to a large class of 
exactly solvable models. We find that in most cases, the $\mathbb{Z}_2$ Higgs phases describe either topologically trivial phases
of matter, or possess conventional topological order and correspond to multiple copies of a conventional $\mathbb{Z}_2$ toric 
code phase. Nevertheless, the exactly solvable models that arise are new and reminiscent of the color code models\cite{ColorCodes}; it might thus
be interesting to consider these models from the perspective of quantum error correction. 

Our results provide a number of general lessons regarding fractons and higher rank gauge theories:

First, we have expanded the set of gapless higher rank field theories, whose matter fields have restricted, subdimensional dynamics. 
In $d=2$, while many or all of these theories may not correspond to stable phases of matter, it appears they can at least correspond to (multi)-critical points. 
However the Higgs phases of all these theories possess either trivial or conventional topological order, which confirms the expectation that
gapped fracton phases cannot exist in $d = 2$. 

Moreover, we found that the Higgs phases of models with continuous rotational symmetry, such as the $(1,1)$ scalar and $(2,1)$ vector charge theories 
do not give rise to gapped fracton phases. However the $\mathbb{Z}_2$ Higgs phase of the $(1,1)$ scalar charge theory in $d = 3$
does admit a transition to X-cube fracton order in the limit of a strong ``Zeeman'' field that breaks the continuous rotational symmetry of the continuum theory
down to a discrete subgroup. The existence of a possible transition between the conventional $\mathbb{Z}_2^4$ topological order and X-cube fracton order
in $d= 3$ may be of a qualitatively new type of quantum phase transition that requires further study. 

We also found that the X-cube fracton order can emerge as the Higgs phase of the $(2r, 2s+1)$ scalar charge theories in $d = 3$. It
is not clear whether the $(0,1)$ scalar charge theory corresponds to a stable gapless phase of matter. However, it appears that the $(2r, 2s+1)$ 
theories with $r > 0$ do correspond to stable gapless phases. Remarkably, this suggests the existence of stable gapless higher rank gauge theories
whose Higgs phases yield fracton order. 

Notably, the above results suggest that X-cube fracton order cannot arise from a theory that is invariant under continuous rotations. 

Our results demonstrate that the gapped X-cube fracton order can indeed be described within the framework of quantum field theory, albeit
with a novel type of gauge theory. In particular, we can consider a continuum version of the $(m,n)$ scalar charge theories, coupled to a charge 
$p$ complex scalar field, whose condensation breaks the $U(1)$ rank-2 gauge symmetry down to $\mathbb{Z}_p$. It would be
interesting to understand the relation between this continuum Higgs theory and an alternative continuum field theory description 
of the X-cube phase, presented in Ref. \onlinecite{SlagleFieldTheory}. 

The considerations presented here raise the question of whether all gapped fracton phases can emerge as Higgs phases of stable gapless
higher rank gauge theories. For example, there is a natural generalization of the higher rank scalar and vector charge theories presented so far, defined by the Gauss Law terms:
\begin{align}
\sum_{\{i\},\{\alpha\}} M_{i_1 \cdots i_n}^{a; \alpha_1 \cdots \alpha_k} \partial_{i_1} \partial_{i_2} \cdots \partial_{i_n} E_{\alpha_1, \alpha_2, \cdots, \alpha_k}^a = \rho_a ,
\end{align}
where $a$ labels distinct flavors of charges and the constraint is on a rank-k electric field $E_{\alpha_1, \alpha_2, \cdots, \alpha_k}^a $. Any discrete or continuous 
rotational invariance imposes constraints on the types of tensors $M$ that can appear in the above. In fact, although for brevity we do not consider them in this paper, cubic symmetry allows another term $\sum_{i\neq j} \Delta_i^2 E_{jj}$ in Gauss' Law. A natural question is whether Higgs phases of such
generalized higher rank gauge theories can describe all consistent types of subdimensional dynamics for particles that arise in gapped fracton phases. 
In particular it would be interesting to understand whether the Chamon model\cite{ChamonGlass} and Haah's code\cite{HaahsCode} can possibly arise from such a
construction. 

\textit{Note:} During the completion of this paper, we learned of closely related work by Ma \textit{et al}\cite{MaHiggs}.

\begin{acknowledgments}
DB is supported by the Laboratory for Physical Sciences and Microsoft. MB is supported by 
NSF CAREER (DMR-1753240) and JQI-PFC-UMD. 
\end{acknowledgments}

\appendix 

\section{Review of the Standard Higgs Mechanism}
\label{app:HiggsReview}

We briefly define the lattice theory of a compact (rank-1) $U(1)$ lattice gauge field coupled to charge-$p$ bosonic matter. We place canonically conjugate rotor variables $\theta$ and $L$ on the sites of a cubic lattice; this will be our matter field. Define $\theta \sim \theta +2\pi$ so that $L$ has integer eigenvalues and $e^{i\theta}$ is a raising operator for $L$. Next, orient the links of the lattice and place canonically conjugate rotor variables $A_i$ and $E_i$ on the links, where $i$ labels the direction of the link. Here 
\begin{equation}
[A_i(\bv{r}), E_j(\bv{r'})] = i\delta_{ij}\delta(\bv{r}-\bv{r}')
\end{equation}
Compactness means we identify $A_i \sim A_i+2\pi$, which quantizes $E_i$ to integer eigenvalues. The standard $U(1)$ gauge transformation is $A_i \rightarrow A_i - \Delta_i \alpha$, leading to the Gauss' Law operator
\begin{equation}
G(E) = \sum_i \Delta_i E_i
\end{equation} 

The gauge transformation leads to a Hamiltonian of the usual form given by Eq. \eqref{eqn:HStructure}. The only difference is that $\tilde{h}_s E_{ii}^2$ and $\tilde{h}_p E_{ij}^2$ are replaced by a single term $\tilde{h}E_i^2$ on the links of the lattice. Note that we are enforcing Gauss' Law energetically on a local rotor model rather than as a strict constraint on the Hilbert space. The magnetic field takes its usual form
\begin{equation}
B_i = \sum_{j,k}\epsilon_{ijk}\Delta_j A_k
\end{equation}
for $i=z$ in $d=2$ and $i=x,y,z$ in $d=3$. The Higgs coupling is
\begin{equation}
H_{Higgs} = \sum_{\bv{r}}\frac{L(\bv{r})^2}{2M} - V\sum_{i,\text{sites}} \cos(\Delta_i \theta + p A_i)
\end{equation}

We take $V \rightarrow \infty$ and restrict our attention to the low-energy subspace from now on.

In said subspace, we require
\begin{equation}
\Delta_i \theta + p A_i = 2\pi n
\end{equation}
for $n \in \mathbb{Z}$. We may choose a gauge where $\theta = 0$ (mod $2\pi$) at every point; in this gauge,
\begin{equation}
A_i = \frac{2\pi}{p} n
\end{equation}
for $n \in \mathbb{Z}$. For simplicity, we specialize to $p=2$; the larger $p$ case is a straightforward generalization. Then $e^{iA_i} = \pm 1$ on each link. Furthermore, since $e^{iA_i}$ is a raising operator for $E_i$, its action flips the eigenvalue of $(-1)^{E_i}$. Therefore, the spectrum and the commutation relations of $e^{iA_i}$ and $(-1)^{E_i}$ in the low-energy subspace are reproduced by the identification $e^{iA_i} = Z$ and $(-1)^{E_i} = X$ where $X,Z$ are the usual Pauli matrices.

In this language, $\cos(B_i)$ is exactly the operator $\prod_{\square} Z$, where the product is around a plaquette perpendicular to the $i$ direction.  The $E_i^2$ term of $H_{Maxwell}$ penalizes fluctuations in $A_i$; after Higgsing, this means we should be penalizing fluctuations in $Z$. Hence the Hamiltonian should have a term $-h X$, where $h \propto \tilde{h}$.

Finally, because charge is condensed, Gauss' Law only a well-defined constraint modulo 2. In the constraint language (i.e. in the $\tilde{U} \rightarrow \infty$ limit) there is a strict constraint
\begin{equation}
1 = (-1)^{\Delta_i E_i - 2L} = (-1)^{\Delta_i E_i} = \prod_{\text{star}}X
\end{equation}
where the product is around the usual star operator. This is the $\mathbb{Z}_2$ gauge constraint. Taking $\tilde{U}$ finite gives an energetic penalty to states violating this condition. The Hamiltonian is therefore
\begin{equation}
H = -U\sum_{\text{sites}} \prod_{\text{star}}X - \frac{1}{g^2}\sum_{\text{plaquettes}} \prod_{\square} Z - h \sum_{\text{links}} X
\end{equation}
where $U \propto \tilde{U}$.

We have thus produced $\mathbb{Z}_2$ gauge theory by Higgsing the $U(1)$ lattice gauge theory. The generalization to $p>2$ is straightfoward and produces $\mathbb{Z}_p$ lattice gauge theory. The model with finite charge gap and $h=0$ is also known as the $d$-dimensional $\mathbb{Z}_p$ toric code, and has topological order; its ground state degeneracy is $p^d$ on the $d$-torus.

\section{Theories With Trivial Higgs Phases}
\label{app:trivialHiggs}

Several of the theories produce topologically trivial Higgs phases, which we study in this appendix. We use the word ``trivial" to describe three different physical cases. First, the model can have a unique classical ground state (i.e. the Hamiltonian is a sum of trivial, commuting terms such that the exact ground state is a product state). Second, it can be mapped a transverse field quantum Ising model, such that the ground state spontaneously breaks a global symmetry. Third, the precise model for the Higgs phase that we study can have a sub-extensive ground state degeneracy, which must be split by additional perturbations; in these cases the perturbations drive the system to a topologically trivial phase, while the patterns of any possible global symmetry breaking will depend on the type of perturbations added. 

We will examine each of these cases in turn.

\subsection{Classical Ground States}

\begin{figure}
\includegraphics[width=2cm]{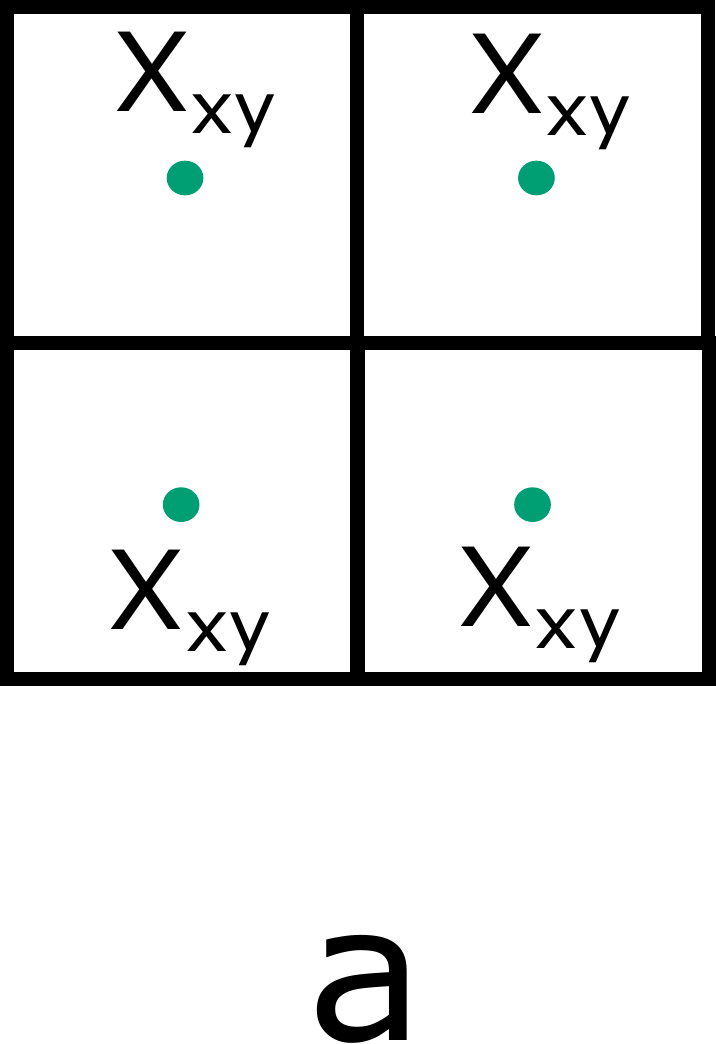}
\caption{$a$ operator for the Higgsed $(0,1)$ scalar charge model in $d=2$.}
\label{fig:Scalar01d2}
\end{figure}

Whenever the parent $U(1)$ theory does not admit a magnetic field, its Higgs phase has a classical ground state. To see how this occurs, consider (for example) the $(0,1)$ scalar charge theory in $d=2$. Its Higgsed Hamiltonian is
\begin{equation}
H = -U \sum_{\text{sites}} a - h_p \sum_{\text{plaquettes}} X_{xy}
\end{equation}
where $a$ is a four-spin product of $X_{xy}$ operators shown in Fig. \ref{fig:Scalar01d2}. Since only $X$ operators appear in this theory, at all $h_s \neq 0$ the model has a unique, classical ground state consisting of all spins in the $X=+1$ eigenstate (assuming $U,h_p>0$).

Clearly whenever a theory fails to admit a magnetic field, the Hamiltonian will only involve $X$ operators and will be 
a sum of Gauss' Law operators and onsite $X_{ij}$ terms. In all such cases, the model is trivial in the sense 
that it is fully classical and has a unique ground state.

The other theories in which this occurs are the $(1,0)$ and $(0,1)$ vector charge theories in both $d=2$ and $d=3$.

\subsection{Mappings to Ising Models}

Several Higgsed theories map onto standard models of spontaneous symmetry breaking. We now list each such theory.

\subsubsection{$d=2$ $(2r+1,2s)$ Scalar Charge}

\begin{figure}
\subfigure[]{
\includegraphics[width=7cm]{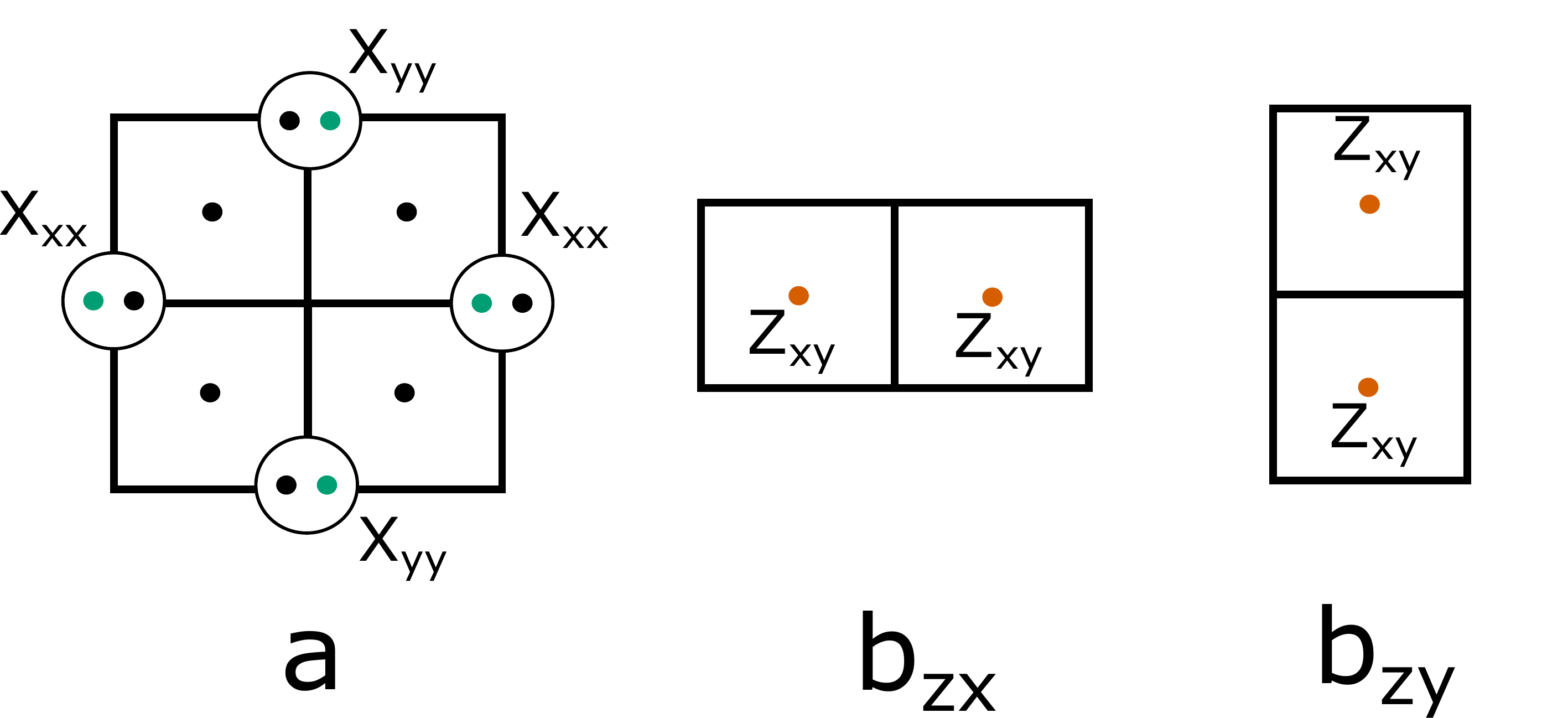}
\label{fig:2DScalarEHam}
}
\subfigure[]{
\includegraphics[width=4cm]{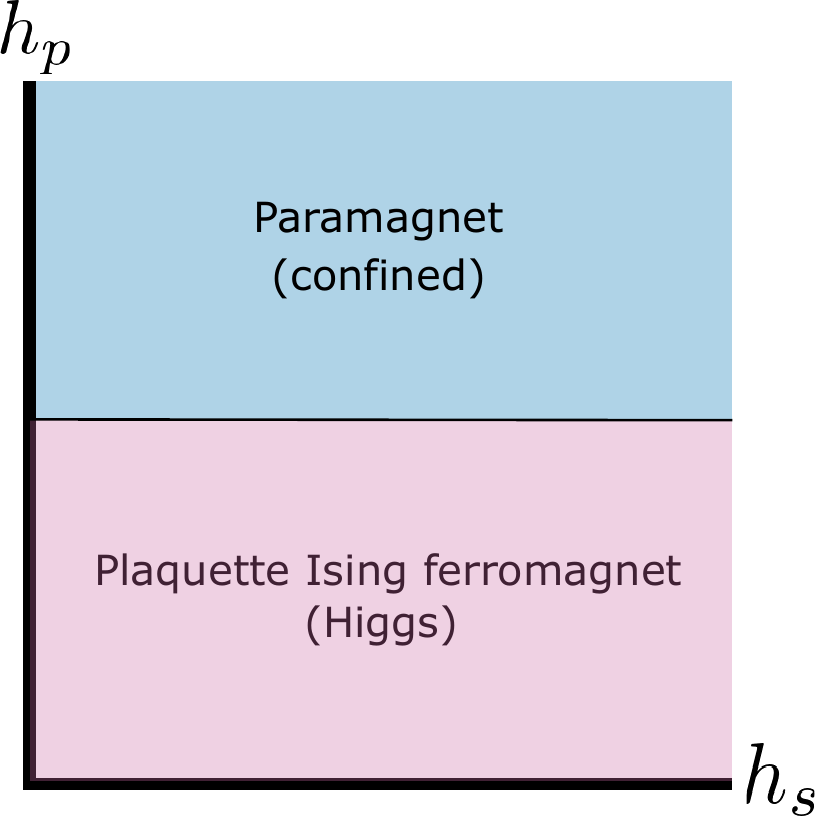}
\label{fig:PhaseDiagram2DScalarE}
}
\caption{(a) Terms in the Hamiltonian and (b) phase diagram for the $d=2$ Higgsed $(2r+1,2s)$ scalar charge theory. 
The electric term $a$ is a four-spin interaction on the sites and the magnetic terms $b_{zi}$ are Ising interactions on the $i$-directed bonds.}
\end{figure}

The Higgsed $(2r+1,2s)$ scalar charge Hamiltonian has the same form Eq. \eqref{eqn:2DScalarHiggs} as the Higgsed $(2r+1,2s+1)$ scalar theory in $d=2$, but $a$ and $b_{zi}$ take new forms, shown in Fig. \ref{fig:2DScalarEHam}. 

We see that the site and plaquette spins have decoupled. Both sectors are topologically trivial. The plaquette spin Hamiltonian is, by inspection, the transverse field Ising model on the dual lattice. The site spins are trivial; they are classical and have a unique ground state at all $h_s \neq 0$. 

\subsubsection{$d=3$ $(2r+1,2s+2)$ Scalar Charge}

\begin{figure}
	\includegraphics[width=7cm]{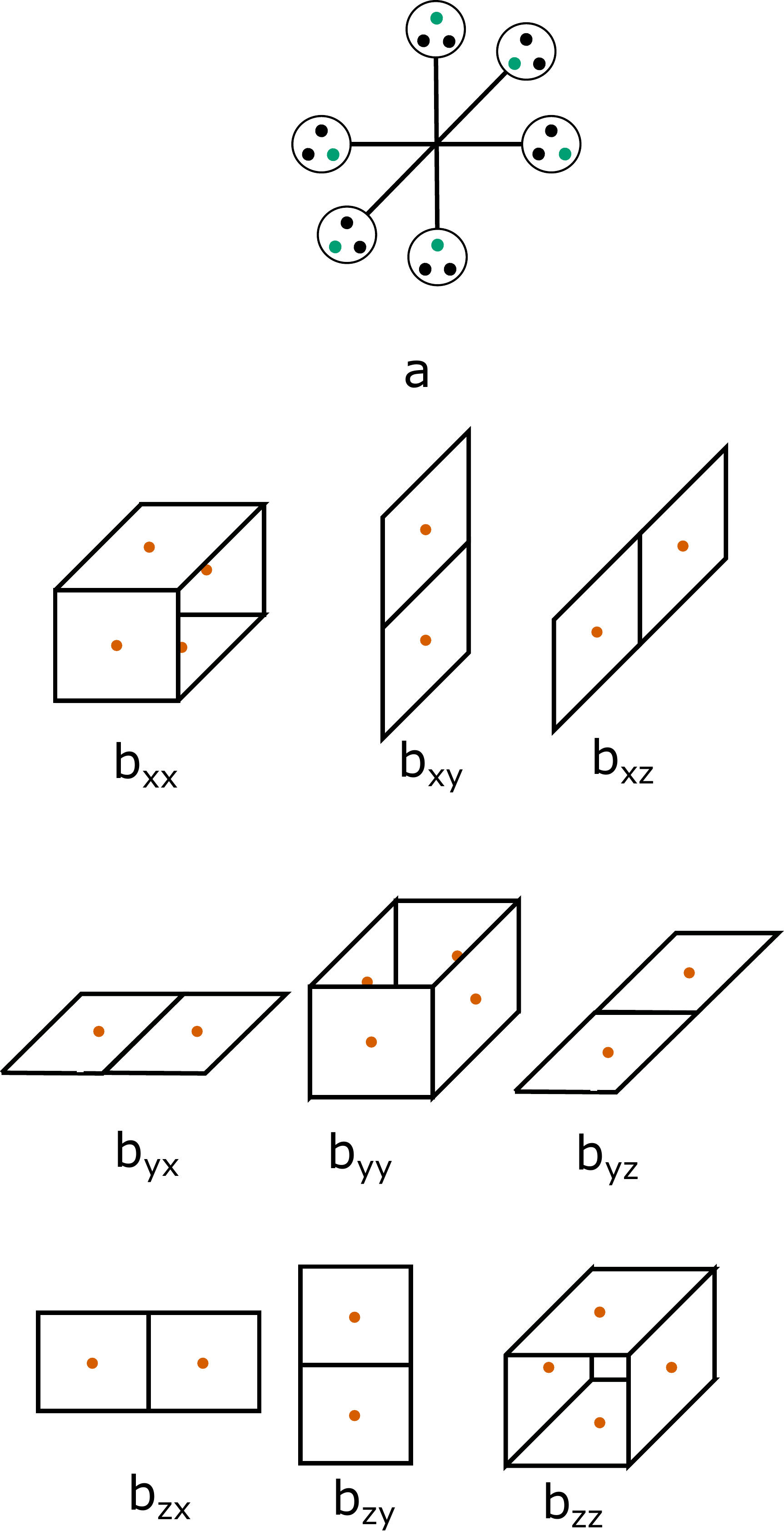}
\caption{Terms in the Hamiltonian of the Higgsed  $(2r+1,2s+2)$ scalar charge theory. The $a$ operators are associated with the sites and are products of six $X_{ii}$ operators acting on the green spins. The $b_{ij}$ operators are products of two (off-diagonal terms, associated with links) or four (diagonal terms, associated with cubes) Pauli $Z_{ij}$ operators acting on the orange spins.}
	\label{fig:ScalarEHam3D}
\end{figure}

The Hamiltonian of the $d=3$  $(2r+1,2s+2)$ scalar charge theory takes the same schematic form as the $d=3$  $(2r+1,2s+1)$ scalar charge theory Eq. \eqref{eqn:3DScalarHam} but with different operators, shown in Fig. \ref{fig:ScalarEHam3D}.

The site and face spins decouple and the site spins form a classical model with a unique ground state at $h_s \neq 0$. The face spins have a 3D transverse field Ising model Hamiltonian with additional four-body terms (the $b_{ii}$) which maintain the Ising symmetry. At $h_p=0$, Ising ferromagnetic states are favored by all the $b_{ij}$; the four-body terms simply modify details of the excitations. This Higgs phase is therefore stable to small $h_p$.

Although the Higgs phase is topologically trivial, this model does have a topologically nontrivial phase in its phase diagram, in the regime
$1/g_{ij}^2 \ll h_S \ll g_{ii}^2$, for $i \neq j$. To see this, we set $1/g_{ij}^2 = 0$ for $i \neq j$ and take $h_s \ll 1/g_{ii}^2$. The effect of 
$h_s$ is incorporated by degenerate perturbation theory; the resulting model is precisely the $d=3$ $\mathbb{Z}_2$ toric code Hamiltonian, which is stable to weak $1/g_{ij}^2 > 0$.

\subsection{Theories With Non-Universal Trivial Higgs Phases}
\label{subsec:illDefined}

As discussed in Sec. \ref{sec:intuition}, our operational definition of ``Higgs phase" is the regime $h_s,h_f \ll 1/g_{ij}^2$ for all $i,j$. As we saw in the $(4r+2,2s+1)$ vector charge model in $d=3$, examined in Sec. \ref{sec:evenOddVector_d3}, in this regime our treatment of the Higgs mechanism can produce a fine-tuned Hamiltonian. That is, there is non-topological degeneracy which is split by local operators which are not present in the Hamiltonian we derive but which are allowed by the residual $\mathbb{Z}_2$ gauge invariance. The precise ground state, including any possible patterns of global symmetry breaking, 
depends on what operators are added, but the topological order of any such ground state can still be determined.

 In this appendix, we list the theories which have this fine-tuned property but which have trivial topological order. Many of our arguments from Sec. \ref{sec:evenOddVector_d3} carry over to these theories.

\subsubsection{$d=2$ $(2r+2,2s+1)$ Scalar Charge}

\begin{figure}
\includegraphics[width=7cm]{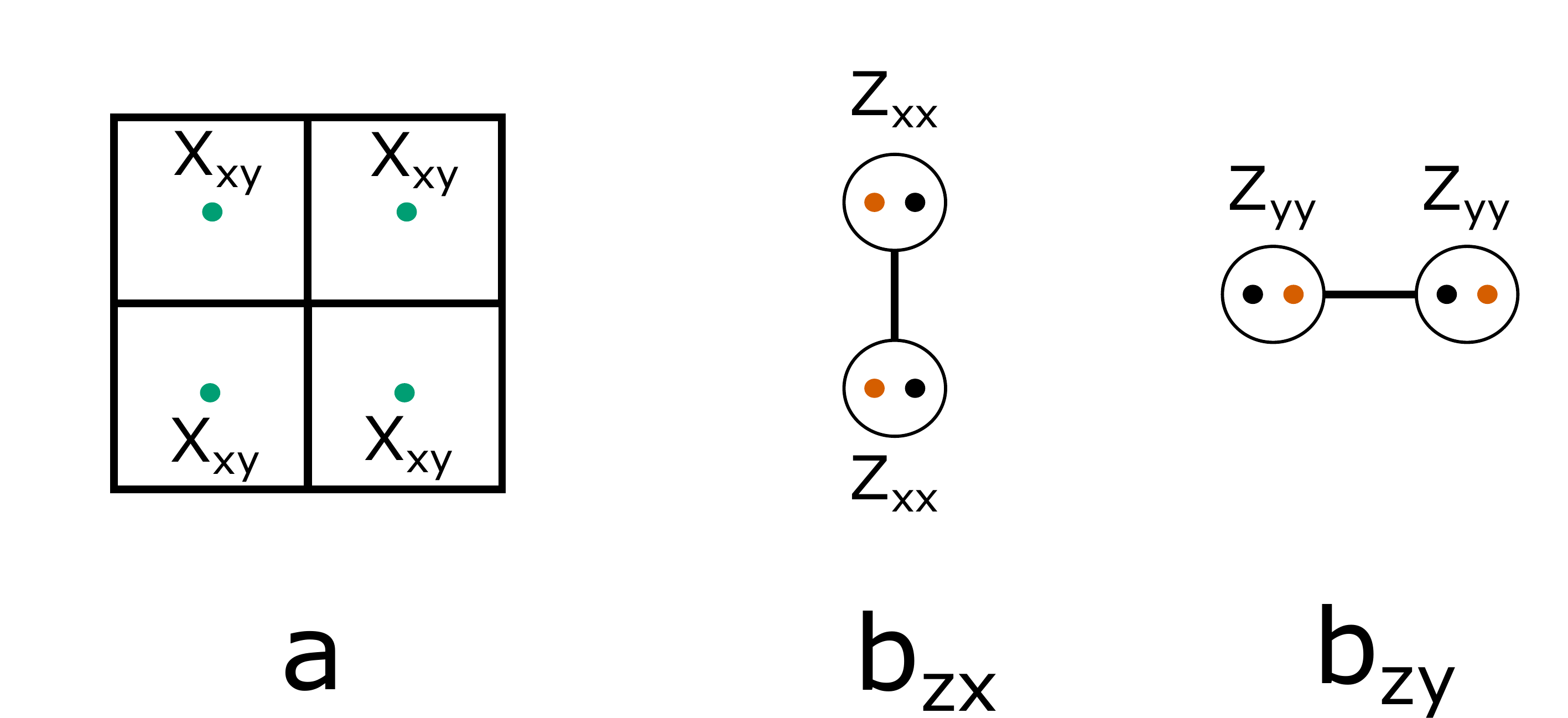}
\caption{Terms in the Hamiltonian and for the $d=2$ Higgsed $(2r+2,2s+1)$ scalar charge theory. The electric term $a$ is a four-spin interaction on the sites and the magnetic terms $b_{zi}$ are Ising interactions on the $i$-directed bonds.}
\label{fig:ScalarEvenOdd2DHam}
\end{figure}

The Hamiltonian has the same structure Eq. \eqref{eqn:2DScalarHiggs} as the other scalar charge models, but with $a$ and $b_{zi}$ as shown in Fig. \ref{fig:ScalarEvenOdd2DHam}. The site and plaquette spins decouple. The plaquette spins are classical and have a unique ground state at all $h_p \neq 0$. The site spin sector consists of decoupled transverse field Ising chains, one for each row (involving only the $yy$ spins) and each column (involving only the $xx$ spins) of the lattice.

 At $h_s=0$, every $Z_{ii}$ commutes with the Hamiltonian, as do the strings $\prod_{x} X_{yy}(x,y_0)$ and $\prod_y X_{xx}(x_0,y)$ for each $x_0,y_0$. At $h_s \neq 0$, the strings still commute with the Hamiltonian, and $h_s$ only contributes at $L$th order in perturbation theory. Therefore, $h_s$ only splits the degeneracy by an amount exponentially small in the system size; the same argument holds for any local operator involving $X_{ii}$ operators.  Hence only products of $Z_{ii}$ operators can split the degeneracy in the thermodynamic limit, but adding any such term to the $h_s=0$ model simply chooses some particular ground state for each Ising chain. Different operators can lead to different ground states. If any degeneracy remains, it still can be split completely by adding $\sum_i Z_{ii}$ to the Hamiltonian, so no topological order can be present; therefore, any possible ground state is topologically trivial.

\subsubsection{$d=2$ $(2r+1,2s+2)$ Vector Charge}

\begin{figure}
	\includegraphics[width=7cm]{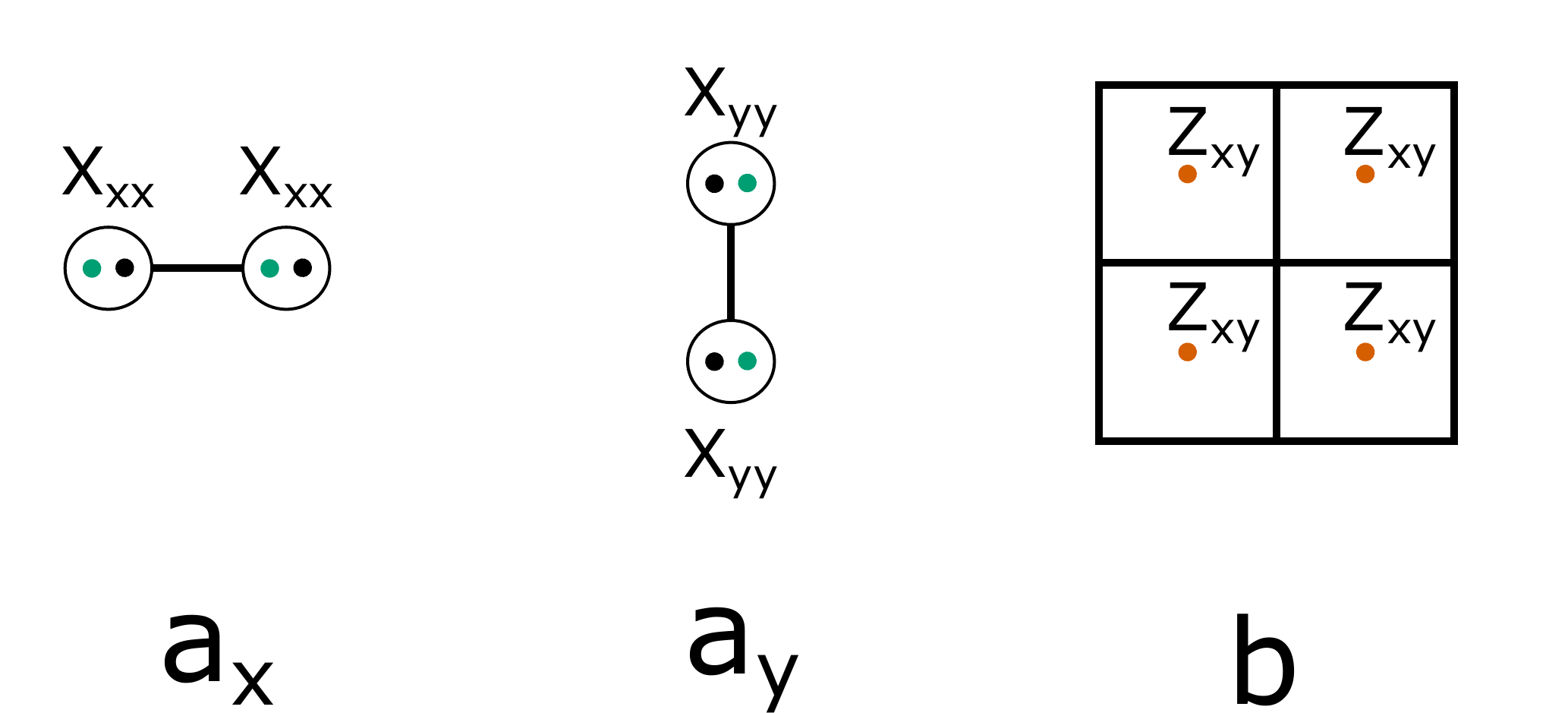}
\caption{ Terms in the Hamiltonian of the $d=2$ Higgsed $(2r+1,2s+2$) vector charge theory Hamiltonian. $a_i$ are associated with the bonds and form Ising interactions acting on the green spins. The $b$ operator is associated with the center site and is a product of the indicated Pauli $Z$ operators on the four orange plaquette spins.}
	\label{fig:VectorOddEven2D}
\end{figure}

The Hamiltonian has the usual form Eq. \eqref{eqn:vectorA2DHam} for vector charge models, but with the forms of $a_i$ and $b$ are given in Fig. \ref{fig:VectorOddEven2D}.

Again the site and plaquette spins decouple. The site spins form decoupled classical Ising chains. At $h_p=0$, all the $Z_{xy}$ commute with the Hamiltonian, as do the string operators $\prod_{x}Z_{xy}(x,y_0)$ and $\prod_{y}Z_{xy}(x_0,y)$ for fixed positions $x_0$ and $y_0$. The $Z_{xy}$ and the string operators anticommute, so there is non-topological ground state degeneracy. Making $h_p$ nonzero but weak will only contribute at $L$th order in degenerate perturbation theory, so $h_p$ only splits the degeneracy by an exponentially small amount. Similarly to the $d=2$ $(2r+2,2s+1)$ scalar charge theory discussed above, the degeneracy is fine-tuned, as local operators can split it. For the same reasons as in the $d=2$ $(2r+2,2s+1)$ scalar charge model discussed above, the resulting ground state depends on what operators are chosen, but said state will always be topologically trivial.

\subsubsection{$d=3$ $(4r+4,2s+1)$ Vector Charge}

\begin{figure}
	\includegraphics[width=6.5cm]{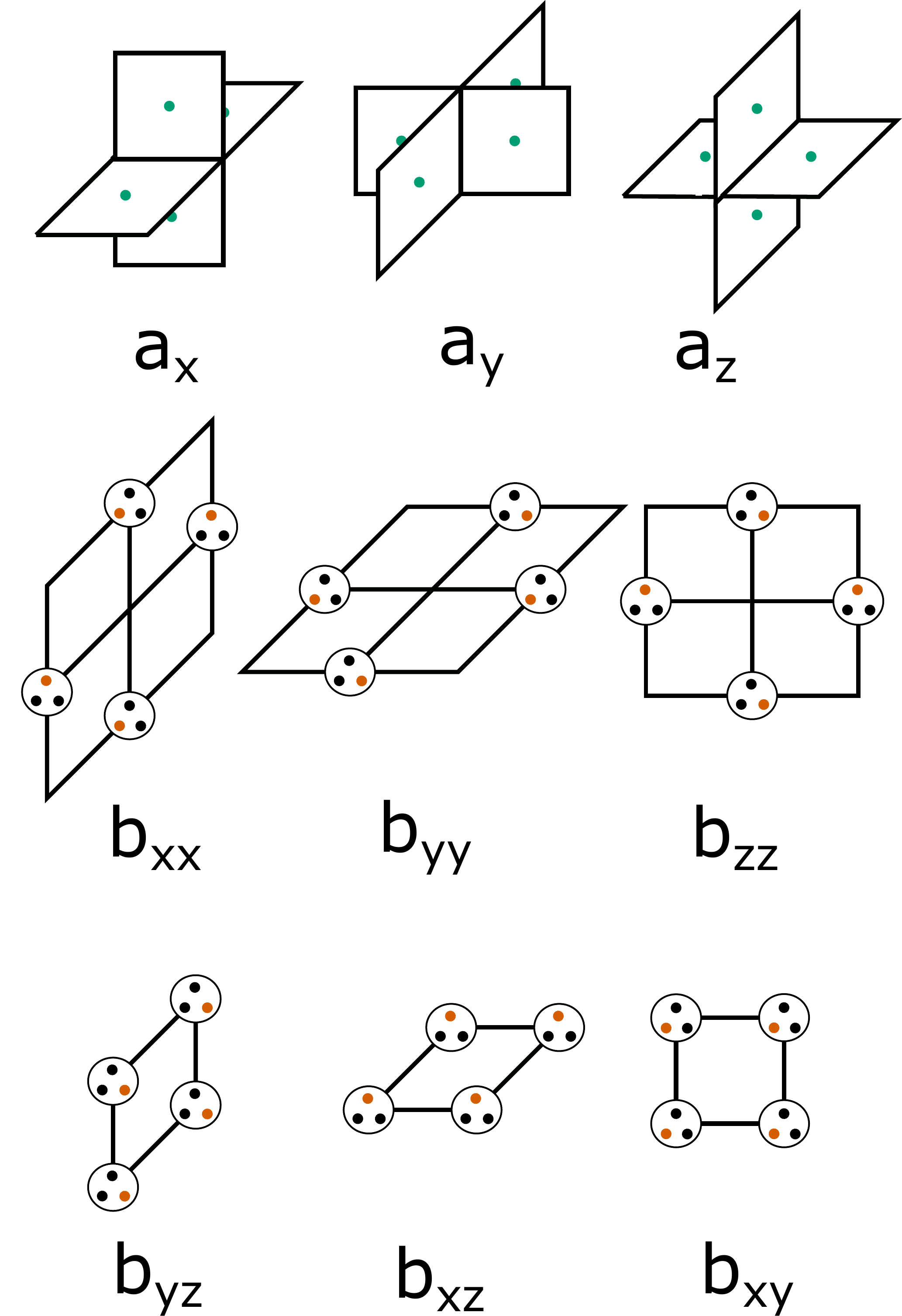}

\caption{Terms in the Hamiltonian of the Higgsed  $(4r+4,2s+1)$ vector charge theory. The $a$ operators are associated with the links and are products of four $X_{ij}$ operators acting on the green spins. The $b_{ij}$ operators are products of four Pauli $Z_{ij}$ operators acting on the orange spins.}
	\label{fig:Vector41Ham3D}
\end{figure}

The Hamiltonian of the $d=3$ $(4r+4,2s+1)$ vector charge theory takes the same schematic form as the $d=3$ $(2r+1,2s+1)$ vector charge theory Eq. \eqref{eqn:OddOddVectorHam3D} but with different operators, shown in Fig. \ref{fig:Vector41Ham3D}.

The sites and faces decouple. The face sector is a classical model with a unique ground state. 

In the site sector at $h_s=0$, obviously every $Z_{ii}$ commutes with the Hamiltonian. There are also $\mathcal{O}(L^2)$ independent string operators which commute with the Hamiltonian and anticommute with the $Z_{ii}$; an example is $\prod_y X_{xx}(x_0+1,y,z_0) X_{xx}(x_0+1,y,z_0+2)X_{zz}(x_0,y,z_0+1)X_{zz}(x_0+2,y,z_0+1)$ where $x_0$ and $z_0$ are fixed positions. Rotations of this operator are also valid. These string operators are the lowest-order operators that enter in the effective Hamiltonian when $h_s$ is taken weak but nonzero; they appear at order $4L$ in degenerate perturbation theory. Therefore, $h_s$ nonzero only splits the ground state degeneracy by an exponentially small amount, but there are local operators (e.g. $Z_{ii}$) which commute with the $a_i$ and split the degeneracy at first order in perturbation theory. For the same reasons as in the other models in this section, the resulting ground state depends on what local operators are chosen, but said state will always be topologically trivial.

Much like the $d=3$  $(2r+1,2s+2)$ scalar charge model, there is also nontrivial topological order in the phase diagram when $1/g_{ij}^2 \ll h_s \ll 1/g_{ii}^2$. Using similar arguments, the effective model at $1/g_{ij}^2 = 0$ obtained from degenerate perturbation theory is the Hamiltonian Eq. \eqref{eqn:EightToricCodesHam}, which is eight decoupled copies of the toric code.

\subsubsection{$d=3$  $(2r+1,2s+2)$ Vector Charge}
\begin{figure}
	\includegraphics[width=6.5cm]{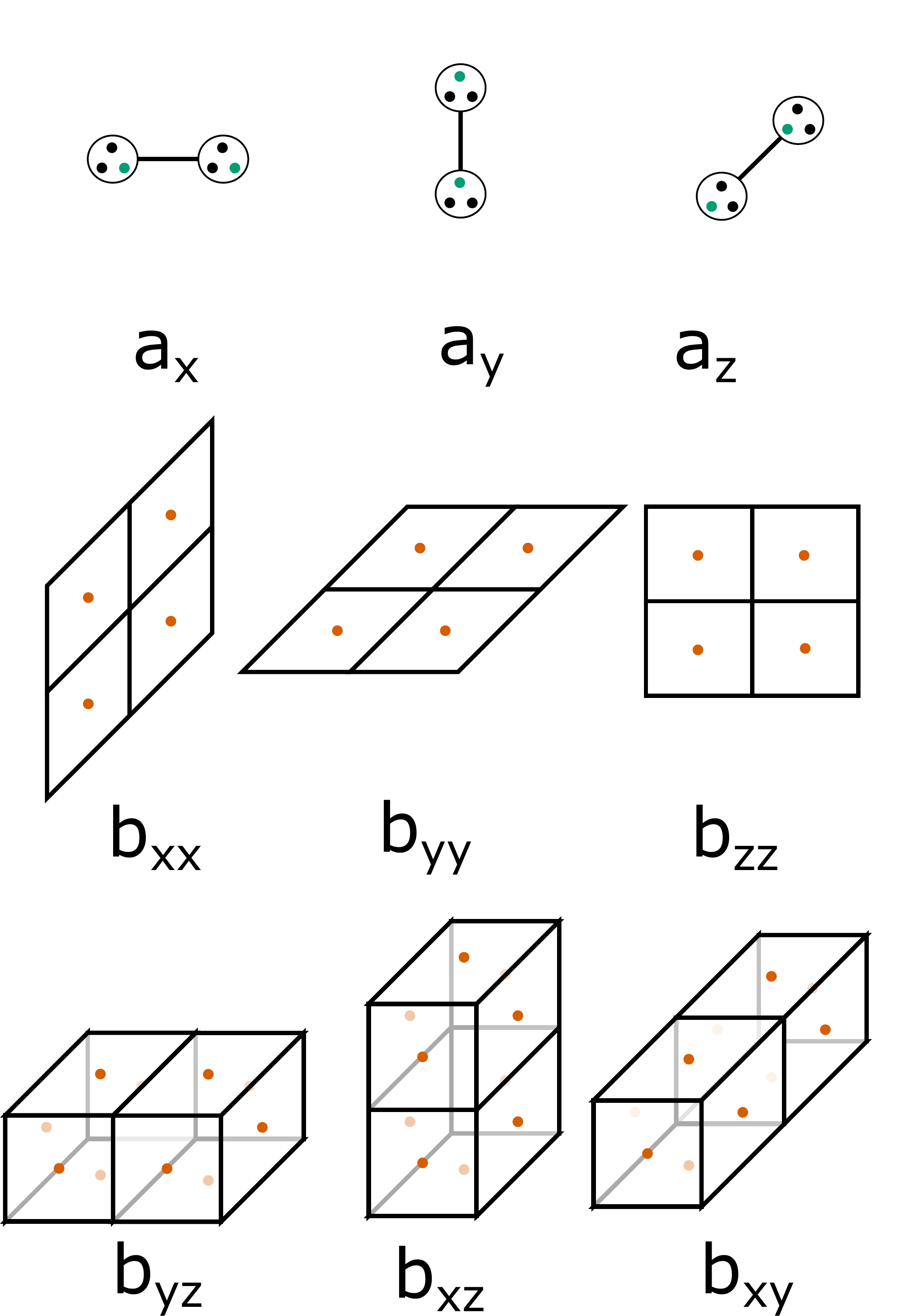}
\caption{Terms in the Hamiltonian of the Higgsed $(2r+1,2s+2)$ vector charge theory. The $a$ operators are associated with the links and are Ising-like products of $X_{ii}$ operators acting on the green spins. The $b_{ij}$ operators are products of ten (off-diagonal terms, associated with faces) or four (diagonal terms, associated with sites) Pauli $Z_{ij}$ operators acting on the orange spins.}
\label{fig:VectorOddEven3D}
\end{figure}

The Hamiltonian of the $d=3$ $(2r+1,2s+2)$ vector charge theory also takes the same schematic form as the $d=3$ $(2r+1,2s+1)$ vector charge theory Eq. \eqref{eqn:OddOddVectorHam3D} but with different operators, shown in Fig. \ref{fig:VectorOddEven3D}.

Again, the sites and faces decouple. The site sector has a unique classical ground state. The face sector is fine-tuned in a similar fashion to the Higgsed $d=3$ $(4r+4,2s+1)$ vector charge theory. This can be seen by examining the face sector at $h_f = 0$. In this limit, every $Z_{ij}$ commutes with the Hamiltonian, but there are also $\mathcal{O}(L^2)$ independent string operators which commute with the Hamiltonian but anticommute with various $Z_{ij}$. An example is the operator $\prod_{x}X_{xy}(x,y_0,z_0)X_{xz}(x,y_0,z_0)$, where $y_0$ and $z_0$ are fixed spatial positions. By standard degenerate perturbation theory arguments, setting $h_f$ small but nonzero only contributes at $2L$th order in perturbation theory, which leads to an exponentially small splitting of the degeneracy. As in the rest of this section, the resulting ground state depends on what local operators are chosen, but said state will always be topologically trivial.

\bibstyle{apsrev4-1} \bibliography{references}

\end{document}